\documentclass[twocolumn]{aastex63}

\accepted{for Publication in ApJ}

\usepackage{natbib,aas_macros,amsmath}
\usepackage{multirow,color}
\usepackage{natbib}

\newcommand{\carcsec}{$\!\!\arcsec$}
\newcommand{\m}[1]{\mathrm{#1}}
\newcommand{\bluec}[1]{\textcolor{blue}{#1}}
\newcommand{\redc}[1]{\textcolor{black}{#1}}

\usepackage{booktabs}

\begin{document}
\shortauthors{Harikane et al.}

\title{
JWST \& ALMA Joint Analysis with [OII]$\lambda\lambda$3726,3729, [OIII]$\lambda$4363, [OIII]88$\mu$m, and [OIII]52$\mu$m: \\
Multi-Zone Evolution of Electron Densities at $\mathbf{z\sim0-14}$ and Its Impact on Metallicity Measurements
}

\shorttitle{
JWST \& ALMA Joint Analysis at High Redshifts
}

\email{hari@icrr.u-tokyo.ac.jp}
\author[0000-0002-6047-430X]{Yuichi Harikane}
\affiliation{Institute for Cosmic Ray Research, The University of Tokyo, 5-1-5 Kashiwanoha, Kashiwa, Chiba 277-8582, Japan}

\author[0000-0003-4792-9119]{Ryan L. Sanders}
\affiliation{Department of Physics and Astronomy, University of Kentucky, 505 Rose Street, Lexington, KY 40506, USA}

\author[0000-0001-7782-7071]{Richard Ellis} 
\affiliation{Department of Physics \& Astronomy, University College London, Gower St., London WC1E 9BT, United Kingdom}

\author[0000-0001-5860-3419]{Tucker Jones} 
\affiliation{Department of Physics and Astronomy, University of California, Davis, 1 Shields Avenue, Davis, CA 95616, USA}

\author[0000-0002-1049-6658]{Masami Ouchi}
\affiliation{National Astronomical Observatory of Japan, 2-21-1 Osawa, Mitaka, Tokyo 181-8588, Japan}
\affiliation{Institute for Cosmic Ray Research, The University of Tokyo, 5-1-5 Kashiwanoha, Kashiwa, Chiba 277-8582, Japan}
\affiliation{Department of Astronomical Science, SOKENDAI (The Graduate University for Advanced Studies), Osawa 2-21-1, Mitaka, Tokyo, 181-8588, Japan}
\affiliation{Kavli Institute for the Physics and Mathematics of the Universe (WPI), University of Tokyo, Kashiwa, Chiba 277-8583, Japan}

\author[0000-0001-7459-6335]{Nicolas Laporte}
\affiliation{Aix-Marseille Universit\'e, CNRS, CNES, LAM (Laboratoire d’Astrophysique de Marseille), UMR 7326, 13388 Marseille, France}

\author[0000-0002-4140-1367]{Guido Roberts-Borsani} 
\affiliation{Department of Physics \& Astronomy, University College London, Gower St., London WC1E 6BT, UK}

\author[0000-0003-1561-3814]{Harley Katz} 
\affiliation{Department of Astronomy \& Astrophysics, University of Chicago, 5640 S Ellis Avenue, Chicago, IL 60637, USA}
\affiliation{Kavli Institute for Cosmological Physics, University of Chicago, Chicago, IL 60637, USA}

\author[0000-0003-2965-5070]{Kimihiko Nakajima}
\affiliation{Institute of Liberal Arts and Science, Kanazawa University, Kakuma-machi, Kanazawa, Ishikawa, 920-1192, Japan}

\author[0000-0001-9011-7605]{Yoshiaki Ono}
\affiliation{Institute for Cosmic Ray Research, The University of Tokyo, 5-1-5 Kashiwanoha, Kashiwa, Chiba 277-8582, Japan}

\author{Mansi Gupta}
\affiliation{Department of Physics and Astronomy, University of Kentucky, 505 Rose Street, Lexington, KY 40506, USA}

\begin{abstract}
We present a JWST and ALMA detailed study of the ISM properties of high-redshift galaxies. Our JWST/NIRSpec IFU spectroscopy targeting three galaxies at $z=6-7$ detects key rest-frame optical emission lines, allowing us to derive {\sc[Oii]}$\lambda\lambda$3726,3729-based electron densities of $n_\mathrm{e,optical}\sim1000$ cm$^{-3}$ on average and {\sc[Oiii]}$\lambda$4363-based metallicities of $\mathrm{12+log(O/H)}=8.0-8.2$ in two galaxies. New ALMA Band 9/10 observations detect the {\sc[Oiii]}52$\mu$m line in one galaxy but do not in the others, resulting in FIR-based densities of $n_\mathrm{e,FIR}\lesssim500$ cm$^{-3}$ from the {\sc[Oiii]}52$\mu$m/{\sc[Oiii]}88$\mu$m ratios, systematically lower than the optical {\sc[Oii]}-based measurements. These low FIR-based densities are comparable to those at both $z\sim0$ and $z>6$ in the literature, including JADES-GS-z14-0 at $z=14.18$, suggesting little evolution up to $z\sim14$, in contrast to the increasing trend of optical-based densities with redshift. By conducting a JWST and ALMA joint analysis using emission lines detected with both telescopes, we find that the observed FIR {\sc[Oiii]}52,88$\mu$m luminosities are too high to be explained by the optical-based densities at which they would be significantly collisionally de-excited. Instead, a 2-zone model with distinct high- and low-density regions is required to reproduce all observed lines, indicating that FIR {\sc[Oiii]} emission arises predominantly from low-density gas, while optical {\sc[Oiii]} and {\sc[Oii]} lines trace both regions. We further demonstrate that the direct-$T_\mathrm{e}$ method can sometimes significantly underestimate metallicities up to 0.8 dex due to the presence of the low-density gas not fully traced by optical lines alone, highlighting the importance of combining optical and FIR lines to accurately determine gas-phase metallicities in the early universe.
\end{abstract}


\section{Introduction}\label{ss_intro}

An understanding of galaxy assembly at high redshifts is essential to form a comprehensive picture of galaxy formation and evolution.  
The gas-phase metallicity (O/H) of the interstellar medium (ISM) is one of the most powerful tools for understanding galaxy growth because it is sensitive to the past and current star formation rate (SFR), the gas reservoir mass, and gaseous inflows and outflows (e.g., \citealt{1980FCPh....5..287T,2013ApJ...772..119L}).  
Considerable effort has been invested to characterizing the stellar mass-ISM metallicity relation and its evolution over the redshift range of $z=0-10$ (e.g., \citealt{2004ApJ...613..898T,2008A&A...488..463M,2013ApJ...765..140A,2021ApJ...914...19S,2023NatAs...7.1517H,2023ApJ...950...67M,2023ApJS..269...33N,2023MNRAS.518..425C,2024A&A...684A..75C,2023ApJ...957...39L,2024ApJ...962...24S}).  
These studies have been fundamental in establishing our current theoretical picture of galaxy growth and have placed key constraints on gas content and inflow/outflow rates at low and high redshifts.

The first JWST NIRSpec spectra from the Early Release Observations \citep{2022ApJ...936L..14P} surprisingly revealed clear detections of the auroral {\sc[Oiii]}$\lambda$4363 line in galaxies at redshifts up to $z=8.5$.  
These novel detections of the weak {\sc[Oiii]}$\lambda$4363 line, which is sensitive to the electron temperature $T_\m{e}$, were enabled by the transformational sensitivity of JWST/NIRSpec, allowing the use of the temperature-based direct method to derive metallicities of galaxies in the early universe.  
This direct-temperature (direct-$T_\m{e}$) method is widely regarded as a robust technique for inferring galaxy gas-phase metallicities, significantly more accurate than the ``strong-line method," which relies on calibrations between metallicity and flux ratios of the brightest lines.  
Many existing strong-line calibrations, based on samples of local galaxies and {\sc Hii} regions \citep[e.g.,][]{2004MNRAS.348L..59P}, are not valid for high-redshift galaxies because of the significant evolution of ISM ionization conditions and global galaxy properties between $z\sim0$ and $z>2$ \citep[e.g.,][]{2018ApJ...859..175B,2020MNRAS.491.1427S}.  
Consequently, the direct-$T_\m{e}$ method is an essential tool for robustly measuring the metallicity of galaxies at $z\gtrsim6$, and indeed, many JWST studies have measured metallicities of high-redshift galaxies using {\sc[Oiii]}$\lambda$4363 and discussed the redshift evolution of the mass-metallicity relation \citep[e.g.,][]{2023ApJ...942L..14R,2022A&A...665L...4S,2023MNRAS.525.2087B,2023ApJ...945...35T,2023MNRAS.518..592K,2023MNRAS.518..425C,2023MNRAS.526.1657T,2024MNRAS.529.3301T,2024MNRAS.529.4087T,2025arXiv250810099S}.

The electron density of the H{\sc ii} region ($n_\m{e}$) is also a key quantity in describing the physical state of the ISM.  
Many studies have measured the electron densities of galaxies at both low and high redshifts using emission-line doublets such as {\sc[Oii]}$\lambda$3729/{\sc[Oii]}$\lambda$3726, {\sc [Sii]}$\lambda$6716/{\sc [Sii]}$\lambda$6731, {\sc [Ciii]}$\lambda$1907/{\sc Ciii]}$\lambda$1909, and {\sc[Oiii]}52$\mu$m/{\sc[Oiii]}88$\mu$m, whose flux ratios depend on the electron density due to the different critical densities of the doublet lines \citep[e.g.,][]{2006agna.book.....O}.  
These studies have revealed redshift evolution, with increasing densities toward higher redshifts \citep[e.g.,][]{2023ApJ...956..139I,2025arXiv250208712T}.  
Some studies have also investigated correlations between electron density and galaxy properties \citep[e.g.,][]{2015MNRAS.451.1284S,2023ApJ...952..167R}.  
Although the densities derived with the optical {\sc[Oii]} and {\sc[Sii]} lines agree well with each other \citep[e.g.,][]{2016ApJ...816...23S}, there is a trend that densities measured with the rest-frame UV {\sc Ciii]} lines are systematically higher than those measured with optical lines \citep[e.g.,][]{2022ApJ...939..110M,2025arXiv250208712T}.  
These results indicate that the ISM structure is not homogeneous and that a multi-zone ISM model must be considered to fully understand the properties of galaxies \citep[e.g.,][]{2021ApJ...922..170B}.

The direct-$T_\m{e}$ method mentioned above usually assumes a homogeneous ISM to measure electron temperatures and metallicities. 
However, {\sc Hii} regions are not homogeneous, and as a result the $T_\m{e}$ method may be biased. 
Temperature fluctuations are often invoked as an explanation of the so-called abundance discrepancy factor (ADF), whereby O/H measured from recombination lines is found to be higher by $\sim0.2-0.3$ dex compared to the $T_\m{e}$ method \citep[e.g.,][]{2017PASP..129h2001P}.
While the cause of the ADF is not firmly established \citep[e.g.,][]{2025arXiv250408933C,2007A&A...471..193S,2023NatAs...7..771C}, it is thought to arise from inhomogeneities.
For galaxy-integrated measurements at high redshift, the ISM structure presumably has a similar effect on temperature and metallicity estimates obtained using the direct-$T_\m{e}$ method \citep[e.g.,][]{2023MNRAS.522L..89C}, but there are no observational studies addressing the robustness of this method.

\begin{deluxetable*}{lccccccccccc}
\tablecaption{Summary of Our JWST and ALMA Observations}
\label{tab_obs}
\tablehead{\colhead{Name} & \colhead{R.A.} & \colhead{Decl.} & \colhead{$z_\m{spec}$}  & \colhead{Telescope/Instrument} & \colhead{Filter/Grating/Band} & \colhead{Exposure Time} 
} 
\startdata
J0217-0208 & 02:17:21.60 & $-$02:08:52.7 & $6.204$ & JWST/NIRCam  & F150W,F200W,F300M,F410M & 0.2 hours\\
& & & & JWST/NIRSpec IFU & G235H/F170LP & 4.6 hours\\
 & & & & & G395H/F290LP & 1.2 hours\\
& & & & ALMA & Band 10 & 7.3 hours\\
SXDF-NB1006-2 & 02:18:56.54 & $-$05:19:58.9 & $7.212$ & JWST/NIRCam  & F150W,F200W,F360M,F430M & 0.6 hours\\
& & & & JWST/NIRSpec IFU & G395H/F290LP & 2.5 hours\\
& & & & & G395M/F290LP & 0.3 hours \\
& & & & ALMA & Band 9 & 17.7 hours\\
J1211-0118 & 12:11:37.09 & $-$01:18:16.5 & $6.031$ & JWST/NIRCam  & F150W,F200W,F300M,F410M & 0.2 hours\\
 & & & & JWST/NIRSpec IFU & G235H/F170LP & 3.9 hours \\
& & & & & G395H/F290LP & 0.7 hours\\
& & & & ALMA & Band 10 & 12.8 hours\\
\enddata
\end{deluxetable*}

In this study, we investigate the ISM properties of high-redshift galaxies in detail using optical and far-infrared (FIR) emission lines detected with JWST and ALMA, respectively.
In particular, FIR fine-structure lines such as {\sc[Oiii]}88$\mu$m and {\sc[Oiii]}52$\mu$m observed with ALMA provide critical advantages: they are relatively insensitive to the electron temperature ($T_e$), and to first order their luminosities scale with the gas-phase metallicity, making them powerful tools for metallicity measurements.
Moreover, due to their low critical densities, these FIR lines predominantly trace low-density ionized gas, offering a complementary view to that from rest-frame optical lines such as {\sc[Oiii]}$\lambda$4363, which are highly $T_e$-sensitive and typically probe higher-density regions \citep{2020ApJ...903..150J,2020MNRAS.499.3417Y,2021MNRAS.504..723Y,2023ApJ...953..140N}.
We aim to investigate the impact of assuming a 1-zone ISM model on the derived ISM properties by first relaxing the single-density criterion and evaluating the effects on measurements such as metallicity and the mass-metallicity relation.  
This JWST-ALMA partnership probing multi-wavelength emission lines provides an opportunity to explore the multi-zone ISM structure and to test the robustness of the direct-$T_\m{e}$ method, which is widely regarded as the ``gold standard" for high-redshift metallicity studies.
\redc{Similar scientific goals have been independently explored for a $z\sim6.8$  galaxy using JWST and ALMA 
\citep{2025ApJ...991L..38U}.}

This paper is organized as follows.  
We describe our galaxy sample and JWST and ALMA observational datasets in Section \ref{ss_data}.  
Section \ref{ss_sed} describes the SED fitting, and Section \ref{ss_ism} presents the basic ISM properties derived from commonly used optical and FIR line diagnostics.  
In Section \ref{ss_jwst_alma}, we conduct a JWST and ALMA joint analysis using emission lines detected with both telescopes.  
In Section \ref{ss_dis}, we discuss the ISM structure of high-redshift galaxies and metallicity estimators using optical and FIR lines.  
Section \ref{ss_summary} summarizes our findings.  
Throughout this paper, we adopt the Planck cosmological parameter sets from the TT, TE, EE+lowP+lensing+BAO results \citep{2020A&A...641A...6P}: $\Omega_\m{m}=0.3111$, $\Omega_\Lambda=0.6899$, $\Omega_\m{b}=0.0489$, $h=0.6766$, and $\sigma_8=0.8102$.  
All magnitudes are given in the AB system \citep{1983ApJ...266..713O}.  
We define the solar metallicity as $\m{12+log(O/H)}=8.69$ and $Z_\odot=0.0142$ \citep{2009ARA&A..47..481A}.

\section{Galaxy Sample and Observational Dataset}\label{ss_data}

\subsection{Galaxy Sample}\label{ss_sample}

As targets for our JWST and ALMA observations, we selected three galaxies at $z=6-7$, J0217-0208 at $z_\m{spec}=6.204$, SXDF-NB1006-2 at $z_\m{spec}=7.212$, and J1211-0118 at $z_\m{spec}=6.031$ (Table \ref{tab_obs}).  
This sample was chosen based on existing {\sc[Oiii]}88$\mu$m detections from ALMA, as well as their compact morphologies in the ground-based Subaru/Hyper Suprime-Cam (HSC; \citealt{2018PASJ...70S...1M}) images obtained from the HSC-Subaru Strategic Program (HSC-SSP) survey \citep{2018PASJ...70S...4A}, allowing us to conduct a JWST and ALMA joint analysis as simply as possible, based on the total line fluxes rather than a spatially resolved analysis.  
J0217-0208 and J1211-0118 were spectroscopically confirmed in \citet{2018PASJ...70S..35M} via their Ly$\alpha$ emission line and the Lyman break feature, respectively, with Gran Telescopio Canarias (GTC)/Optical System for Imaging and low-Intermediate-Resolution Integrated Spectroscopy (OSIRIS; \citealt{2000SPIE.4008..623C}) and Subaru/Faint Object Camera and Spectrograph (FOCAS; \citealt{2002PASJ...54..819K}), and their {\sc[Oiii]}88$\mu$m emission lines were detected with ALMA in \citet{2020ApJ...896...93H}.  
SXDF-NB1006-2 was spectroscopically confirmed via its Ly$\alpha$ emission line with Subaru/FOCAS in \citet{2012ApJ...752..114S}, and its {\sc[Oiii]}88$\mu$m emission line was reported in \citet{2016Sci...352.1559I} and \citet{2023ApJ...945...69R}.

\subsection{JWST Datasets}

We conducted JWST NIRCam \citep{2023PASP..135b8001R} and NIRSpec/IFU \citep{2022A&A...661A..80J} observations targeting the three galaxies (J0217-0208, SXDF-NB1006-2, and J1211-0118) in GO-1657 (PIs: Y. Harikane \& R. Sanders).  
The details of the observations are described below.

\begin{figure*}
\centering
\begin{minipage}{0.9\hsize}
\includegraphics[width=0.99\hsize, bb=0 0 720 165]{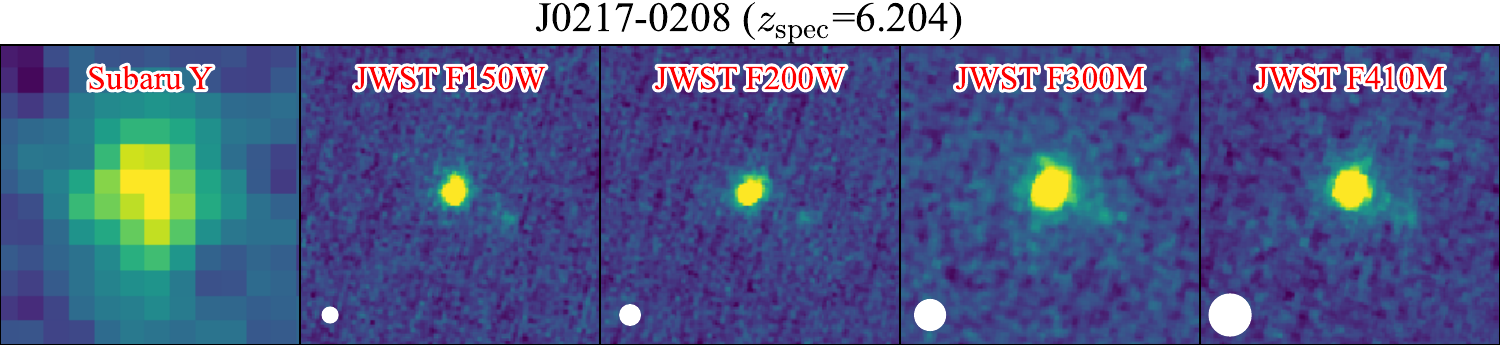}
\includegraphics[width=0.99\hsize, bb=0 0 720 170]{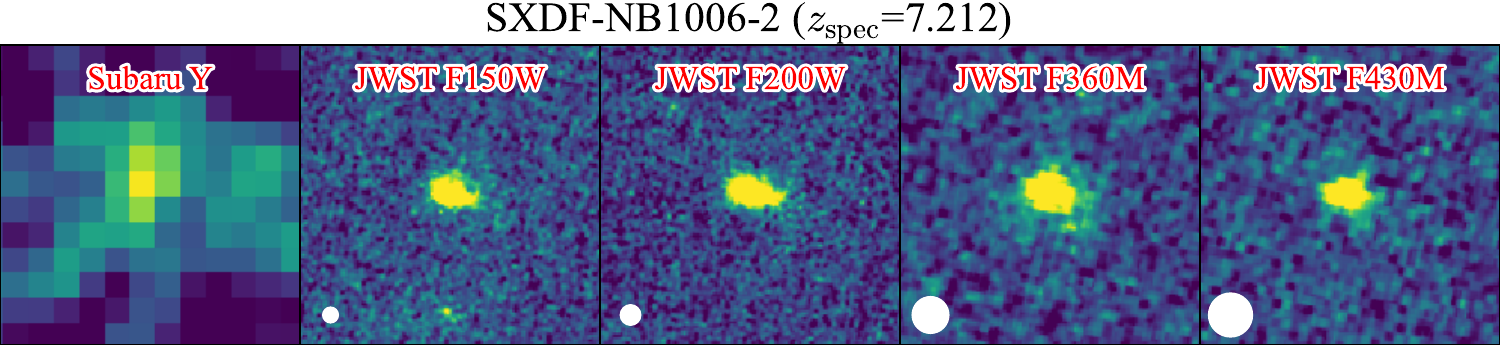}
\includegraphics[width=0.99\hsize, bb=0 0 720 170]{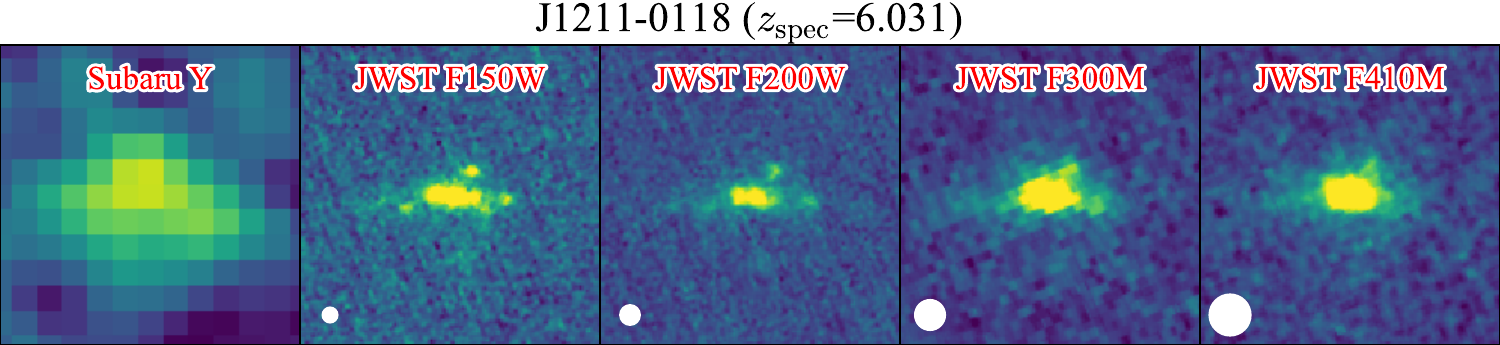}
\end{minipage}
\caption{
$2\arcsec \times 2\arcsec$ cutout images of J0217-0208 (top panel), SXDF-NB1006-2 (middle panel), and J1211-0118 (bottom panel).
The leftmost panels display Subaru/HSC Y-band images, while the other four panels show JWST/NIRCam images obtained in this study.
\redc{The white circles in the JWST cutouts show the PSF FWHMs.}
J1211-0118 is resolved into multiple clumps in the JWST/F150W image whose \redc{PSF} FWHM is $\sim0.\carcsec07$.
}
\label{fig_snap_hscjwst}
\end{figure*}

\subsubsection{NIRCam Image}

NIRCam images of the three galaxies were taken to investigate their morphologies as well as stellar populations.  
In our program, we chose four filters, F150W, F200W, F300M, and F410M (F150W, F200W, F360M, and F430M) for J0217-0208 and J1211-0118 (for SXDF-NB1006-2) to probe the stellar continuum in bandpasses that avoid strong emission lines, and further to provides measurements on both sides of the Balmer and 4000 $\m{\AA}$ breaks, such that SED fitting will yield robust stellar population parameters.  
The total exposure time for each filter was 10 (35) minutes for J0217-0208 and J1211-0118 (for SXDF-NB1006-2).  
In addition to the NIRCam data obtained in our program, we also use NIRCam images obtained in GO-1840 (PIs: J. Alvarez-Marquez \& T. Hashimoto).  
These images were taken with the six filters F115W, F150W, F200W, F250M, F300M, and F444W (F115W, F150W, F200W, F277W, F356W, and F444W) for J0217-0208 and J1211-0118 (for SXDF-NB1006-2).  
We reduced these data using the JWST Calibration Pipeline (versions 1.12.5) and the Calibration Reference Data System context file {\tt jwst\_1193.pmap} with custom modifications described in \citet{2023ApJS..265....5H}.

Figure \ref{fig_snap_hscjwst} shows the comparison of the Subaru/HSC Y-band images and JWST/NIRCam images obtained in this study.  
None of the three galaxies are spatially resolved in the Subaru/HSC images, but the high-resolution JWST images allow us to investigate their morphologies.  
J0217-0208 and SXDF-NB1006-2 do not show clumpy morphologies in either the rest-frame UV or optical images.  
In contrast, J1211-0118 is resolved into multiple sub-components, similar to UV-bright galaxies at $z\sim7$ reported in \citet{2025ApJ...980..138H}.

\begin{figure*}
\centering
\begin{minipage}{0.99\hsize}
\includegraphics[width=0.20\hsize, bb=6 -50 240 251]{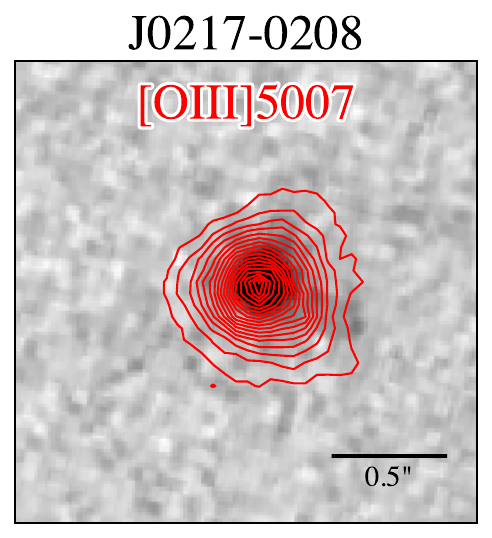}
\includegraphics[width=0.79\hsize, bb=13 0 848 281]{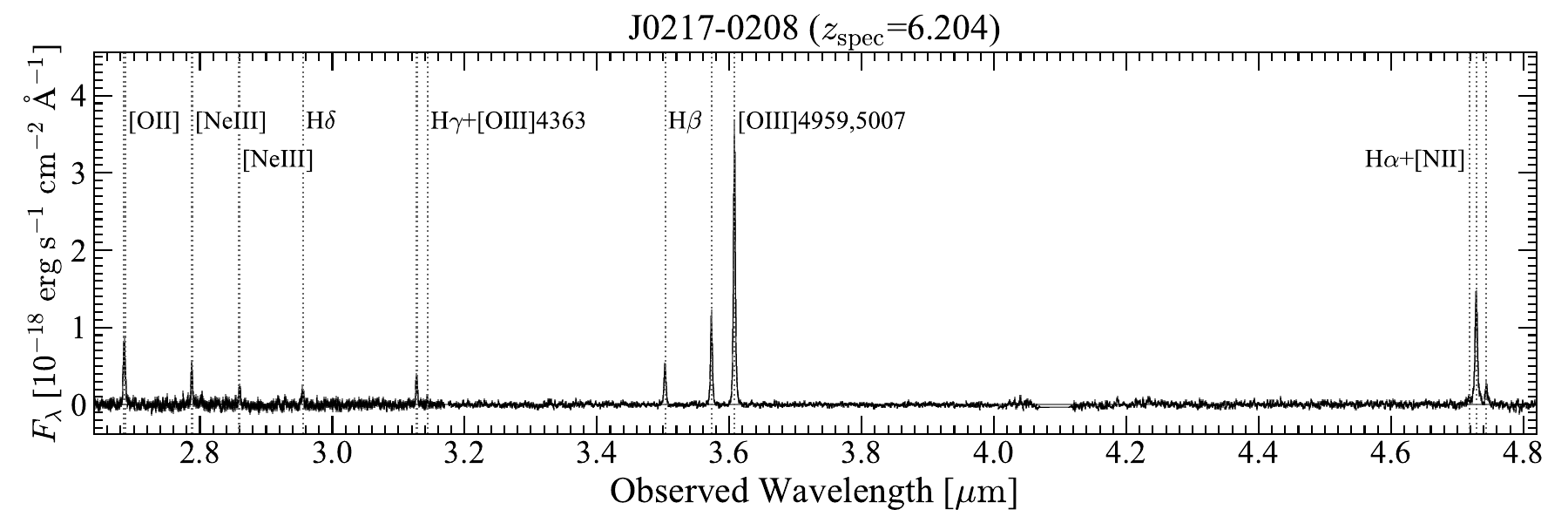}
\end{minipage}
\begin{minipage}{0.99\hsize}
\includegraphics[width=0.20\hsize, bb=6 -50 240 251]{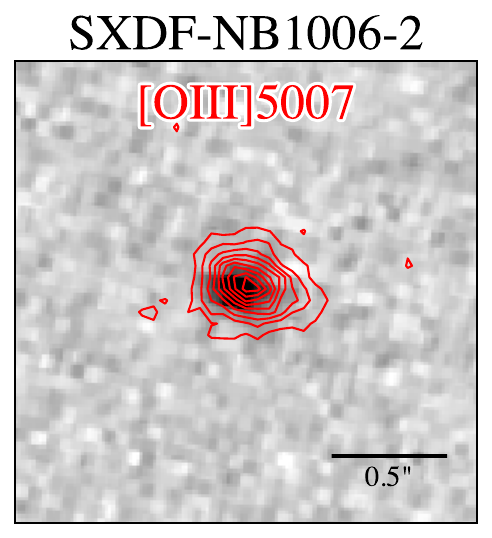}
\includegraphics[width=0.79\hsize, bb=13 0 848 281]{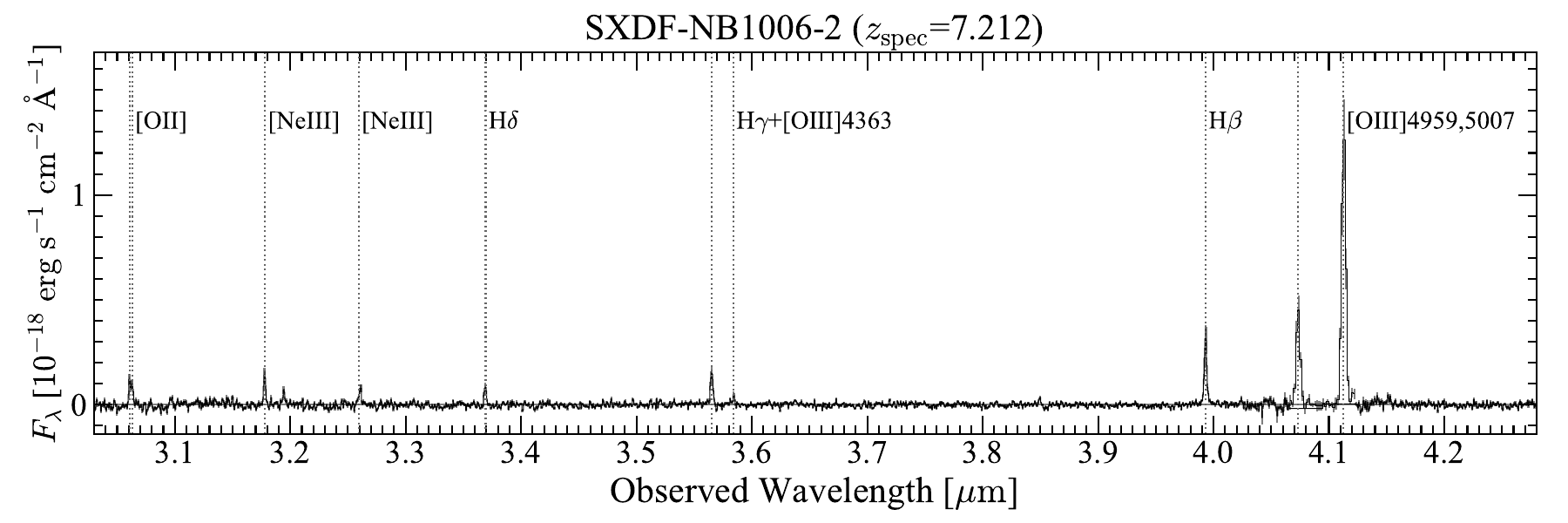}
\end{minipage}
\begin{minipage}{0.99\hsize}
\includegraphics[width=0.20\hsize, bb=6 -50 240 251]{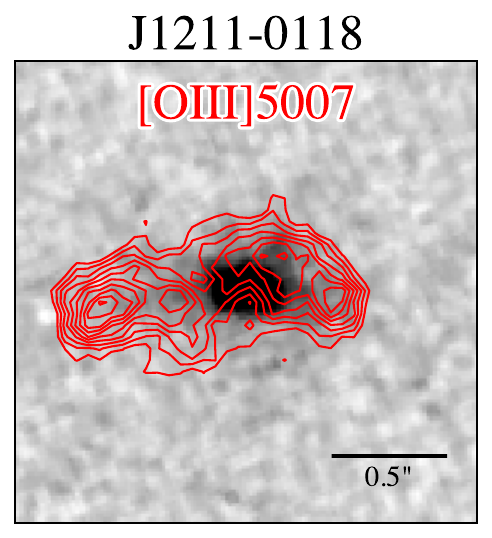}
\includegraphics[width=0.79\hsize, bb=13 0 848 281]{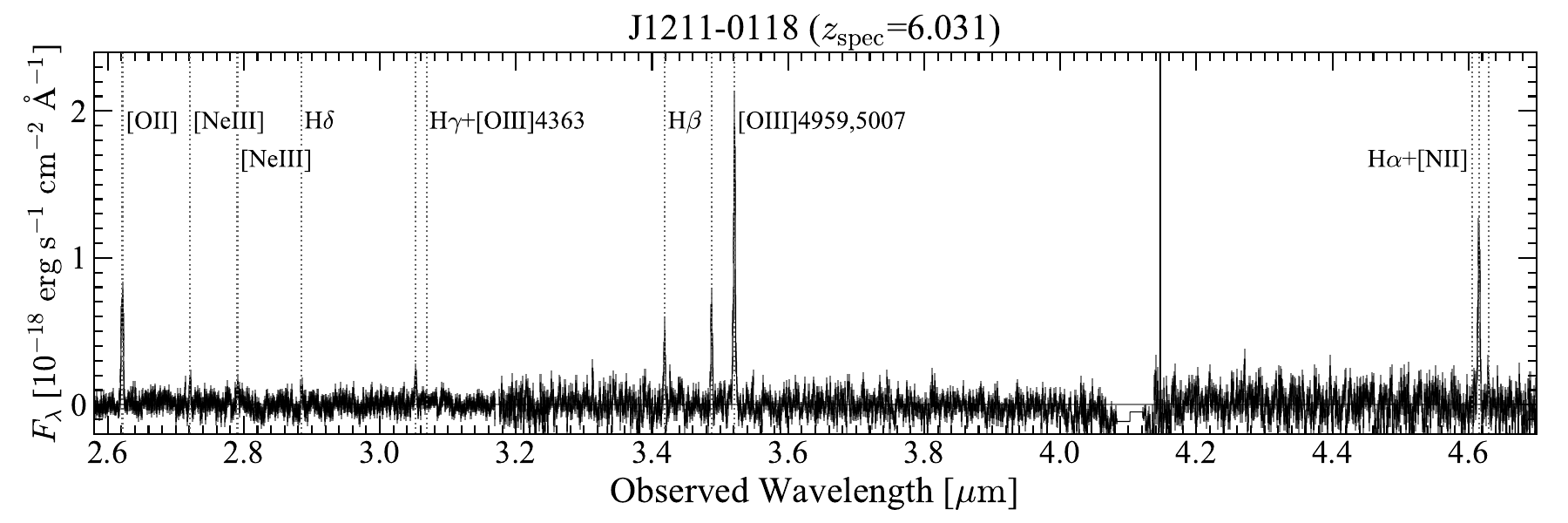}
\end{minipage}
\caption{
JWST/NIRSpec IFU data for J0217-0208 (top panels), SXDF-NB1006-2 (middle panels), and J1211-0118 (bottom panels) obtained in this study.
The left panels show the $2\arcsec \times 2\arcsec$ {[\sc Oiii]$\lambda$5007} map integrated over 300 km s$^{\redc{-1}}$.
The contours are drawn from $\pm5\sigma$ ($\pm3\sigma$) with an interval of $4\sigma$ ($1\sigma$) for J0217-0208 and SXDF-NB1006-2 (J1211-0118).
The background image is the rest-frame UV (JWST/NIRCam F150W) image.
The right panels display the continuum-subtracted spectra, with the observed wavelengths of several emission lines indicated by the gray dotted lines.
}
\label{fig_spec_jwst}
\end{figure*}

\subsubsection{NIRSpec Spectrum}\label{ss_nirspec}

NIRSpec IFU spectroscopic observations of the three galaxies were designed to detect key rest-frame optical emission lines necessary for standard gas-phase metallicity diagnostics, including {\sc[Oii]}$\lambda\lambda$3726,3729, H$\gamma$, H$\beta$, {\sc[Oiii]}$\lambda$4363, {\sc[Oiii]}$\lambda\lambda$4959,5007, and H$\alpha$, where accessible.
To resolve the {\sc[Oii]}$\lambda\lambda$3726,3729 doublet and enable reliable electron density measurements, high-resolution gratings were employed for the {\sc[Oii]} observations.
For J0217$-$0208 and J1211$-$0118 at $z=6.0-6.2$, we used the grating-filter pairs G235H/F170LP for {\sc[Oii]}$\lambda\lambda$3726,3729 and G395H/F290LP for H$\beta$, {\sc[Oiii]}$\lambda\lambda$4959,5007, and H$\alpha$.
For SXDF-NB1006-2 at $z=7.2$, we used G395H/F290LP to cover {\sc[Oii]}$\lambda\lambda$3726,3729 and H$\beta$, and additionally employed G395M/F290LP to fill the wavelength gap present in the G395H/F290LP coverage and observe {\sc[Oiii]}$\lambda\lambda$4959,5007.

We reduced the NIRSpec IFU data using the JWST Calibration Pipeline (versions 1.12.5) and the Calibration Reference Data System context file of {\tt jwst\_1189.pmap} with several modifications \citep[see also][]{2024ApJ...976..142X}.  
We downloaded raw data from the MAST archive and ran Stage 1 calibration with {\tt calwebb\_detector1}.  
The resulting count-rate frames were corrected for 1/f noise using NSClean \citep{2024PASP..136a5001R}.  
We then ran Stage 2 calibration with {\tt calwebb\_spec2} and produced the 3D data cube for each exposure with a pixel scale of $0.\carcsec05$.  
Following the procedure used by the Q3D team \citep[e.g.,][]{2022ApJ...940L...7W,2023ApJ...955...92V,2023ApJ...953...56V}, we used the {\sf reproject} package \citep{2023zndo...7584411R} to combine the 3D data cubes taken in different dither positions into a single data cube using the {\tt reproject\_interp} routine.  
We then removed artifacts by sigma clipping and created a combined cube by median stacking.  
The background was estimated by calculating the median at each wavelength with smoothing over 100 channels in the wavelength direction and was subtracted from the data cube.  
Finally, we corrected for astrometric offsets by creating a continuum image from the data cube and comparing it with the NIRCam F150W image, which has the highest spatial resolution.

\begin{figure*}
\centering
\begin{minipage}{0.9\hsize}
\begin{center}
\includegraphics[width=0.32\hsize, bb=4 6 283 279]{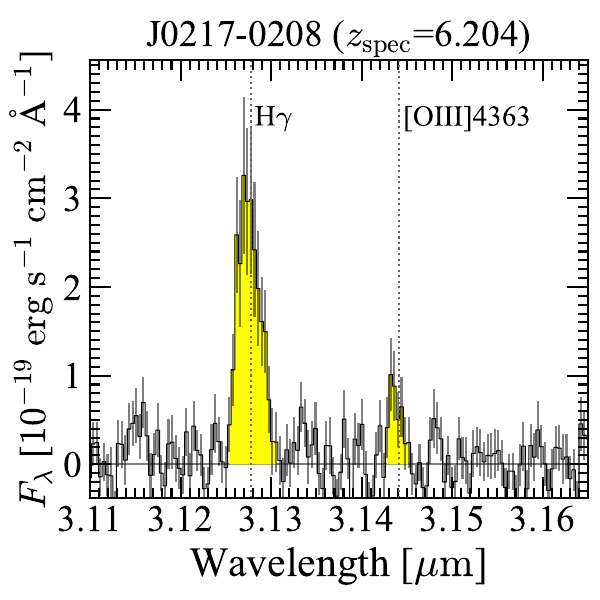}
\includegraphics[width=0.32\hsize, bb=4 6 283 279]{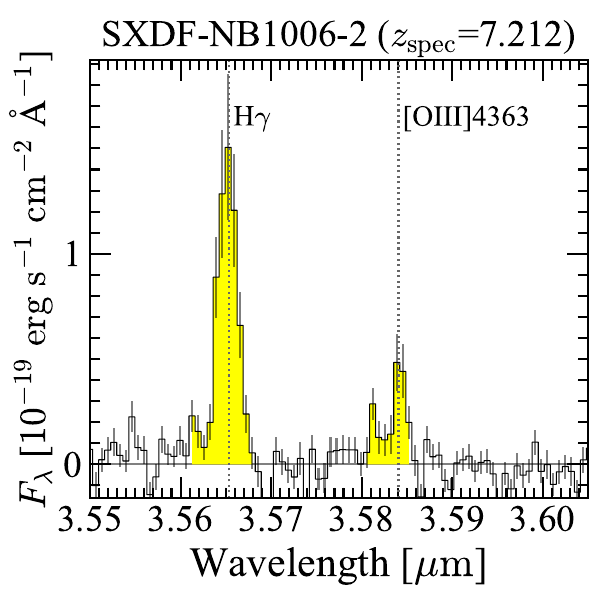}
\includegraphics[width=0.32\hsize, bb=4 6 283 279]{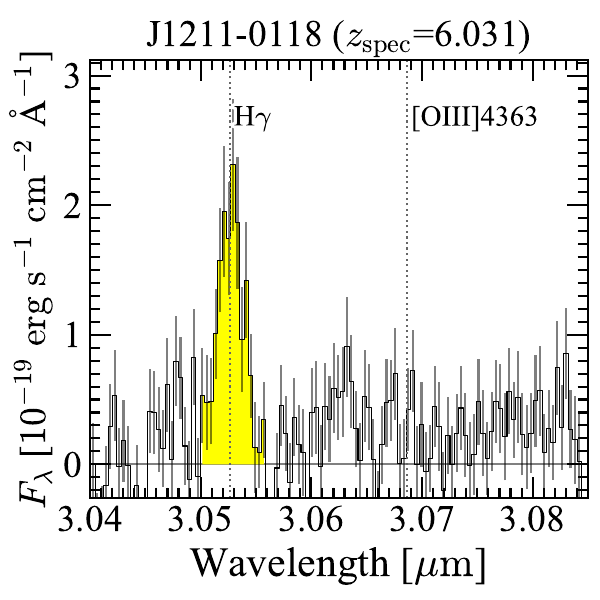}
\end{center}
\end{minipage}
\caption{
JWST/NIRSpec spectra around the H$\gamma$ and {[\sc Oiii]$\lambda$4363} emission lines after continuum subtraction.
The {[\sc Oiii]$\lambda$4363} emission line is detected (tentatively identified) at the $6\sigma$ ($4\sigma$) significance level in SXDF-NB1006-2 (J0217-0208).
}
\label{fig_spec_jwst_OIII4363}
\end{figure*}

\begin{figure*}
\centering
\begin{minipage}{0.9\hsize}
\begin{center}
\includegraphics[width=0.32\hsize, bb=4 6 283 279]{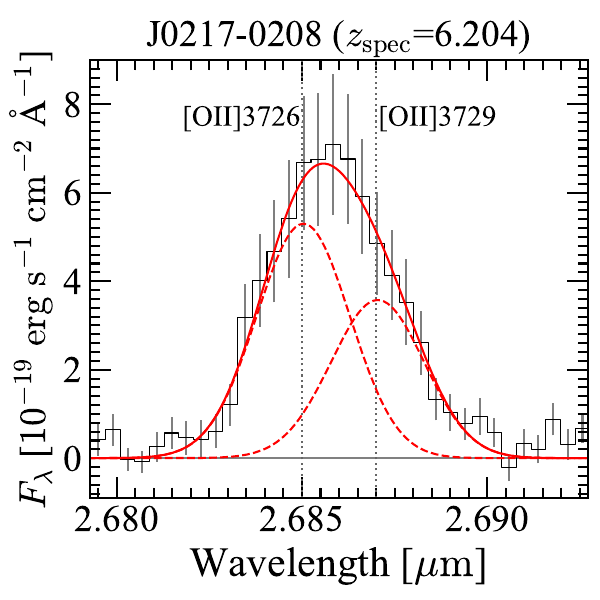}
\includegraphics[width=0.32\hsize, bb=4 6 283 279]{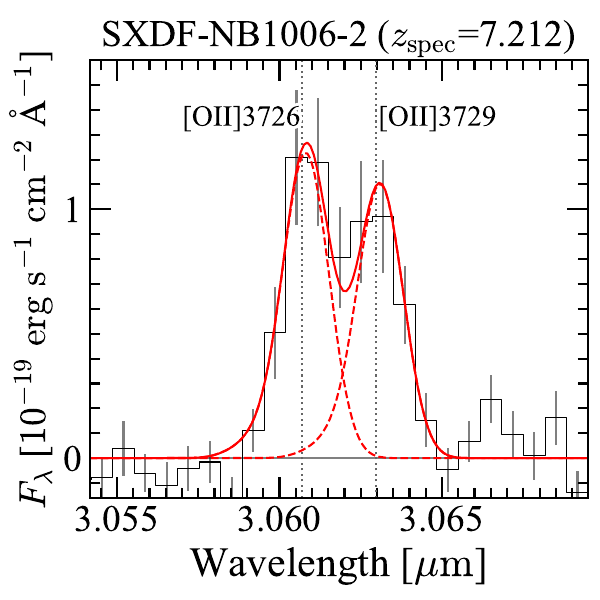}
\includegraphics[width=0.32\hsize, bb=4 6 283 279]{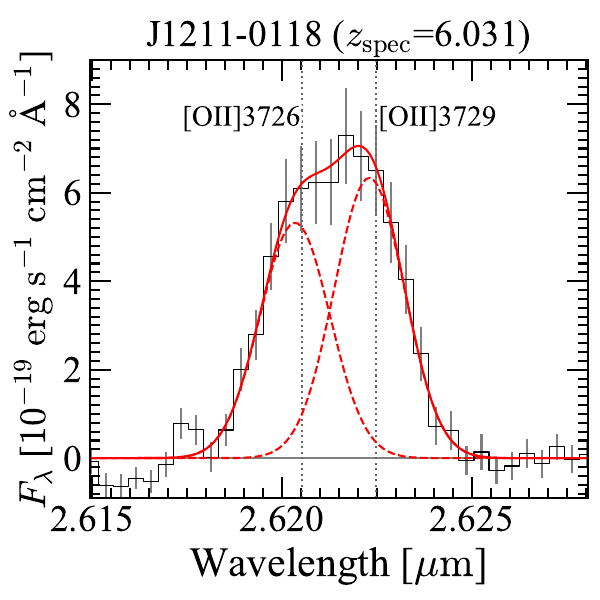}
\end{center}
\end{minipage}
\caption{
JWST/NIRSpec spectra around the {[\sc Oii]$\lambda\lambda$3726,3729} emission lines after continuum subtraction.
The red dashed lines represent the best-fit Gaussian functions for each emission line, convolved with the line-spread function, while the red solid line shows the total profile.
Thanks to the resolution ($R \sim 2700$) of the G235H (for J0217-0208 and J1211-0118) and G395H (for SXDF-NB1006-2) gratings, we can measure the {[\sc Oii]}$\lambda$3729/{[\sc Oii]}$\lambda$3726 line ratios, allowing us to constrain the electron densities of these galaxies.
}
\label{fig_spec_jwst_OIId}
\end{figure*}

We extracted a 1D spectrum from the data cube using a circular aperture whose diameter was $0.\carcsec6$ ($1.\carcsec0$) for J0217-0208 and J1211-0118 (for SXDF-NB1006-2).  
Aperture correction for each spectrum obtained with a filter-grating pair was calculated by comparing emission line fluxes in the spectrum with those in a large $2.\carcsec0$-diameter aperture.  
\redc{By fitting line-free regions of each spectrum with a cubic polynomial, we estimated the continuum emission and subtracted it to make a continuum-subtracted spectrum.}
Following \citet{2023A&A...677A.145U}, we scaled an error spectrum from the pipeline to match the standard deviation in line-free regions, taking into account the spatial correlations of the noise due to the PSF size.

Figure \ref{fig_spec_jwst} shows the galaxy-integrated NIRSpec IFU spectra for our targets.
In the 1D spectra, we clearly detect rest-frame optical strong emission lines such as {\sc[Oii]}$\lambda\lambda$3726,3729, H$\gamma$, H$\beta$, {\sc[Oiii]}$\lambda\lambda$4959,5007.  
We also identify {\sc[Oiii]}$\lambda$4363 emission lines in J0217-0208 and SXDF-NB1006-2 (Figure \ref{fig_spec_jwst_OIII4363}).  
Furthermore, as shown in Figure \ref{fig_spec_jwst_OIId}, the moderately-high resolution of the NIRSpec spectra ($R\sim2700$) allows us to measure the emission line ratio of the {\sc[Oii]} doublet.  
We will present the electron temperature and density measured with these lines in Section \ref{ss_ism}.  
In the {\sc[Oiii]}$\lambda$5007 maps, we find that the emission of J0217-0208 and SXDF-NB1006-2 are cospatial with their rest-frame UV stellar continuum and {\sc[Oiii]}88$\mu$m emission, while J1211-0118 shows an extended and complex {\sc[Oiii]}$\lambda$5007 emission.  
Spatially resolved analyses will be presented in a forthcoming paper.

We measured emission line fluxes by fitting each emission line with a Gaussian.  
First, we determined the redshift by fitting the H$\beta$ and {\sc[Oiii]}$\lambda\lambda$4959,5007 lines with an assumed line ratio of $f_\m{\sc[OIII]5007}/f_\m{[OIII]4959}=2.98$, based on theoretical calculations \citep{2000MNRAS.312..813S}.  
Then, we fit other emission lines with Gaussian functions with fixing the redshift to the value determined with the H$\beta$ and {\sc[Oiii]}$\lambda\lambda$4959,5007 lines.
In the fitting of the {\sc[Nii]}$\lambda\lambda$6548,6584 lines for J0217-0208 and J1211-0118, we assumed the line ratio of $f_\m{\sc[NII]6584}/f_\m{[NII]6548}=2.94$, \redc{based on a theoretical value in \citet{2020MNRAS.495.1016R}}.  
\redc{We simultaneously fit the {\sc[Oii]}$\lambda\lambda$3726,3729 lines with two Gaussian functions to estimate the line ratio and its error with taking the correlated uncertainties into account.}
We assumed the same line width for each set of the doublet lines, i.e., $\m{FWHM}_\m{[OII]3726}=\m{FWHM}_\m{\sc[OII]3729}$, $\m{FWHM}_\m{[OIII]4959}=\m{FWHM}_\m{\sc[OIII]5007}$, and $\m{FWHM}_\m{[NII]6548}=\m{FWHM}_\m{\sc[NII]6584}$.
The line widths of SXDF-NB1006-2 and J1211-0118 are narrow, $\m{FWHM}\sim100-200\ \m{km\ s^{-1}}$, comparable to the {\sc[Oiii]}88$\mu$m line \citep{2016Sci...352.1559I,2020ApJ...896...93H}.
Because the {\sc[Oiii]}$\lambda\lambda$4959,5007 and H$\alpha$ lines in J0217-0208 show a weak moderately broad component ($\m{FWHM}\sim500-800\ \m{km\ s^{-1}}$; see also \citealt{2025arXiv250504825M}) in addition to the narrow component  ($\m{FWHM}\sim300\ \m{km\ s^{-1}}$), we fit these lines with two Gaussian components.  
These broad components are probably due to another ionized gas component or outflow driven by star formation, not a signature of broad-line AGN activity, given their line widths and the fact that the broad component is seen in the forbidden {\sc[Oiii]}$\lambda\lambda$4959,5007 line.  
In the subsequent analyses, we use a total flux including both the narrow and broad components for these lines, but the conclusions do not change if we use only the narrow-component flux given the small fraction of the broad-component flux ($\sim10\%$).  
The measured emission line fluxes are summarized in Table \ref{tab_lineflux}.

\begin{deluxetable}{lccc}
\tablecaption{Optical Line Fluxes of Our Targets}
\label{tab_lineflux}
\tablehead{\colhead{} & \colhead{J0217-0208} & \colhead{SXDF-NB1006-2} & \colhead{J1211-0118}}
\startdata
{\sc[Oii]}$\lambda$3726 & $89.3\pm8.5$ & $26.6\pm3.7$ & $97.6\pm7.5$\\
{\sc[Oii]}$\lambda$3729 & $60.3\pm6.9$ & $23.9\pm3.3$ & $116.3\pm8.5$\\
{[Ne\sc iii]}$\lambda$3869 & $64.6\pm6.1$ & $37.5\pm4.6$ & $32.1\pm4.5$\\
H$\delta$ & $33.9\pm3.9$ & $21.9\pm2.9$ & $20.3\pm3.9$\\
H$\gamma$ & $51.0\pm4.8$ & $47.7\pm4.5$ & $47.4\pm4.6$\\
{\sc[Oiii]}$\lambda$4363 & $7.5\pm1.9$ & $11.3\pm1.9$ & $<6.2$\\
H$\beta$ & $100.0\pm7.5$ & $100.0\pm7.4$ & $100.0\pm9.2$\\
{\sc[Oiii]}$\lambda$4959 & $220.4\pm16.0$ & $226.2\pm20.9$ & $160.5\pm11.8$\\
{\sc[Oiii]}$\lambda$5007 & $656.7\pm40.7$ & $674.1\pm49.6$ & $478.3\pm27.1$\\
{\sc[Nii]}$\lambda$6549 & $18.7\pm3.0$ & \nodata & $<22.8$\\
H$\alpha$ & $374.7\pm29.7$ & \nodata & $348.2\pm20.9$\\
{\sc[Nii]}$\lambda$6584 & $55.1\pm4.9$ & \nodata & $<23.5$\\
\enddata
\tablecomments{Observed line flux ratios relative to H$\beta$ (×100). The observed H$\beta$ line fluxes are $1.88\times10^{-17}$, $8.55\times10^{-18}$, and $1.26\times10^{-17}$ erg s$^{-1}$ cm$^{-2}$ for J0217-0208, SXDF-NB1006-2, and J1211-0118, respectively.
Errors are $1\sigma$ and upper limits are $2\sigma$.}
\end{deluxetable}

\begin{figure*}
\centering
\begin{minipage}{0.75\hsize}
\centering
\begin{minipage}{0.3\hsize}
\begin{center}
\includegraphics[width=0.99\hsize, bb=6 -30 230 251]{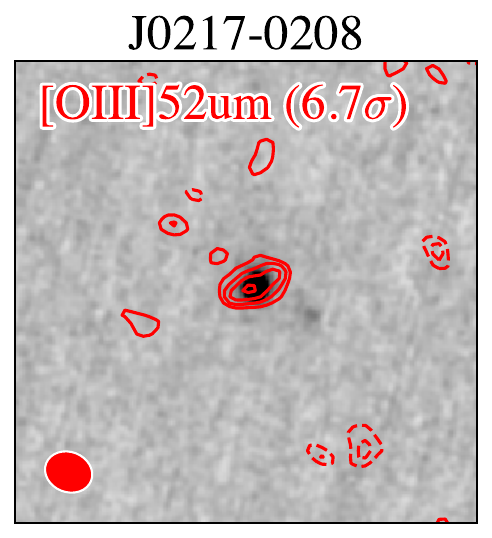}
\end{center}
\end{minipage}
\begin{minipage}{0.68\hsize}
\begin{center}
\includegraphics[width=0.99\hsize, bb=5 2 500 286]{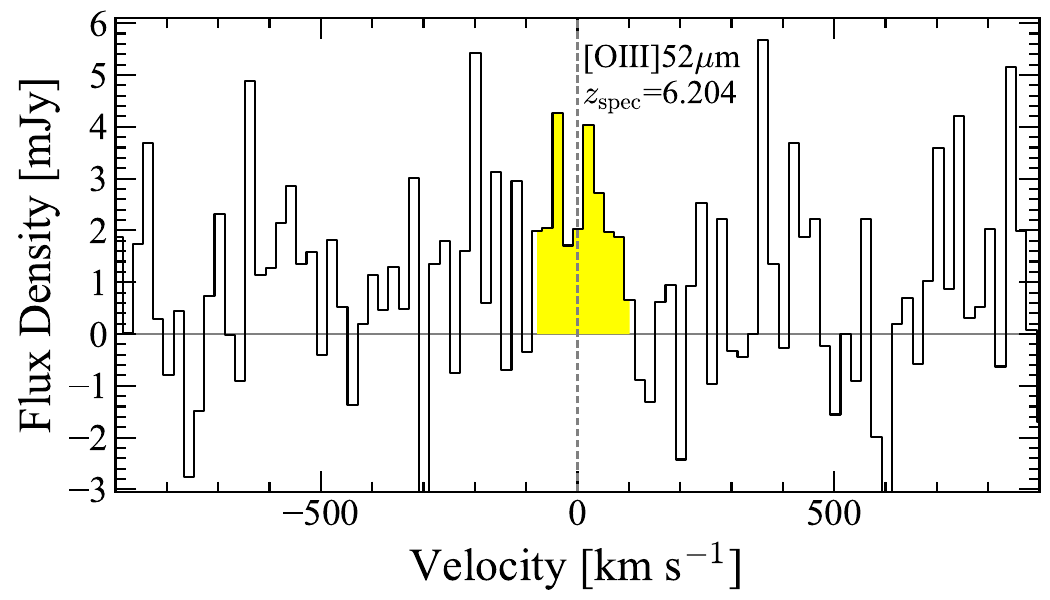}
\end{center}
\end{minipage}
\end{minipage}
\centering
\begin{minipage}{0.75\hsize}
\centering
\begin{minipage}{0.3\hsize}
\begin{center}
\includegraphics[width=0.99\hsize, bb=6 -30 230 251]{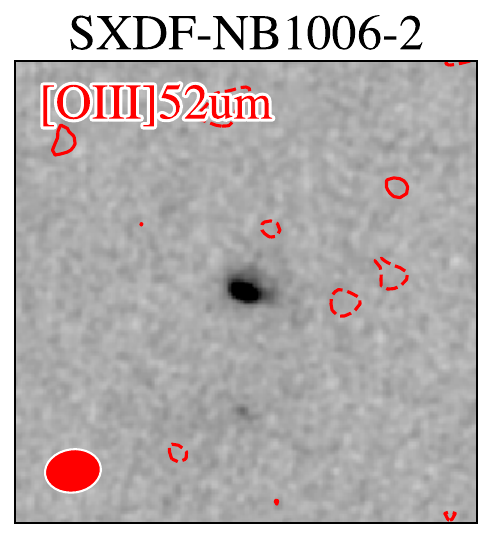}
\end{center}
\end{minipage}
\begin{minipage}{0.68\hsize}
\begin{center}
\includegraphics[width=0.99\hsize, bb=5 2 500 286]{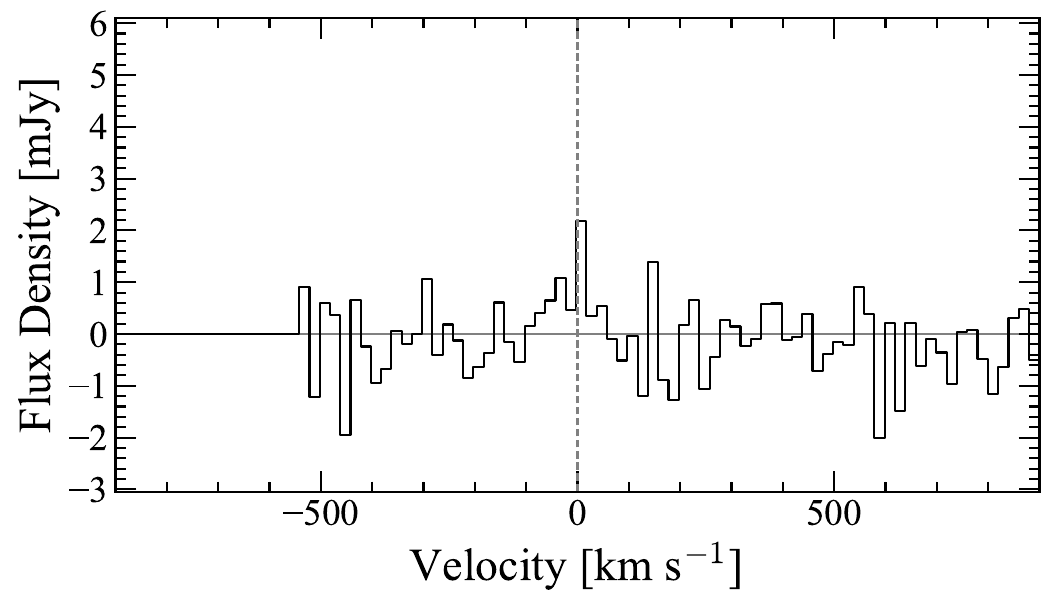}
\end{center}
\end{minipage}
\end{minipage}
\centering
\begin{minipage}{0.75\hsize}
\centering
\begin{minipage}{0.3\hsize}
\begin{center}
\includegraphics[width=0.99\hsize, bb=6 -30 230 251]{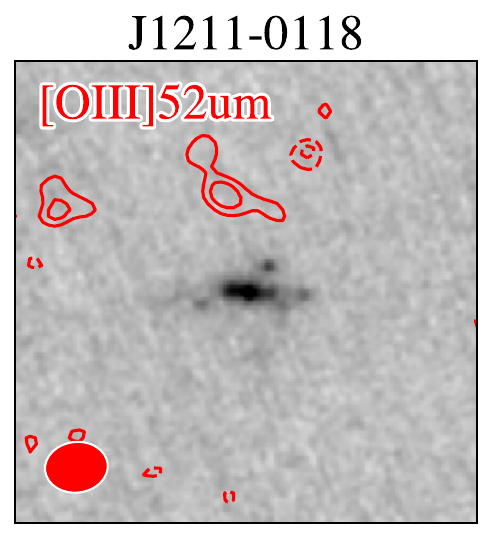}
\end{center}
\end{minipage}
\begin{minipage}{0.68\hsize}
\begin{center}
\includegraphics[width=0.99\hsize, bb=5 2 500 286]{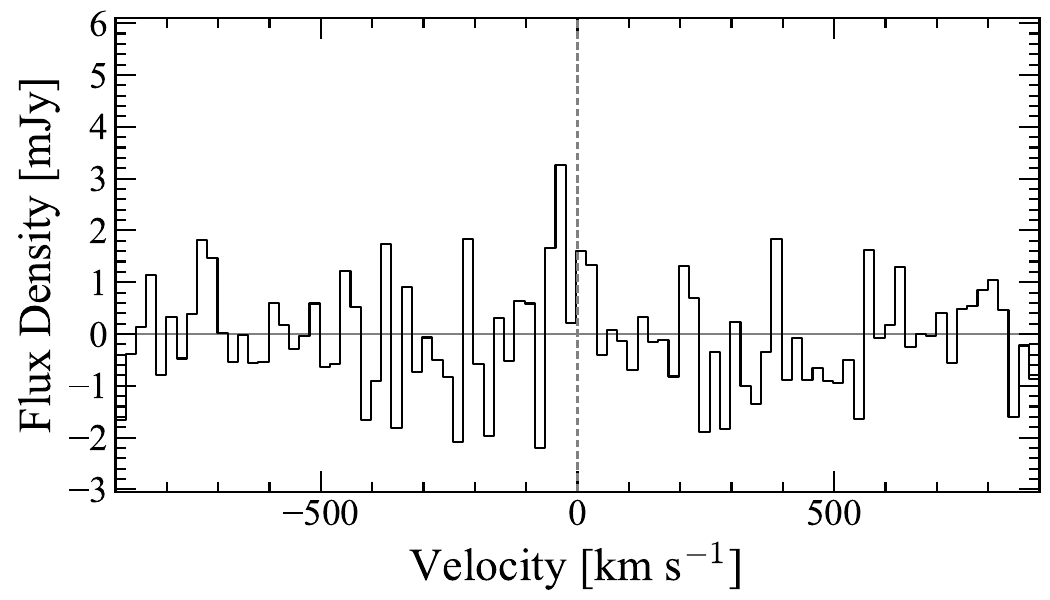}
\end{center}
\end{minipage}
\end{minipage}
\caption{
ALMA {[\sc Oiii]}52$\mu$m data for J0217-0208 (top panels), SXDF-NB1006-2 (middle panels), and J1211-0118 (bottom panels) obtained in this study.
The left-hand panels display the $3\arcsec\times3\arcsec$ {[\sc Oiii]}52$\mu$m maps, with contours drawn from $\pm3\sigma$ with an interval of $1\sigma$.
Positive and negative contours are shown as solid and dashed lines, respectively.
The background is the JWST/NIRCam F150W image, and the red ellipse at the lower left corner indicates the synthesized beam sizes of ALMA.
The right-hand panels present the continuum-subtracted spectra, with the observed wavelength of the {[\sc Oiii]}52$\mu$m line indicated by the gray dotted line.
We clearly detect the {[\sc Oiii]}52$\mu$m line in J0217-0208 at the $6.7\sigma$ significance level.
}
\label{fig_spec_ALMA}
\end{figure*}

\subsection{ALMA Dataset}\label{ss_alma}

As described in Section \ref{ss_sample}, our three targets, J0217-0208, SXDF-NB1006-2, and J1211-0118, have previous detections of the {\sc[Oiii]}88$\mu$m line.
To gain additional insight into the physical conditions of the ionized medium, particularly electron density, we observed the {\sc[Oiii]}52$\mu$m line, which has a higher critical density and thus provides complementary constraints to the {\sc[Oiii]}88$\mu$m line.
The {\sc[Oiii]}52$\mu$m observations were carried out during ALMA Cycles 9 and 10 (2022.1.00012.S and 2023.1.00022.S, PI: Y. Harikane) using Band 9 for SXDF-NB1006-2 and Band 10 for J0217-0208 and J1211-0118.
The antenna configurations were C43-1, C43-2, and C43-3, achieving beam sizes of $0.\carcsec3-0.\carcsec4$, and total integration times were 7.3, 17.7, and 12.8 hours for J0217-0208, SXDF-NB1006-2, and J1211-0118, respectively.  
We used four spectral windows (SPWs) with 1.875 GHz bandwidths in Frequency Division Mode (FDM), yielding a total bandwidth of 7.5 GHz.  
The velocity resolution was set to $\sim3$ km/s.
One of the SPWs was centered on the {\sc[Oiii]}52$\mu$m line frequency, expected from the redshift of the {\sc[Oiii]}88$\mu$m emission reported in \citet{2020ApJ...896...93H} and \citet{2023ApJ...945...69R}.

The data were processed using the Common Astronomy Software (CASA; \citealt{2007ASPC..376..127M}) pipeline version 6.4.1.12.
Using the task {\tt tclean}, we produced images and cubes with natural weighting without taper to maximize point-source sensitivities.  
To create a pure line image, we subtracted the continuum using off-line channels in the line cube with the CASA task {\tt uvcontsub}.
However, the measured line fluxes remain consistent within the uncertainties even without applying {\tt uvcontsub}.  
We generated {\sc[Oiii]}52$\mu$m maps by integrating the ALMA data cubes over the velocity range corresponding to $2\times \m{FWHM}_\m{[OIII]88\mu m}$, where $\m{FWHM}_\m{[OIII]88\mu m}$ is the FWHM of the {\sc[Oiii]}88$\mu$m line reported in \citet{2020ApJ...896...93H} and \citet{2023ApJ...945...69R}.  
1D spectra were extracted from the data cube using a $0.\carcsec3$-diameter circular aperture.  
We also generated a dust continuum image but did not detect any dust continuum emission, resulting in $3\sigma$ upper limits of $0.61$, $0.25$, and $0.60$ mJy/beam for J0217-0208, SXDF-NB1006-2, and J1211-0118, respectively.  

Figure \ref{fig_spec_ALMA} shows the ALMA {\sc[Oiii]}52$\mu$m maps and spectra obtained in this study.  
We clearly detect the {\sc[Oiii]}52$\mu$m emission line in J0217-0208 at the $6.7\sigma$ significance level.  
We do not detect {\sc[Oiii]}52$\mu$m in the other two galaxies.  
The total line flux or the upper limits are measured on the {\sc[Oiii]}52$\mu$m maps in a large $0.\carcsec8$-diameter circular aperture (Table \ref{tab_target}).  
These measurements are useful for constraining the electron densities of our galaxies, as presented in Section \ref{ss_den_fir}.

\section{SED Fitting}\label{ss_sed}

We conduct SED fitting to understand the physical properties of our galaxies.
We calculate the total magnitude in each band from an aperture magnitude measured in a 0.\carcsec3-diameter circular aperture with an aperture correction, following the method of \citet{2023ApJS..265....5H}.  
For the SED fitting, we use \textsc{prospector} \citep{2021ApJS..254...22J}, varying the dust optical depth in the $V$-band, metallicity, star formation history, and total stellar mass as free parameters, while fixing the redshift to the spectroscopically determined value.  
Model spectra are generated using the Flexible Stellar Population Synthesis \citep[FSPS;][]{2009ApJ...699..486C,2010ApJ...712..833C} package with the Modules for Experiments in Stellar Astrophysics Isochrones and Stellar Tracks (MIST; \citealt{2016ApJ...823..102C}).  
The boost in ionizing flux production from massive stars due to rotation is included in the MIST isochrones \citep{2017ApJ...838..159C}.  
We assume the stellar IMF from \citet{2003PASP..115..763C}, the dust \redc{attenuation} law of \citet{2000ApJ...533..682C}, and the intergalactic medium (IGM) attenuation model by \citet{1995ApJ...441...18M}.  
We adopt a flexible star formation history with five bins: the first bin is fixed at 0-10 Myr, and the other bins are spaced equally in logarithmic time between 10 Myr and a lookback time corresponding to $z=30$, assuming a constant SFR within each bin.  
A continuity prior is applied to the star formation history, with flat priors for other parameters in the ranges $0<\tau_\m{V}<2$, $-2.0<\m{log}(Z/Z_\odot)<0.4$, and $6<\m{log}(M_*/M_\sun)<12$.  
We search for the best-fit model to the observed photometric data points using the Markov Chain Monte Carlo (MCMC) method with {\sc emcee} \citep{2013PASP..125..306F}.  
Table \ref{tab_sed} summarizes the results of the SED fitting.  

\begin{figure*}
\centering
\begin{minipage}{0.9\hsize}
\centering
\begin{minipage}{0.55\hsize}
\begin{center}
\includegraphics[width=0.99\hsize, bb=19 10 427 314]{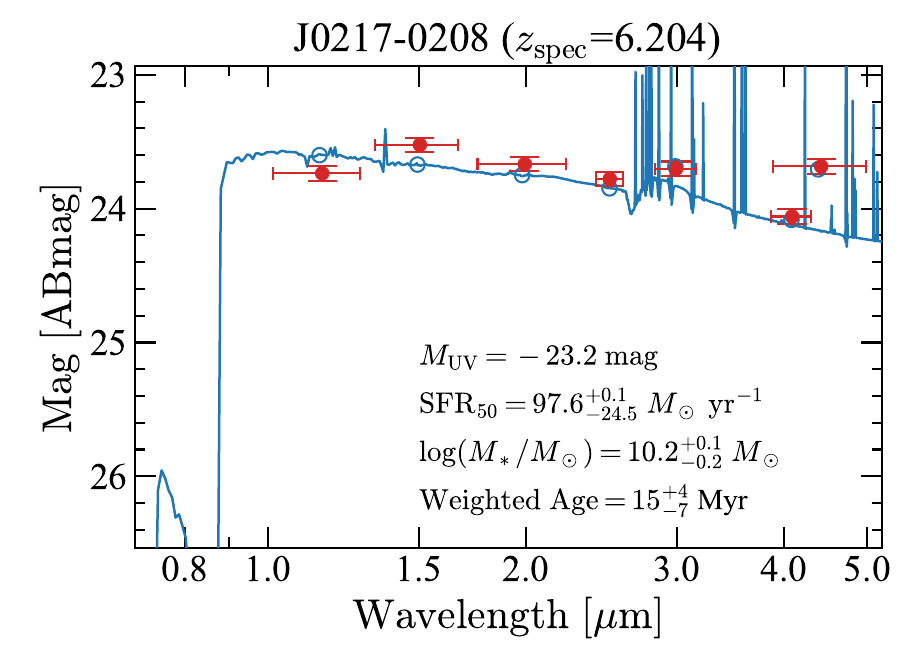}
\end{center}
\end{minipage}
\begin{minipage}{0.39\hsize}
\begin{center}
\includegraphics[width=0.99\hsize, bb=2 1 283 314]{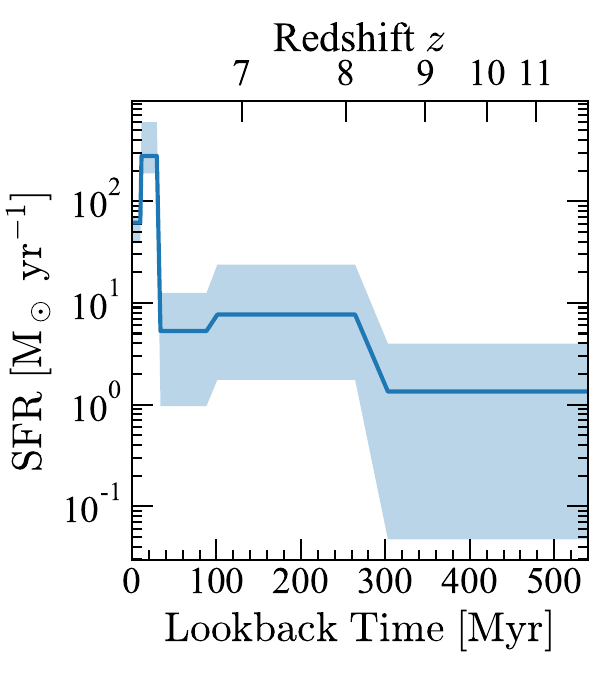}
\end{center}
\end{minipage}
\centering
\begin{minipage}{0.55\hsize}
\begin{center}
\includegraphics[width=0.99\hsize, bb=19 10 427 314]{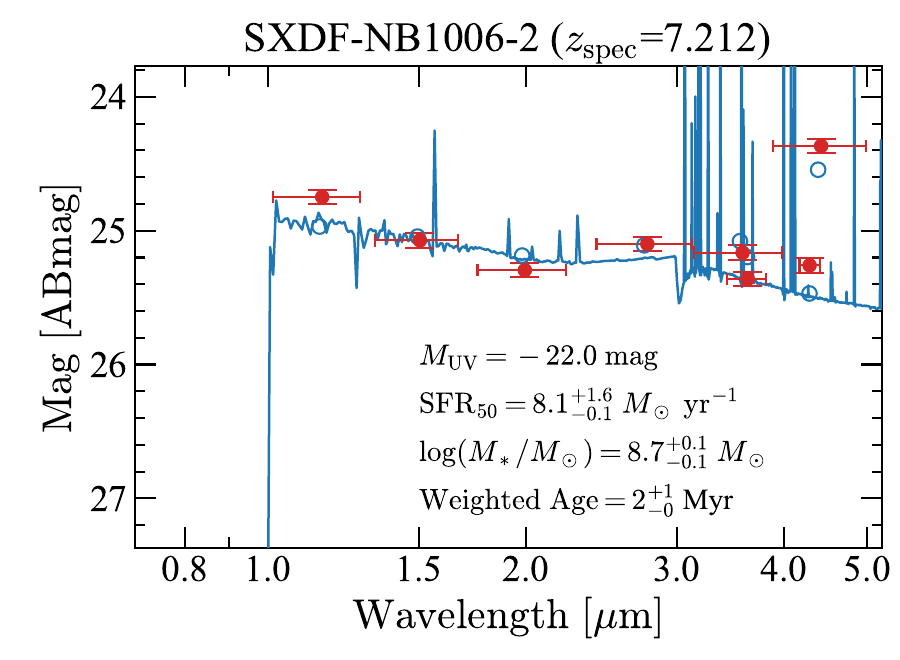}
\end{center}
\end{minipage}
\begin{minipage}{0.39\hsize}
\begin{center}
\includegraphics[width=0.99\hsize, bb=2 1 283 314]{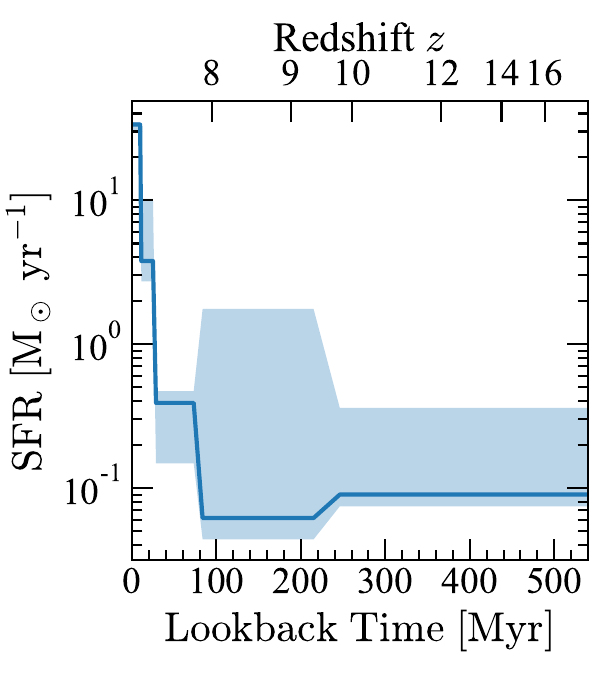}
\end{center}
\end{minipage}
\centering
\begin{minipage}{0.55\hsize}
\begin{center}
\includegraphics[width=0.99\hsize, bb=19 10 427 314]{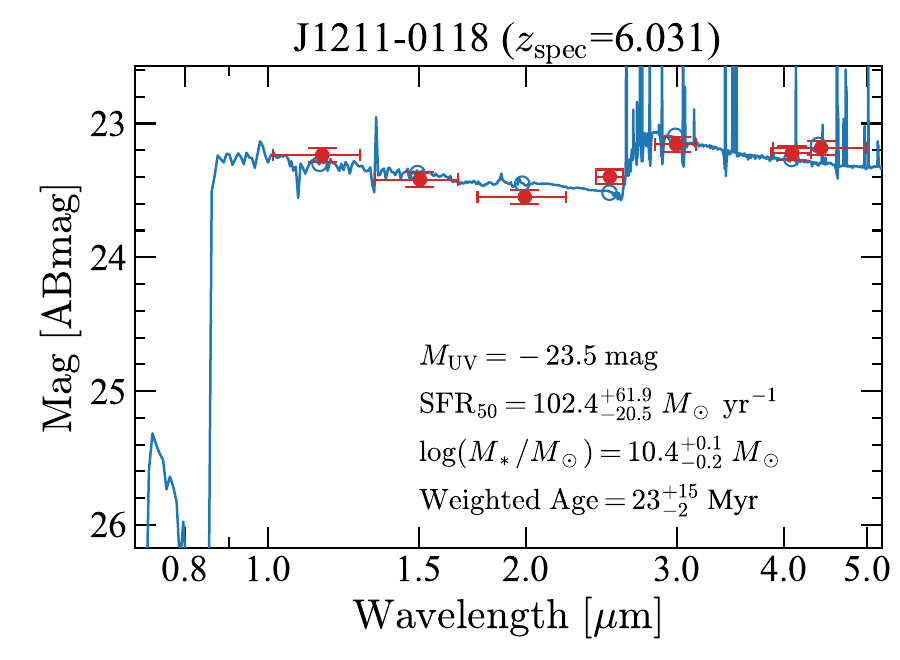}
\end{center}
\end{minipage}
\begin{minipage}{0.39\hsize}
\begin{center}
\includegraphics[width=0.99\hsize, bb=2 1 283 314]{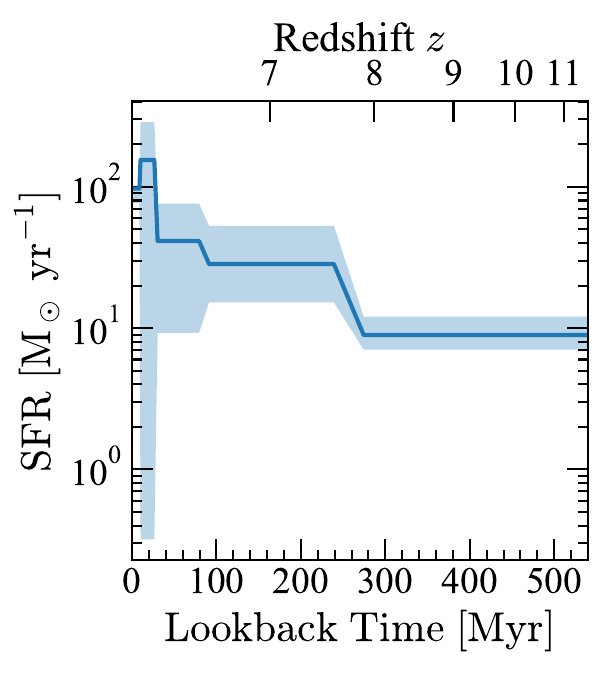}
\end{center}
\end{minipage}
\end{minipage}
\caption{
SED fitting results for J0217-0208 (top panels), SXDF-NB1006-2 (middle panels), and J1211-0118 (bottom panels).
The left-hand panels show the SEDs, where the red circles represent observed magnitudes, and the blue line along with open circles denote the best-fit model SED obtained with {\sf Prospector}.
The right-hand panels display the star formation history constrained by the SED fitting.
}
\label{fig_sed}
\end{figure*}

\begin{figure}
\centering
\begin{center}
\begin{minipage}{0.99\hsize}
\begin{center}
\includegraphics[width=0.9\hsize, bb=3 5 350 281,clip]{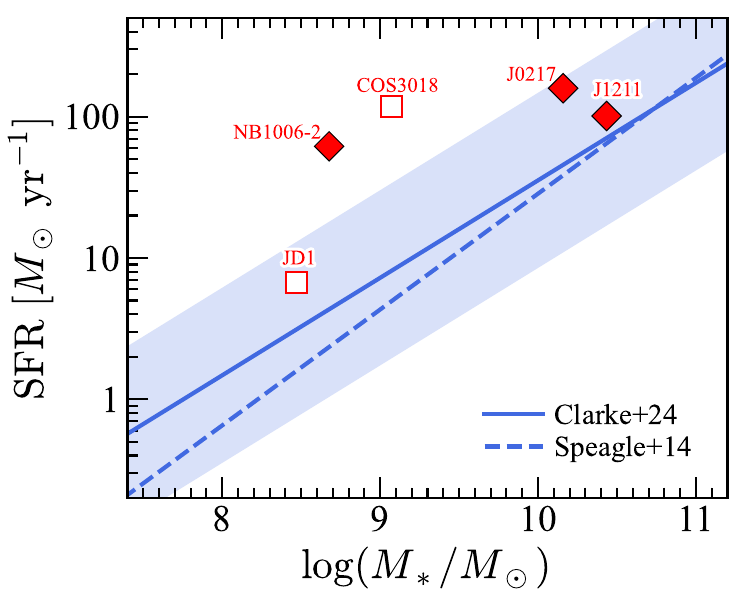}
\end{center}
\end{minipage}
\end{center}
\caption{
SFR and stellar mass ($M_*$) of the galaxies analyzed in this study.
The red diamonds represent J0217-0208, SXDF-NB1006-2, and J1211-0118, whose stellar masses are estimated in this study.
The red open squares represent COS-3018555981 and MACS1149-JD1, which are analyzed in Section \ref{ss_jwst_alma}.
Their SFRs and stellar masses are estimated in the literature \citep{2024arXiv241107695S,2024MNRAS.533.2488M}.
The SFRs are measured from the dust-corrected H$\beta$ luminosities.
The blue solid line represents the star-formation main sequence at $z\sim6-7$ from \citet{2024ApJ...977..133C}, with the shaded region indicating its intrinsic scatter at the high-mass end.
The blue dashed line shows the main sequence at $z=7$ from \citet{2014ApJS..214...15S}, shifted by 0.32 dex to match the low-metallicity calibration adopted in \citet{2024ApJ...977..133C}.
}
\label{fig_Ms_SFR}
\end{figure}

Figure \ref{fig_sed} presents the SED fitting results for our galaxies.  
J0217-0208 exhibits a rising star formation history, with its SFR having increased by a factor of 10 over the last 50 Myr, consistent with its strong {\sc[Oiii]}$\lambda$5007 emission line, whose rest-frame {\sc[Oiii]}$\lambda$5007 equivalent width (EW) is $\m{EW_0}\sim500\ \m{\AA}$ in the NIRSpec spectrum.  
SXDF-NB1006-2 is much younger, with a weighted age of $\sim2$ Myr, and its rest-frame {\sc[Oiii]}$\lambda$5007 EW is $\m{EW_0}\sim900\ \m{\AA}$.  
J1211-0118 shows a prominent Balmer break in the SED, indicating past star formation activity over the last 500 Myr.  
Figure \ref{fig_Ms_SFR} shows the SFRs and stellar masses of our targets.  
J0217-0208 and J1211-0118 lie on the star-forming main sequence, while SXDF-NB1006-2 is located above the sequence, consistent with its young stellar population and strong emission lines.

\begin{deluxetable}{lcccc}
\tablecaption{SED Fitting Results}
\label{tab_sed}
\setlength{\tabcolsep}{3pt}
\tablehead{\colhead{Name} & \colhead{SFR} & \colhead{$\m{log}M_*$} & \colhead{$\m{Age}$} & \colhead{$E(B-V)$}\\
\colhead{}& \colhead{($M_\odot\ \m{yr^{-1}}$)}& \colhead{($M_\odot$)}& \colhead{(Myr)}& \colhead{ABmag}} 
\startdata
J0217-0208 & $97.6_{-24.5}^{+0.1}$  & $10.2_{-0.2}^{+0.1}$  & $14.6_{-6.6}^{+4.3}$  & $0.05_{-0.02}^{+0.01}$ \\
SXDF-NB1006-2 & $8.1_{-0.1}^{+1.6}$  & $8.7_{-0.1}^{+0.1}$  & $1.8_{-0.4}^{+1.2}$  & $0.01_{-0.01}^{+0.01}$ \\
J1211-0118 & $102.4_{-20.5}^{+61.9}$  & $10.4_{-0.2}^{+0.1}$  & $23.3_{-1.5}^{+14.5}$  & $0.01_{-0.01}^{+0.02}$ \\
\enddata
\tablecomments{Errors are $1\sigma$. The SFR presented here is the SFR averaged over the past 50 Myr, and the stellar age is the mass-weighted age calculated from the star formation history.
}
\end{deluxetable}

\section{ISM Property}\label{ss_ism}

\subsection{Dust \redc{Attenuation}}\label{ss_dust}

We evaluate the dust attenuation of our galaxies using the color excess, $E(B-V)$, estimated from the Balmer decrement.  
We use the H$\alpha$, H$\beta$, and H$\gamma$ emission lines, which are detected with the $>10\sigma$ significance levels, and calculate the color excess assuming the \citet{2000ApJ...533..682C} dust \redc{attenuation} law.  
We assume intrinsic line ratios for a temperature of 10,000 K from \citet{1989agna.book.....O}, namely $\m{H\alpha/H\beta=2.86}$ and $\m{H\gamma/H\beta=0.468}$, although the conclusions of this paper do not change for the range of temperatures inferred in Section \ref{ss_2zone}.  
The resulting color excesses are summarized in Table \ref{tab_target}.  
We find small but non-negligible dust \redc{attenuation}, with $E(B-V)\simeq0.1$, in J0217-0208 and J1211-0118.  
SXDF-NB1006-2 shows almost no dust \redc{attenuation}, consistent with the non-detections of dust continuum emission reported by ALMA observations in Section \ref{ss_alma} and in the previous studies \citep{2016Sci...352.1559I,2023ApJ...945...69R}.  
\redc{The estimated $E(B-V)$ value for each galaxy is consistent with the $\m{H\alpha/H\beta}$ and $\m{H\gamma/H\beta}$ ratios within errors, and we do not find any significant signature of a Balmer decrement anomaly beyond errors, sometimes seen in high redshift galaxies \citep[e.g.,][]{2024ApJ...969...90P,2024arXiv240409015S,2024ApJ...974..180Y}.}
In the following sections, we use emission line fluxes corrected for dust \redc{attenuation} based on the $E(B-V)$ values measured from the Balmer decrement.  
The obtained color excesses are comparable to those estimated from the SED fitting in Section \ref{ss_sed}, although they can differ because $E(B-V)$ from the Balmer decrement and SED fitting probe nebular and stellar \redc{attenuation}, respectively.

\begin{deluxetable*}{lccc}
\tablecaption{Properties of Our Targets Measured with Emission Lines}
\label{tab_target}
\tablehead{\colhead{} & \colhead{J0217-0208} & \colhead{SXDF-NB1006-2} & \colhead{J1211-0118}}
\startdata
$L_\m{H\beta}$$^\m{\bluec{a}}$ [$L_\odot$] & $(3.4\pm0.3)\times10^{9}$ & $(1.3\pm0.1)\times10^{9}$ & $(2.2\pm0.2)\times10^{9}$\\
$L_\m{[OIII]52\mu m}$ [$L_\odot$] & $(1.4\pm0.2)\times10^{10}$ & $<1.0\times10^{9}$ & $<3.8\times10^{9}$\\
$L_\m{[OIII]88\mu m}$ [$L_\odot$] & $(8.5\pm2.0)\times10^{9}$$^\m{\bluec{b}}$ & $(1.3\pm0.3)\times10^{9}$$^\m{\bluec{c}}$ & $(4.8\pm0.7)\times10^{9}$$^\m{\bluec{b}}$\\
$L_\m{[CII]158\mu m}$ [$L_\odot$] & $(1.4\pm0.2)\times10^{9}$$^\m{\bluec{b}}$ & $(1.2\pm0.5)\times10^{8}$$^\m{\bluec{c}}$ & $(1.4\pm0.1)\times10^{9}$$^\m{\bluec{b}}$\\
$\m{EW_0}$(H$\beta$) & $74.8\pm5.6$ & $134.7\pm10.0$ & $22.9\pm2.1$\\
$\m{EW_0}$({\sc[Oiii]$\lambda$5007}) & $498.8\pm30.9$ & $895.1\pm65.8$ & $110.5\pm6.3$\\
$\m{SFR}_\m{H\beta}$$^\m{\bluec{d}}$ [$M_\odot\ \m{yr^{-1}}$] & $159\pm12$ & $62\pm5$ & $101\pm9$\\
$E(B-V)$$^\m{\bluec{e}}$ [ABmag] & $0.10\pm0.08$ & $0.00\pm0.11$ & $0.10\pm0.08$\\
$T_\m{e}$$^\m{\bluec{f}}$ [K] & $12300^{+1100}_{-1200}$ & $13900^{+1000}_{-1100}$ & $<12900$\\
$12+\m{log(O/H)}$$^\m{\bluec{f}}$ & $8.20^{+0.15}_{-0.11}$ & $7.99^{+0.10}_{-0.09}$ & (\redc{$8.51^{+0.13}_{-0.15}$})\\
$n_\m{e,optical}$$^\m{\bluec{g}}$ [cm$^{-3}$] & \redc{$1782^{+1092}_{-584}$} & \redc{$774^{+695}_{-360}$} & \redc{$258^{+171}_{-127}$}\\
$n_\m{e,FIR}$$^\m{\bluec{h}}$ [cm$^{-3}$] & $505^{+253}_{-228}$ & $<87$ & $<97$\\
\enddata
\tablecomments{Errors are $1\sigma$ and upper limits are $2\sigma$.\\
$^\m{a}$ Dust-corrected H$\beta$ luminosity.\\
$^\m{b}$ \citet{2020ApJ...896...93H}.\\
$^\m{c}$ \citet{2023ApJ...945...69R}.\\
$^\m{d}$  SFR estimated from $L_\m{H\beta}$ adopting the conversion factor in \citet{2024ApJ...977..133C}.
$^\m{e}$ Dust \redc{attenuation} calculated with the Balmer decrement.\\
$^\m{f}$ For J0217-0208 and SXDF-NB1006-2, the electron temperatures and metallicities are calculated using the {\sc[Oiii]}$\lambda$4363 emission line.
As shown in Figure \ref{fig_1zone}, these parameters cannot reproduce the FIR {\sc[Oiii]}88$\mu$m and {\sc[Oiii]}52$\mu$m emission lines.
For J1211-0118, since we do not detect the {\sc[Oiii]}$\lambda$4363 emission line, the metallicity (in parentheses) is estimated based on the strong-line R23 calibrator (see text).\\
$^\m{g}$ Electron density calculated using the {\sc[Oii]}$\lambda$3729/{\sc[Oii]}$\lambda$3726 ratio.\\
$^\m{h}$ Electron density calculated using the {\sc[Oiii]}52$\mu$m/{\sc[Oiii]}88$\mu$m ratio.
}
\end{deluxetable*}

\subsection{Electron Temperature and Metallicity}\label{ss_Te}

We calculate the electron temperature and metallicity of our galaxies.  
We identify the {\sc[Oiii]}$\lambda$4363 emission line in J0217-0208 and J1211-0118, allowing us to measure metallicities using the direct-$T_\m{e}$ method.  
Although the {\sc[Oiii]}$\lambda$4363/{\sc[Oiii]}$\lambda$5007 ratio depends on both electron temperature and density, the density dependence is negligible as long as the density is well below the critical densities of {\sc[Oiii]}$\lambda$4363 and {\sc[Oiii]}$\lambda$5007 ($2.8\times10^7$ and $6.4\times10^5$ cm$^{-3}$, respectively).  
Since the electron densities of our galaxies measured in Section \ref{ss_den} are much lower than these critical densities, the derived electron temperatures are not significantly affected by different density assumptions.  
Here, we adopt the electron densities measured with the optical {\sc[Oii]} lines from Section \ref{ss_den_opt}.  
If we instead use the densities measured with the FIR {\sc[Oiii]} lines from Section \ref{ss_den_fir}, the obtained temperatures increase by up to 1\%.  
We calculate the electron temperatures based on the {\sc[Oiii]}$\lambda$4363/{\sc[Oiii]}$\lambda$5007 line ratios using {\sf PyNeb} \citep{2015A&A...573A..42L}.  
We then estimate the ionic oxygen abundance ratios of O$^+$/H$^+$ and O$^{2+}$/H$^+$ using the equations in \citet{2006A&A...448..955I}, based on the observed emission line fluxes of {\sc[Oiii]}$\lambda\lambda$4959,5007, H$\beta$, and {\sc[Oii]}$\lambda\lambda$3726,3729, assuming an extrapolation of the electron temperature relation between O$^+$ and O$^{2+}$.  
\redc{Adopting the more recent $\m{O}^+-\m{O}^{2+}$ electron temperature relation from \citet{2025arXiv250403839C} changes the derived oxygen abundances by less than 5\%.}
We do not account for a higher ionization state of O$^{3+}$/H$^+$, as the ionization potential of O$^{3+}$ is too high for the stellar radiation fields of our galaxies, which do not exhibit high-ionization lines (e.g., He{\sc ii}).  
As shown in Table \ref{tab_target}, we find electron temperatures and metallicities of $\simeq12000-13000$ K and $12+\m{log(O/H)}\simeq8.0-8.2$, respectively, which are comparable to those of other high-redshift galaxies \citep[e.g.,][]{2023ApJS..269...33N,2024ApJ...962...24S}.  
However, as discussed in Section \ref{ss_1zone}, these parameters cannot reproduce the observed luminosities of the FIR {\sc[Oiii]} emission lines detected with ALMA.

\begin{figure*}
\centering
\begin{minipage}{0.9\hsize}
\begin{center}
\includegraphics[width=0.32\hsize, bb=6 10 265 243]{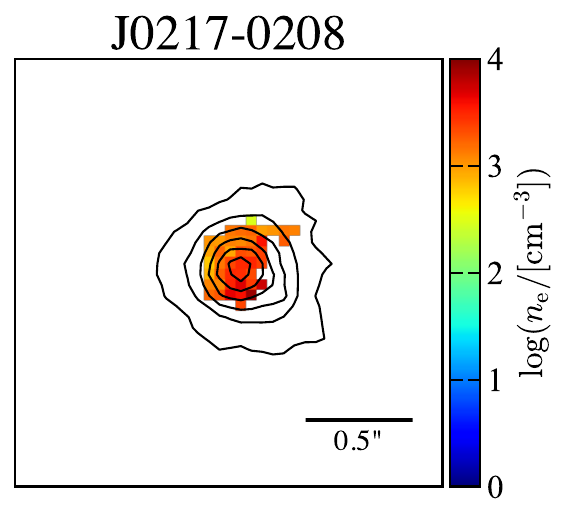}
\includegraphics[width=0.32\hsize, bb=6 10 265 243]{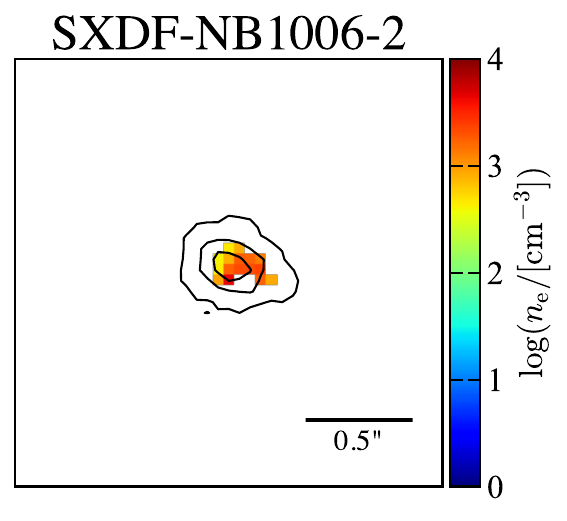}
\includegraphics[width=0.32\hsize, bb=6 10 265 243]{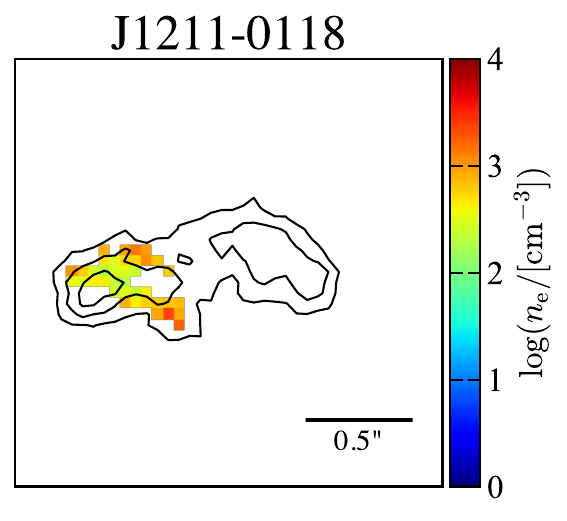}
\end{center}
\end{minipage}
\caption{
Resolved map of electron densities derived from the {\sc[Oii]}$\lambda$3729/{\sc[Oii]}$\lambda$3726 line ratio.
We mask spaxels where the electron density is not reliably estimated, specifically when the ratio of the estimated density to its uncertainty is less than 3.
The black contours represent arbitrary {\sc[Oiii]}$\lambda$5007 flux levels.
In J0217-0208 and SXDF-NB1006-2, most ($>90\%$) of the pixels show densities higher than 500 and 100 cm$^{-3}$, respectively, and no significant fraction of pixels with low densities is observed.
}
\label{fig_map_ne}
\end{figure*}

We do not detect the {\sc[Oiii]}$\lambda$4363 emission line in J1211-0118.  
We estimate an upper limit on the electron temperature of $T_\m{e}<12900$ K, \redc{which corresponds to a metallicity lower limit of $12+\m{log(O/H)}>8.1$}.  
Instead of the direct-$T_\m{e}$ method, we apply the R23 strong-line calibration, with parameters constrained by JWST spectroscopy in \citet{2024ApJ...962...24S}, where $\m{R23=(\textsc{[Oiii]}\lambda\lambda4959,5007+{\textsc{[Oii]}}\lambda\lambda3726,3729)/H\beta}$.  
\redc{Using the measured R23 value of J1211-0118, we obtain metallicity estimates of $12+\m{log(O/H)}\simeq7.6$ and $8.5$ for low-metallicity and high-metallicity branches, respectively.
Given the lower limit of metallicity from {\sc[Oiii]}$\lambda$4363, we adopt the value for the high-metallicity branch as a fiducial value.
Note that if we instead use the O32 calibration in \citet{2024ApJ...962...24S}, we obtain a metallicity estimate of $12+\m{log(O/H)}\simeq8.2$, consistent with the lower limit from {\sc[Oiii]}$\lambda$4363.}

\subsection{Electron Density}\label{ss_den}

\subsubsection{Density from the {\sc [Oii]}$\lambda\lambda$3726,3729 Line Ratio}\label{ss_den_opt}

The {\sc [Oii]}$\lambda\lambda$3726,3729 emission lines have been commonly used as a density indicator \citep[e.g.,][]{2016ApJ...816...23S,2023ApJ...956..139I}, because of their different critical densities ($3300$ and $14000$ cm$^{-3}$ for {\sc[Oii]}$\lambda$3726 and {\sc[Oii]}$\lambda$3729, respectively).
Thanks to the moderately high spectral resolution of the NIRSpec G235H and G395H spectra, we can resolve the {\sc[Oii]} doublet and thus measure the {\sc[Oii]}$\lambda$3729/{\sc[Oii]}$\lambda$3726 line ratio (Figure \ref{fig_spec_jwst_OIId}).
\redc{We calculate the electron densities of our galaxies from the measured {\sc[Oii]}$\lambda$3729/{\sc[Oii]}$\lambda$3726 ratios using {\sf PyNeb}.
We assume the electron temperature of $T_\m{e}=13000$ K, although the {\sc[Oii]}$\lambda$3729/{\sc[Oii]}$\lambda$3726 ratio does not strongly depend on the electron temperature.
For example, using the calibration presented in \citet{2016ApJ...816...23S} assuming $T_\m{e}=10000$ K gives results consistent within errors.} 
As summarized in Table \ref{tab_target}, the calculated electron densities are \redc{$n_\m{e,optical}\simeq1800$, 800, and 300 cm$^{-3}$} for J0217-0208, SXDF-NB1006-2, and J1211-0118, respectively, which are comparable to those measured in other high redshift galaxies with the {\sc [Oii]}$\lambda\lambda$3726,3729 lines \citep[e.g.,][]{2023ApJ...956..139I,2025ApJ...979L..13L}.

Thanks to the IFU data, we can also investigate the spatially-resolved density map.
Figure \ref{fig_map_ne} shows the map of the electron densities of our three galaxies.
We see that the electron densities exhibit spatial variations, and some regions reaching very high densities of $\sim10^4$ cm$^{-3}$.
Nonetheless, in J0217-0208 and SXDF-NB1006-2, most ($>90\%$) of the pixels show densities higher than 500 and 100 cm$^{-3}$, respectively, and we do not find a significant fraction of pixels with low densities.

\subsubsection{Density from the {\sc[Oiii]}52$\mu$m/{\sc[Oiii]}88$\mu$m Line Ratio}\label{ss_den_fir}

\begin{figure}
\centering
\begin{center}
\includegraphics[width=0.99\hsize, bb=5 8 357 354,clip]{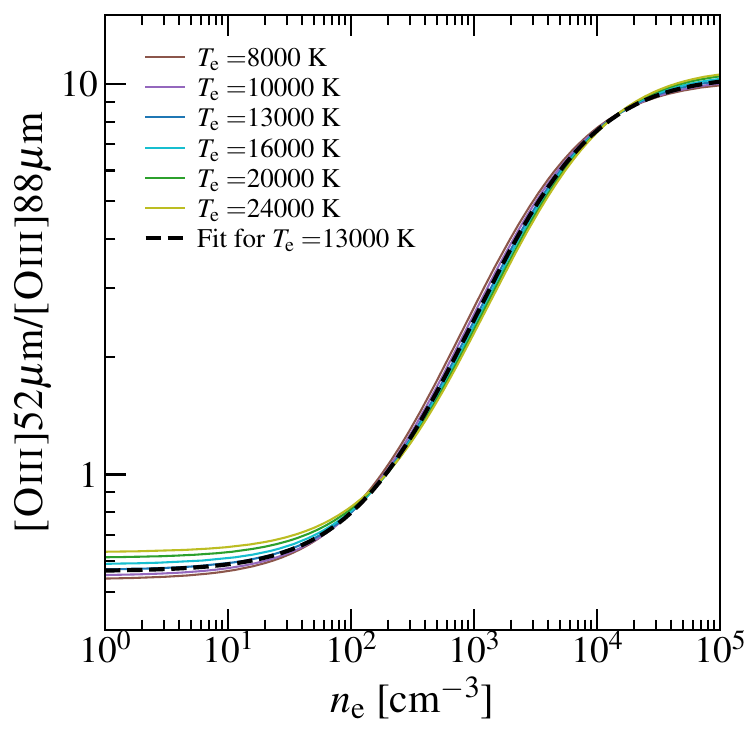}
\end{center}
\caption{
{\sc[Oiii]}52$\mu$m/{\sc[Oiii]}88$\mu$m line ratio as a function of electron density.
The brown, purple, blue, cyan, green, and yellow curves represent the line ratios for electron temperatures of $T_\m{e}=8000$, 10000, 13000, 16000, 20000, and 24000 K, respectively, calculated using {\sf PyNeb}.
The black dashed curve shows the fit for $T_\m{e}=13000$ K (Equation \ref{eq_OIIIr}).
}
\label{fig_OIIIr_ne}
\end{figure}

The line ratio of {\sc[Oiii]}52$\mu$m/{\sc[Oiii]}88$\mu$m is another reliable indicator of the electron density because the emissivities of these two FIR emission lines do not strongly depend on temperature but density \citep[e.g.,][]{2012MNRAS.423L..35P,2020ApJ...903..150J}.  
We use {\sf PyNeb} to calculate the line ratio as a function of electron density and temperature.  
Figure \ref{fig_OIIIr_ne} shows the line ratio as a function of electron density for various temperatures.  
The ratio is largely insensitive to the electron temperature but changes rapidly over the electron density range of $\sim100-10000$ cm$^{-3}$, reflecting the different critical densities of {\sc[Oiii]}52$\mu$m and {\sc[Oiii]}88$\mu$m (3600 and 510 cm$^{-3}$, respectively).  
The ratio for an electron temperature of $T_\m{e}=13000$ K is well fitted by the following function:  
\begin{equation}
f_\m{[OIII]52\mu m}/f_\m{[OIII]88\mu m}=a\frac{b+n_\m{e}}{c+n_\m{e}}, \label{eq_OIIIr}
\end{equation}
where $a=10.53$, $b=224.46$, and $c=4181.45$.  
The electron density measured with the {\sc[Oiii]}52$\mu$m/{\sc[Oiii]}88$\mu$m ratio or their 2$\sigma$ upper limits are $n_\m{e,FIR}\simeq500$, $<90$, and $<100$ cm$^{-3}$ for J0217-0208, SXDF-NB1006-2, and J1211-0118, respectively, as summarized in Table \ref{tab_target}.
\redc{Although \citet{2019ARA&A..57..511K} discuss that the {\sc[Oiii]}52$\mu$m/{\sc[Oiii]}88$\mu$m ratio can exhibit some dependence on the ionization parameter at $n_\m{e}\lesssim10$ cm$^{-3}$, our estimates and upper limits do not significantly change by changing the ionization parameter.}

\begin{figure}
\centering
\begin{center}
\includegraphics[width=0.99\hsize, bb=3 8 350 350,clip]{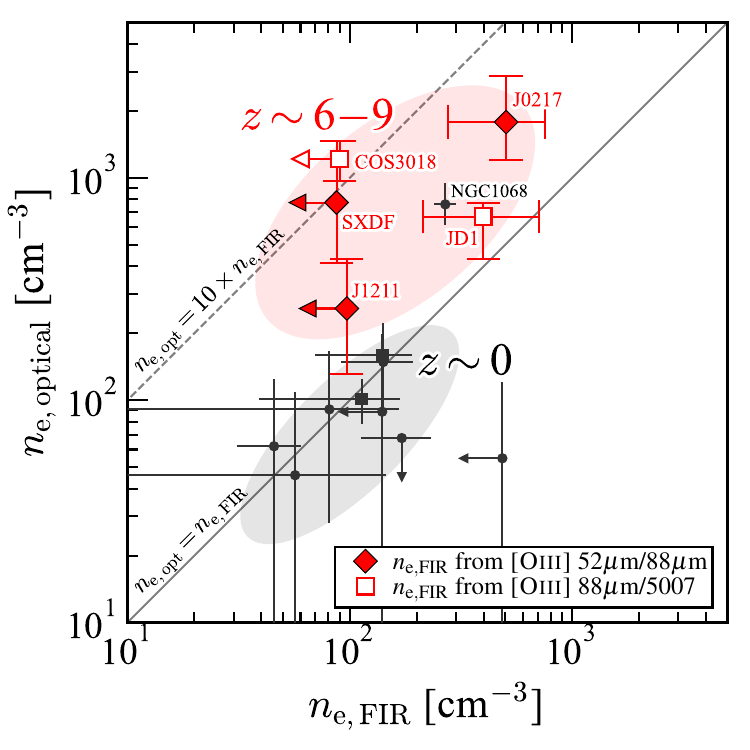}
\end{center}
\caption{
Comparison of electron densities measured using the optical line ratio ($n_\m{e,optical}$) and the FIR {\sc[Oiii]} line ($n_\m{e,FIR}$).
The red diamonds represent our measurements from the {\sc[Oiii]}52$\mu$m/{\sc[Oiii]}88$\mu$m and {\sc[Oii]}$\lambda$3729/{\sc[Oii]}$\lambda$3726 ratios for J0217-0208, SXDF-NB1006-2, and J1211-0118.
The red open squares show densities measured using the {\sc[Oiii]}88$\mu$m/{\sc[Oiii]}$\lambda$5007 and {\sc[Oii]}$\lambda$3729/{\sc[Oii]}$\lambda$3726 ratios for COS-3018555981 \citep{2022MNRAS.515.1751W,2024arXiv241107695S} and MACS1149-JD1 \citep{2023ApJ...957L..18S,2024MNRAS.533.2488M}.
The black circles (squares) represent densities of galaxies at $z\sim0$ measured using the {\sc[Oiii]}52$\mu$m/{\sc[Oiii]}88$\mu$m and {\sc [Sii]}$\lambda$6716/{\sc [Sii]}$\lambda$6731 ({\sc[Oii]}$\lambda$3729/{\sc[Oii]}$\lambda$3726) ratios in \citet{2006ApJS..164...81M} and \citet{2008ApJS..178..280B} \citep{2023NatAs...7..771C,2024arXiv240518476C}.
At $z\sim0$, the electron densities measured with the optical line ratios are comparable to those measured with the FIR {\sc[Oiii]} line ratio, except for a Seyfert galaxy NGC1068.
At $z\sim6-9$, the FIR-based electron densities are systematically lower than those with the optical line ratios.
}
\label{fig_ne_ne}
\end{figure}

\begin{figure}
\centering
\begin{center}
\includegraphics[width=0.99\hsize, bb=8 14 347 354,clip]{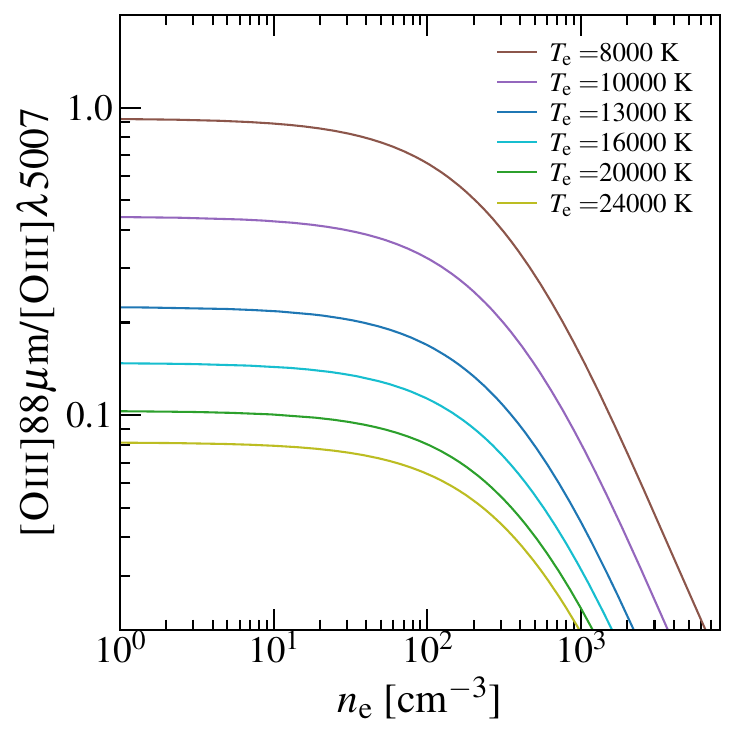}
\end{center}
\caption{
Same as Figure \ref{fig_OIIIr_ne}, but for the {\sc[Oiii]}88$\mu$m/{\sc[Oiii]}$\lambda$5007 ratio.
This ratio depends on both the electron temperature and density.
Therefore, if the temperature is determined from other lines (e.g., {\sc[Oiii]}$\lambda$4363), we can derive the electron density.
}
\label{fig_OIIIr_5007_ne}
\end{figure}

\subsubsection{Comparison of Densities from Optical and FIR}\label{ss_ne_ne}

In Figure \ref{fig_ne_ne}, we compare the electron densities of our galaxies measured using the {\sc[Oiii]}52$\mu$m/{\sc[Oiii]}88$\mu$m ($n_\m{e,FIR}$) and {\sc[Oii]}$\lambda$3729/{\sc[Oii]}$\lambda$3726 ($n_\m{e,optical}$) line ratios.  
The electron densities measured with the {\sc[Oiii]}52$\mu$m/{\sc[Oiii]}88$\mu$m ratios (the red diamonds) are systematically lower than those measured with the {\sc[Oii]}$\lambda$3729/{\sc[Oii]}$\lambda$3726 ratios.  
Interestingly, for J0217-0208 and SXDF-NB1006-2, we do not observe the low-density components suggested by $n_\m{e,FIR}$ (500 and $<100$ cm$^{-3}$, respectively) in the density maps derived from the {\sc[Oii]}$\lambda$3729/{\sc[Oii]}$\lambda$3726 ratios shown in Figure \ref{fig_map_ne}.  
These results indicate that the FIR {\sc[Oiii]} lines are tracing gas whose densities are, on average, lower than  the gas emitting the optical {\sc[Oii]} lines, suggesting the necessity for multi-zone ISM modeling with different densities when interpreting ISM properties using both optical and FIR emission lines.

A similar trend is observed in other high-redshift galaxies (the red open squares in Figure \ref{fig_ne_ne}).  
For these galaxies, although {\sc[Oiii]}52$\mu$m observations are not available, $n_\m{e,FIR}$ can be calculated from the dust-corrected {\sc[Oiii]}88$\mu$m/{\sc[Oiii]}$\lambda$5007 ratio combined with the {\sc[Oiii]}$\lambda$4363-based electron temperature, given that the {\sc[Oiii]}88$\mu$m/{\sc[Oiii]}$\lambda$5007 ratio depends on both electron temperature and density (Figure \ref{fig_OIIIr_5007_ne}; see also \citealt{2023ApJ...957L..18S,2024ApJ...964..146F}).  
\citet{2023ApJ...957L..18S} report the electron density of MACS1149-JD1 at $z=9.1$ to be $n_\m{e,FIR}\simeq400$ cm$^{-3}$ based on the {\sc[Oiii]}88$\mu$m/{\sc[Oiii]}$\lambda$5007 ratio, which is lower than the value of $n_\m{e,optical}\simeq700$ cm$^{-3}$ reported by \citet{2024MNRAS.533.2488M} using the {\sc[Oii]}$\lambda$3729/{\sc[Oii]}$\lambda$3726 ratio.  
Similarly, we estimate the electron density of COS-3018555981 at $z=6.9$ from the {\sc[Oiii]}88$\mu$m/{\sc[Oiii]}$\lambda$5007 ratio, using the {\sc[Oiii]}88$\mu$m and optical emission line fluxes reported by \citet{2022MNRAS.515.1751W} and \citet{2024arXiv241107695S}, respectively.  
We sum the fluxes of the main, north, and south components of COS-3018555981 defined in  \citet{2024arXiv241107695S} to obtain the total optical flux, which we compare with the ALMA {\sc[Oiii]}88$\mu$m line flux.  
We derive a 2$\sigma$ upper limit of $n_\m{e,FIR}<90$ cm$^{-3}$ because the dust-corrected {\sc[Oiii]}88$\mu$m/{\sc[Oiii]}$\lambda$5007 ratio, $f_\m{[OIII]88\mu m}/f_\m{[OIII]\lambda5007}=0.22\pm0.04$, reaches the maximum ratio in the low-density limit with an electron temperature of $T_\m{e}\simeq13000$ K estimated from the {\sc[Oiii]}$\lambda$4363 line (see Figure \ref{fig_OIIIr_5007_ne}).  
This upper limit is significantly lower than the density inferred from the {\sc[Oii]}$\lambda$3729/{\sc[Oii]}$\lambda$3726 ratio for COS-3018555981, $n_\m{e,optical}\simeq1000$ cm$^{-3}$.  
These results clearly indicate that $n_\m{e,FIR}$ values for galaxies at $z=6-9$ are systematically lower than their $n_\m{e,optical}$ values, regardless of whether $n_\m{e,FIR}$ is measured with {\sc[Oiii]}52$\mu$m/{\sc[Oiii]}88$\mu$m or {\sc[Oiii]}88$\mu$m/{\sc[Oiii]}$\lambda$5007 ratios.

These discrepancies are in stark contrast with $z\sim0$ galaxies whose optical and FIR-derived densities are in agreement.
\citet{2023NatAs...7..771C,2024arXiv240518476C} measure the electron densities of local galaxies Mrk71 and Haro3, finding that their densities derived from the {\sc[Oii]}$\lambda$3729/{\sc[Oii]}$\lambda$3726 ratios are $\sim100$ cm$^{-3}$, comparable to those from the {\sc[Oiii]}52$\mu$m/{\sc[Oiii]}88$\mu$m ratios.  
\citet{2006ApJS..164...81M} and \citet{2008ApJS..178..280B} present rest-frame optical and FIR emission line fluxes of nearby galaxies, respectively.  
For these local galaxies, we derive $n_\mathrm{e,optical}$ and $n_\mathrm{e,FIR}$ using the optical {\sc[Sii]}$\lambda$6716/{\sc[Sii]}$\lambda$6731 and FIR {\sc[Oiii]}52$\mu$m/{\sc[Oiii]}88$\mu$m line ratios reported therein.  
\redc{Note that although the ionization potentials of {\sc[Oii]} and {\sc[Sii]} are different (e.g., \citealt{2021ApJ...909...78D}, see also \citealt{2025arXiv250513677B}), \citet{2016ApJ...816...23S} report that the densities from {\sc[Oii]} and {\sc[Sii]} agree well using high-resolution spectra with $R\sim8000-23000$.}
As shown in Figure \ref{fig_ne_ne}, $n_\m{e,optical}$ for most nearby galaxies is comparable to their $n_\m{e,FIR}$, except for a Seyfert galaxy NGC1068, which shows a discrepancy between $n_\m{e,FIR}$ and $n_\m{e,optical}$ similar to galaxies at $z\sim6-9$.

\begin{figure*}
\centering
\begin{center}
\includegraphics[width=0.99\hsize, bb=2 4 859 425,clip]{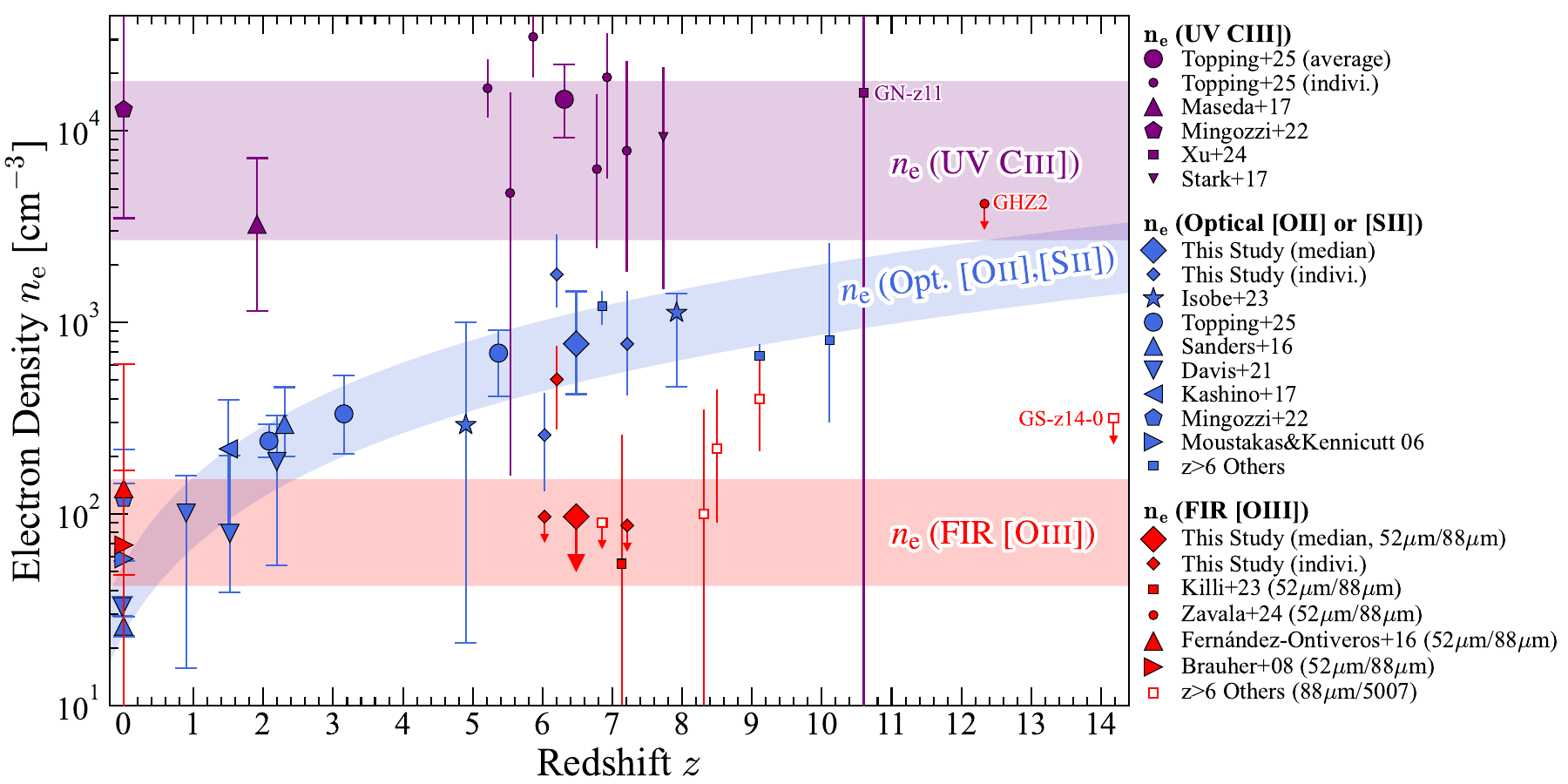}
\end{center}
\vspace{-0.5cm}
\caption{
Redshift evolution of electron densities.
The red, blue, and purple symbols represent electron densities measured using the FIR {\sc[Oiii]} emission line, optical {\sc[Oii]}$\lambda\lambda$3726,3729 or {\sc[Sii]}$\lambda\lambda$6716,6731 emission lines, and UV {\sc Ciii]}$\lambda$1907,1909 emission lines, respectively.
We find that the electron densities measured with the FIR {\sc[Oiii]} emission line are typically $\sim100$ cm$^{-3}$, regardless of redshift.
In contrast, densities measured with the optical lines increase towards higher redshifts.
The densities measured with the UV {\sc Ciii]}$\lambda$1907,1909 lines are higher than 1000 cm$^{-3}$ and exceed those measured with the optical lines at fixed redshift.
The red small diamonds represent our individual measurements for the three galaxies from the {\sc[Oiii]}52$\mu$m/{\sc[Oiii]}88$\mu$m ratio and the red large diamond represents the median of these measurements.
The red small square and circle represent densities measured for A1689-zD1 at $z=7.1$ \citep{2023MNRAS.521.2526K} and GHZ2 at $z=12.3$ \citep{2024ApJ...977L...9Z}, respectively.
The red up-pointing and right-pointing triangles represent the median densities of galaxies at $z\sim0$ measured using the {\sc[Oiii]}52$\mu$m/{\sc[Oiii]}88$\mu$m ratio in \citet{2016ApJS..226...19F} and \citet{2008ApJS..178..280B}, respectively.
The red small open squares represent measurements from the {\sc[Oiii]}88$\mu$m/{\sc[Oiii]}$\lambda$5007 ratio for COS-3018555981 at $z=6.9$ \citep{2022MNRAS.515.1751W,2024arXiv241107695S}, MACS0416-Y1 at $z=8.3$ \citep{2024ApJ...977L..36H}, ID4590 at $z=8.5$ \citep{2024ApJ...964..146F}, MACS1149-JD1 at $z=9.1$ \citep{2023ApJ...957L..18S}, and JADES-GS-z14-0 at $z=14.2$ \citep{2024arXiv240920549S,2024arXiv240920533C}.
The blue small diamonds represent our individual measurements from the {\sc[Oii]}$\lambda$3729/{\sc[Oii]}$\lambda$3726 ratio, and the blue large diamond represents the median of them.
Densities measured with the {\sc[Oii]}$\lambda$3729/{\sc[Oii]}$\lambda$3726 or {\sc[Sii]}$\lambda$6716/{\sc[Sii]}$\lambda$6731 ratios from the literature are shown with the blue symbols (\citealt{2023ApJ...956..139I}: star, \citealt{2025arXiv250208712T}: circle, \citealt{2016ApJ...816...23S}: up-pointing triangle, \citealt{2021ApJ...909...78D}: down-pointing triangle, \citealt{2017ApJ...835...88K}: left-pointing triangle, \citealt{2022ApJ...939..110M}: pentagon, \citealt{2006ApJS..164...81M}: right-pointing triangle).
The blue small squares represent measurements for COS-3018555981 at $z=6.9$ \citep{2024arXiv241107695S}, MACS1149-JD1 at $z=9.1$ \citep{2024MNRAS.533.2488M}, and MACS0647-JD at $z=10.2$ \citep{2024ApJ...973...47A}.
The purple small circles represent measurements from the {\sc[Ciii]}$\lambda$1907/{\sc Ciii]}$\lambda$1909 ratio in \citet{2025arXiv250208712T}, and the purple large circle represents their average.
The purple up-pointing triangle and pentagon are measurements from \citet{2017A&A...608A...4M} and \citet{2022ApJ...939..110M}, respectively, while the purple small square and down-pointing triangle represent densities for GN-z11 at $z=10.6$ \citep{2024ApJ...976..142X} \redc{and for EGS-zs8-1 at $z=7.7$ \citep{2017MNRAS.464..469S}, respectively}.
}
\label{fig_ne_z}
\end{figure*}

\subsubsection{Redshift Evolution}\label{ss_ne_z}

Figure \ref{fig_ne_z} compiles the electron density measurements obtained in this study and from the literature using rest-frame UV, optical, and FIR lines. 
Our densities measured with the JWST optical {\sc[Oii]} lines (the blue diamonds) are consistent with the redshift evolution of electron density reported in previous JWST studies using the optical {\sc[Oii]} and/or {\sc[Sii]} lines \citep{2023ApJ...956..139I,2025arXiv250208712T}. 
Combined with results from the literature, it is clear that the density measured with the optical {\sc[Oii]} or {\sc[Sii]} lines increases toward higher redshifts.

Our densities measured with the ALMA FIR {\sc[Oiii]}52$\mu$m/{\sc[Oiii]}88$\mu$m line ratio at $z\sim6-7$ (the red diamonds) are lower than the redshift evolution trend seen with the optical lines and are comparable to those of galaxies at $z\sim0$ \citep{2016ApJS..226...19F,2008ApJS..178..280B}. 
The density of A1689-zD1 at $z=7.1$, measured with {\sc[Oiii]}52$\mu$m/{\sc[Oiii]}88$\mu$m, also shows a similarly low value, $n_\m{e,FIR}\sim55\ \m{cm}^{-3}$.
Some galaxies at $z>6$, including a galaxy at $z=14.2$, JADES-GS-z14-0, are observed with both JWST and ALMA, with detections of the {\sc[Oiii]}88$\mu$m line \citep{2024arXiv240920533C,2024ApJ...964..146F,2024ApJ...977L..36H,2023ApJ...957L..18S,2024arXiv241107695S,2024arXiv240920549S,2022MNRAS.515.1751W}.
The densities of these galaxies, measured with the {\sc[Oiii]}88$\mu$m/{\sc[Oiii]}$\lambda$5007 ratios (the red open squares), also show lower values than the redshift evolution trend seen with the optical lines.
These results indicate that $n_\m{e,FIR}$ of high-redshift galaxies are low and do not show redshift evolution, regardless of whether {\sc[Oiii]}52$\mu$m/{\sc[Oiii]}88$\mu$m or {\sc[Oiii]}88$\mu$m/{\sc[Oiii]}$\lambda$5007 is used, which is in contrast to the evolution trend seen in the optical-based densities.
We will discuss the physical interpretation of these two different redshift evolution trends in Section \ref{ss_dis_z06}.

In Figure \ref{fig_ne_z}, we also show densities measured with the rest-frame UV {\sc[Ciii]}$\lambda$1907/{\sc Ciii]}$\lambda$1909 ratio.
As reported in \citet{2022ApJ...939..110M} and \citet{2025arXiv250208712T}, the densities measured with the rest-frame UV {\sc Ciii]} lines are very high, $\gtrsim1000$ cm$^{-3}$, higher than those measured with the optical and FIR lines.
These results indicate that the rest-frame {\sc Ciii]} lines trace high-density regions in galaxies at all redshifts.

\section{JWST \& ALMA Joint Analysis}\label{ss_jwst_alma}

\subsection{Galaxy Sample for the Analysis}

To further investigate the ISM properties of high-redshift galaxies in detail and test the validity of the direct-$T_\m{e}$ method, we conduct a JWST and ALMA joint analysis to constrain the ISM parameters using emission lines detected with both telescopes. 
The galaxies analyzed here are limited to those with detections of the {\sc[Oii]}$\lambda\lambda$3726,3729, H$\gamma$, {\sc[Oiii]}$\lambda$4363, H$\beta$, {\sc[Oiii]}$\lambda\lambda$4959,5007, and {\sc[Oiii]}88$\mu$m lines.
From our three galaxies, we select J0217-0208 and SXDF-NB1006-2 for the joint analysis.
We do not include J1211-0118, as its {\sc[Oiii]}$\lambda$4363 line is not detected.
In addition, we analyze COS-3018555981 at $z=6.9$ and MACS1149-JD1 at $z=9.1$, taken from the literature.
Figure \ref{fig_Ms_SFR} shows the stellar masses and SFRs of these two galaxies. 
We use the emission line fluxes of COS-3018555981 reported in \citet{2022MNRAS.515.1751W} and \citet{2024arXiv241107695S} in the same manner as in Section \ref{ss_ne_ne}.
For the rest-frame optical emission lines in MACS1149-JD1, we reduced the NIRSpec IFU data taken in GTO-1262 (PI: N. L\"utzgendorf) and measured the emission line fluxes in the same manner as in Section \ref{ss_nirspec}.
For the {\sc[Oiii]}88$\mu$m line flux, we adopt the value obtained in \citet{2018Natur.557..392H}.
All emission line fluxes are corrected for dust \redc{attenuation} using the Balmer decrement, following the procedure described in Section \ref{ss_dust}.

\subsection{1-Zone Model Fitting}\label{ss_1zone}

First, we test whether the ISM parameters constrained with the JWST optical emission lines can reproduce the luminosity of the ALMA FIR emission lines.
Table \ref{tab_Z1} summarizes the electron densities, temperatures, and metallicities of the four galaxies analyzed here, determined with the {\sc[Oii]}$\lambda\lambda$3726,3729, {\sc[Oiii]}$\lambda$4363, H$\beta$, and {\sc[Oiii]}$\lambda\lambda$4959,5007 lines.
With these parameters, we can predict the luminosities of the {\sc[Oiii]}88$\mu$m and {\sc[Oiii]}52$\mu$m lines using {\sf PyNeb}.
We hereafter refer to this model as a 1-zone model, as we assume only one set of parameters with a homogeneous condition.

\begin{figure*}
\centering
\begin{minipage}{0.95\hsize}
\centering
\begin{center}
\includegraphics[width=0.6\hsize, bb=12 4 571 248,clip]{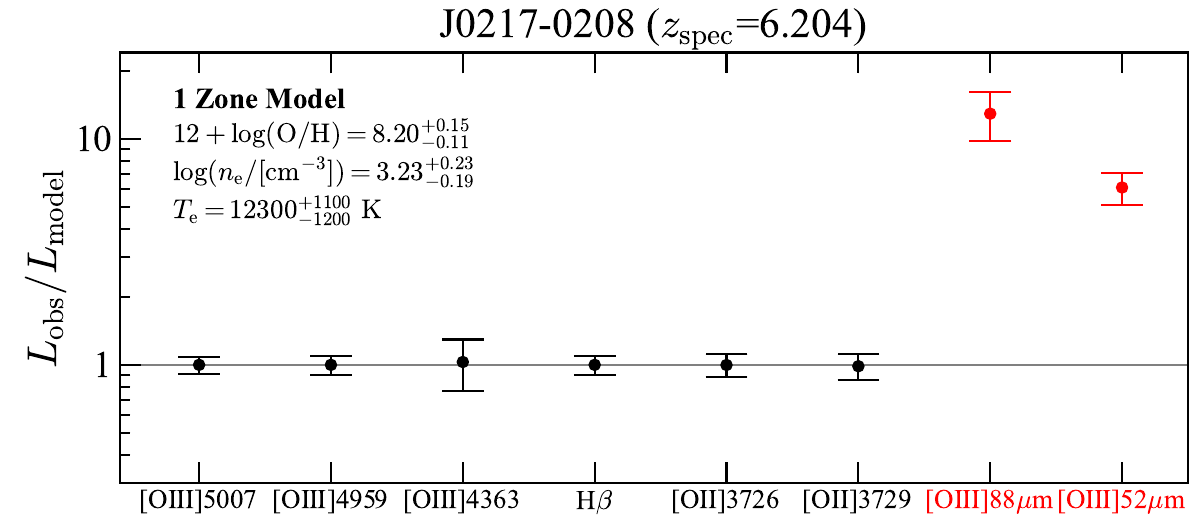}
\end{center}
\begin{center}
\includegraphics[width=0.6\hsize, bb=12 4 571 248,clip]{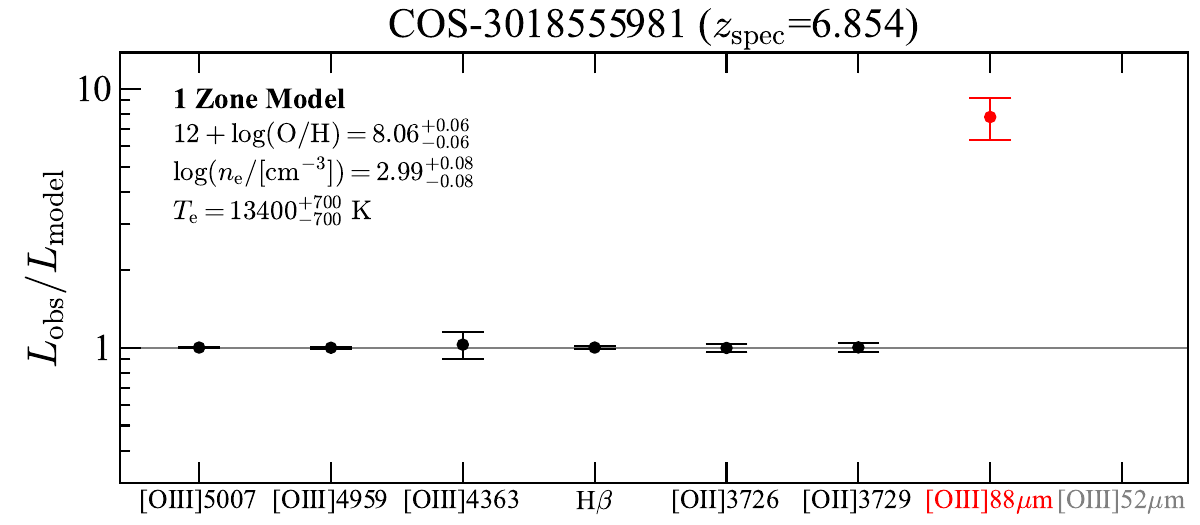}
\end{center}
\begin{center}
\includegraphics[width=0.6\hsize, bb=12 4 571 248,clip]{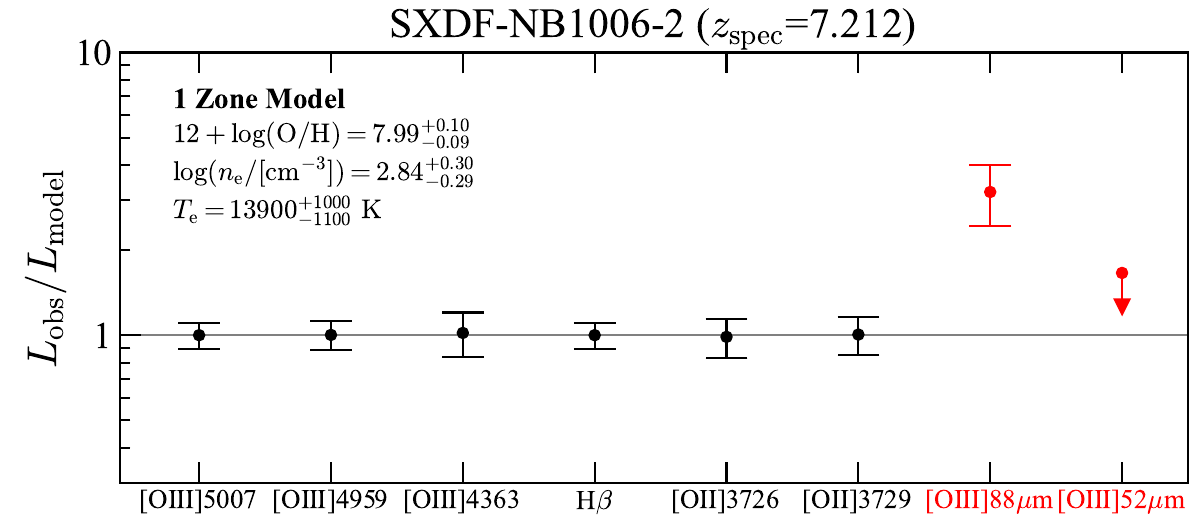}
\end{center}
\begin{center}
\includegraphics[width=0.6\hsize, bb=12 4 571 248,clip]{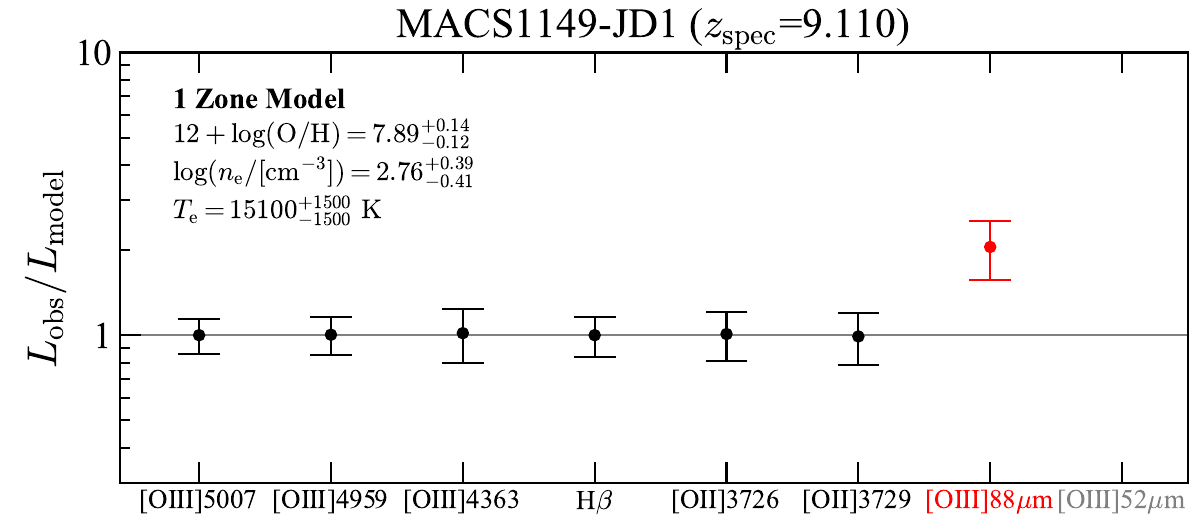}
\end{center}
\end{minipage}
\caption{
Comparison of observed and predicted luminosities with the 1-Zone model for J0217-0208, COS-3018555981, SXDF-NB1006-2, and MACS1149-JD1.
The circles represent the ratios of the observed and predicted luminosities.
Using the electron density and temperature measured from the optical {\sc[Oii]}$\lambda$3729/{\sc[Oii]}$\lambda$3726 and {\sc[Oiii]}$\lambda$4363/{\sc[Oiii]}$\lambda$5007 line ratios, the 1-zone model successfully reproduces the optical emission line luminosities (the black circles), but it significantly underpredicts the luminosities of the FIR {\sc[Oiii]}88$\mu$m and {\sc[Oiii]}52$\mu$m emission lines (the red circles).
}
\label{fig_1zone}
\end{figure*}

Figure \ref{fig_1zone} shows comparisons between the observed and predicted luminosities using the 1-zone model.
We find that the 1-zone model reproduces the observed luminosities of the JWST optical emission lines but significantly underestimates those of the ALMA FIR {\sc[Oiii]}88$\mu$m and {\sc[Oiii]}52$\mu$m lines in all four galaxies analyzed here, with the predicted {\sc[Oiii]}88$\mu$m luminosities being $\sim2-10$ times lower than the observations.
These discrepancies arise because the optical and FIR lines trace different gas components with different densities, as discussed in Section \ref{ss_ne_ne}, again suggesting the necessity of multi-zone ISM modeling with different densities.

Clearly the {\sc[Oiii]}88$\mu$m and {\sc[Oiii]}52$\mu$m emission lines could originate in highly obscured dusty regions for which no optical emission emerges, as discussed in \citet{2024arXiv240518476C}.
We now consider whether this could account for the discrepancies attributed to the 1-zone modeling.
To reconcile the situation we would need dust-obscured star formation rates $2-10$ times higher than those inferred optically.
Among the four galaxies analyzed here, SXDF-NB1006-2 and MACS1149-JD1 do not show dust continuum emission in the deep ALMA data in \citet{2023ApJ...945...69R} and \citet{2018Natur.557..392H} with the $2\sigma$ upper limits on the obscured SFRs of $<3$ and $<0.6\ M_\odot\ \m{yr^{-1}}$, respectively, assuming the dust temperature of $T_\m{dust}=40$ K.
The SFRs estimated from the H$\beta$ luminosities are 62 and 7 $M_\odot\ \m{yr^{-1}}$ for SXDF-NB1006-2 and MACS1149-JD1, respectively.
These SFR constraints indicate that, even if dust-obscured star formation is present, the obscured fraction cannot exceed 10\% which is insufficient to explain the factor of $>2$ discrepancies in the {\sc[Oiii]}88$\mu$m line luminosities between the predictions and observations.
J0217-0208 and COS-3018555981 show dust continuum detections in the ALMA data, and their dust-obscured SFRs are $\sim15$ $M_\odot\ \m{yr^{-1}}$ \citep{2020ApJ...896...93H,2022MNRAS.515.1751W}. 
Given their SFRs estimated from the dust-corrected H$\beta$ luminosities, 160 and 120 $M_\odot\ \m{yr^{-1}}$, respectively, the contributions of dust-obscured star formation to the {\sc[Oiii]}88$\mu$m luminosity can be up to $\sim10\%$, again insufficient to explain the factor of $\sim10$ discrepancies seen in the {\sc[Oiii]}88$\mu$m line luminosities.
Although the dust-obscured SFR estimates depend on the assumed dust temperature, reconciling the FIR and optical line measurements would require extremely high dust temperatures ($>120$ K), which are rarely observed in normal star-forming galaxies \citep[e.g.,][]{2024ApJ...971..161M}.
In addition, the observed {\sc[Oiii]}52$\mu$m/{\sc[Oiii]}88$\mu$m ratios in J0217-0208 and SXDF-NB1006-2 (Figure \ref{fig_1zone}) cannot be reproduced by the dust attenuation that affect both FIR lines equally without altering the line ratio.
These comparisons indicate that heavily obscured dusty gas cannot explain the discrepancies seen in the 1-zone modeling.

\begin{deluxetable*}{lccccccccccc}
\tablecaption{ISM Properties Constrained with the 1-Zone Model Using the JWST Optical Emission Lines}
\label{tab_Z1}
\tablehead{\colhead{Name} & \colhead{$z_\m{spec}$} & \colhead{$\m{log}M_*$} & \colhead{$\m{log}n_\m{e,optical}$}  & \colhead{$T_\m{e}$} & \colhead{$12+\m{log(O/H)}$} & \colhead{$E(B-V)$} \\
\colhead{} & \colhead{} & \colhead{($M_\odot$)} & \colhead{(cm$^{-3}$)} & \colhead{(K)} & \colhead{} & \colhead{(ABmag)}
} 
\startdata
J0217-0208 & $6.204$ & $10.2_{-0.2}^{+0.1}$ & $3.23^{+0.23}_{-0.19}$ & $12300^{+1100}_{-1200}$ & $8.20^{+0.15}_{-0.11}$ & $0.10\pm0.08$\\
COS-3018555981 & $6.854$ & $9.6_{-0.1}^{+0.1}$$^\dagger$ & $2.99^{+0.08}_{-0.08}$ & $13400^{+700}_{-700}$ & $8.06^{+0.06}_{-0.06}$ & $0.17\pm0.01$\\
SXDF-NB1006-2 & $7.212$ & $8.7_{-0.1}^{+0.1}$ & $2.84^{+0.30}_{-0.29}$ & $13900^{+1000}_{-1100}$ & $7.99^{+0.10}_{-0.09}$ & $0.00\pm0.11$\\
MACS1149-JD1 & $9.110$ & $8.5_{-0.1}^{+0.1}$$^*$ & $2.76^{+0.39}_{-0.41}$ & $15100^{+1500}_{-1500}$ & $7.89^{+0.14}_{-0.12}$ & $0.00\pm0.18$\\
\enddata
\tablecomments{Errors are $1\sigma$.
As shown in Figure \ref{fig_1zone}, these parameters cannot reproduce the FIR {\sc[Oiii]}88$\mu$m and {\sc[Oiii]}52$\mu$m emission lines.\\
$^\dagger$ Total stellar mass including the main, north, and south components in \citet{2024arXiv241107695S}.\\
$^*$ Taken from \citet{2024MNRAS.533.2488M} assuming the magnification factor of $\mu=10$.
}
\end{deluxetable*}

Another way to construct the 1-zone model is to determine the parameters purely based on the {\sc[Oiii]} emission lines.
However, as discussed in Section \ref{ss_ne_ne}, the electron density determined with the {\sc[Oiii]} lines (either {\sc[Oiii]}52$\mu$m/{\sc[Oiii]}88$\mu$m or {\sc[Oiii]}88$\mu$m/{\sc[Oiii]}$\lambda$5007) is inconsistent with the {\sc[Oii]}-based density.
Moreover, for J0217-0208, the electron density of $500$ cm$^{-3}$ and temperature of $12000$ K measured with the {\sc[Oiii]}52$\mu$m/{\sc[Oiii]}88$\mu$m and {\sc[Oiii]}$\lambda$4363/{\sc[Oiii]}$\lambda$5007 ratios suggest a {\sc[Oiii]}88$\mu$m/{\sc[Oiii]}$\lambda$5007 ratio of 0.2, which is inconsistent with the observed ratio of 0.4.
Thus, to reconcile the optical and FIR lines, we need a multi-zone model mixing the {\sc[Oiii]} and {\sc[Oii]} emission in both regions, rather than spatially distinct {\sc[Oiii]} and {\sc[Oii]} regions where each ion has its own temperature and density.

\redc{Finally, we test whether photoionization modeling with {\sf Cloudy} \citep{2017RMxAA..53..385F} can reproduce our results, as {\sf Cloudy} is widely used to interpret optical and FIR emission lines \citep[e.g.,][]{2019A&A...626A..23C,2020ApJ...896...93H,2022MNRAS.515.1751W,2022A&A...667A..35R,2022A&A...665A..85B,2022A&A...667A..34L}.
We use {\sf Cloudy} version 17.01 and compute the emission-line luminosities in the same manner as \citet{2020ApJ...896...93H}.
We find that the JWST optical emission lines are well reproduced, whereas the ALMA {\sc [Oiii]}88$\mu$m and {\sc [Oiii]}52$\mu$m line luminosities are underestimated by a factor of $\sim2$-10, consistent with the results obtained using {\sf PyNeb}.
}

\begin{figure*}
\centering
\begin{minipage}{0.85\hsize}
\centering
\begin{minipage}{0.65\hsize}
\begin{center}
\includegraphics[width=0.99\hsize, bb=7 7 571 291,clip]{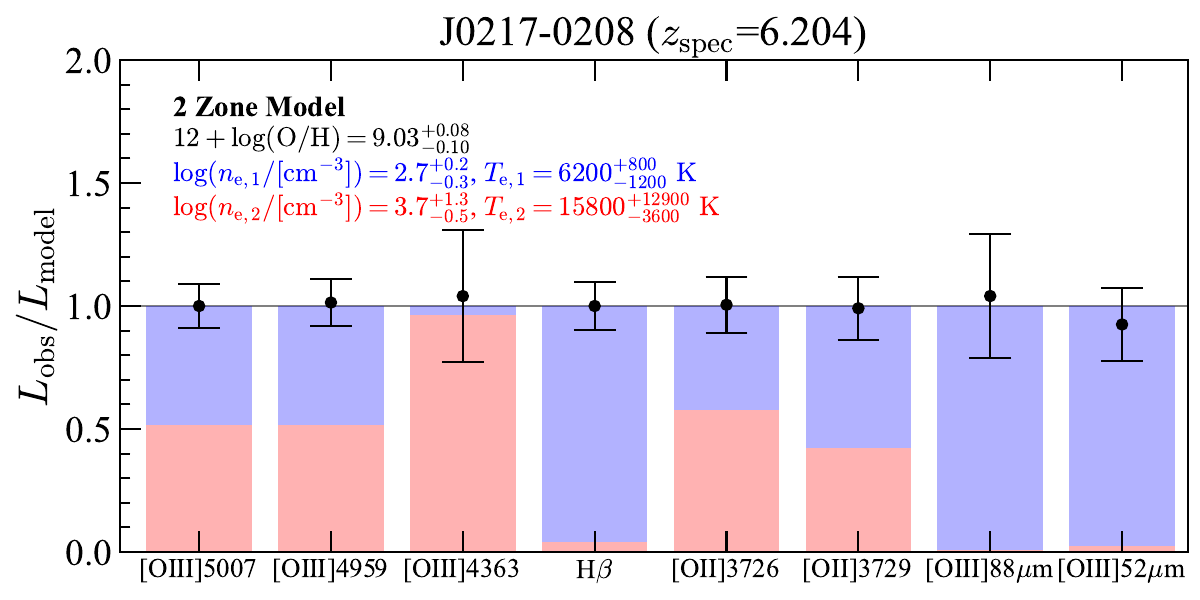}
\end{center}
\end{minipage}
\begin{minipage}{0.3\hsize}
\begin{center}
\includegraphics[width=0.99\hsize, bb=5 24 354 424]{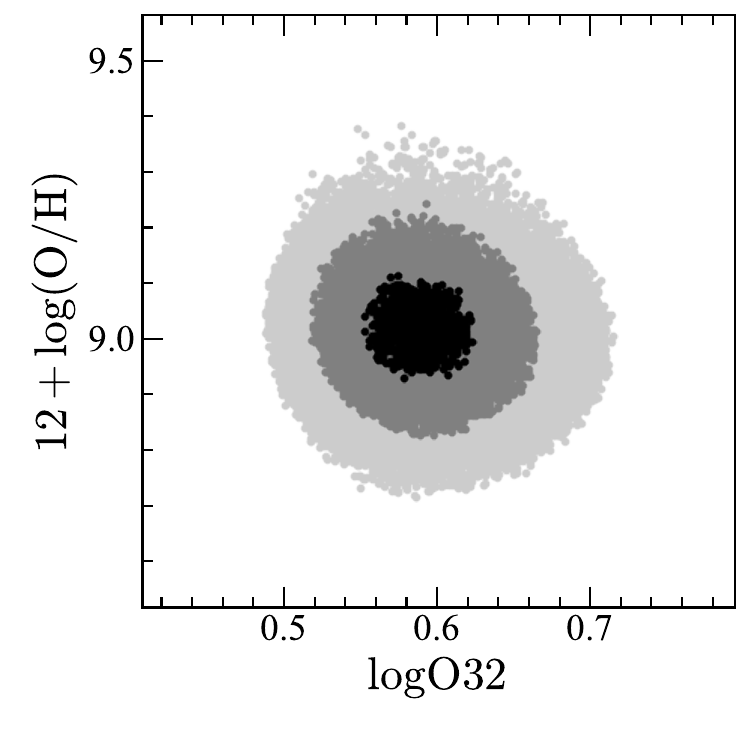}
\end{center}
\end{minipage}
\vspace{-0.2cm}
\end{minipage}
\begin{minipage}{0.85\hsize}
\centering
\begin{minipage}{0.65\hsize}
\begin{center}
\includegraphics[width=0.99\hsize, bb=7 7 571 291,clip]{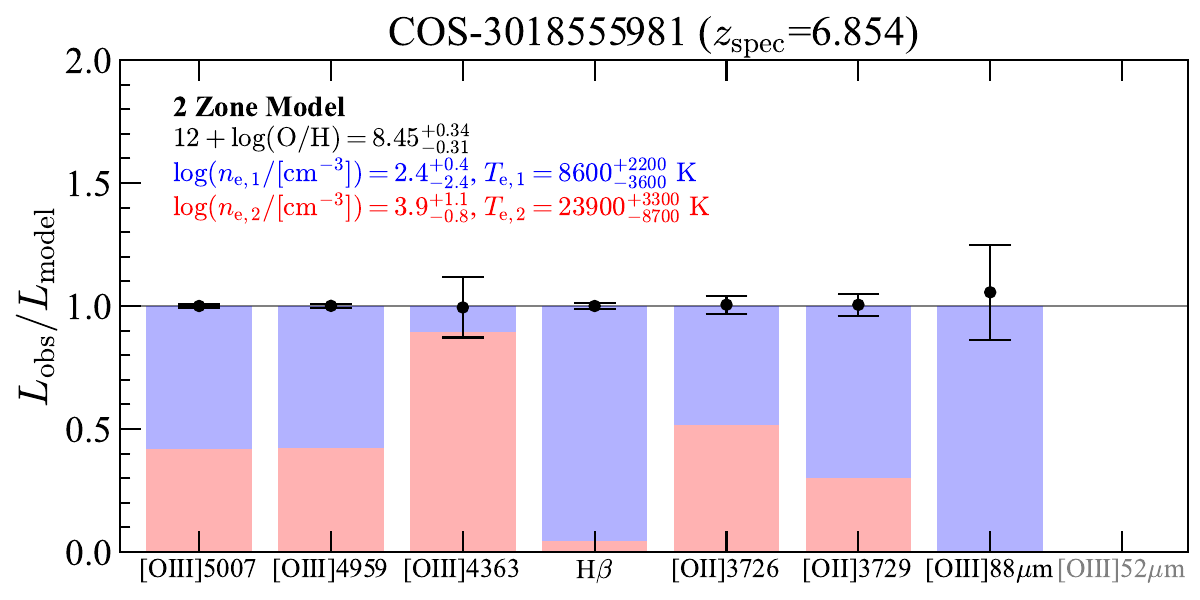}
\end{center}
\end{minipage}
\begin{minipage}{0.3\hsize}
\begin{center}
\includegraphics[width=0.99\hsize, bb=5 24 354 424]{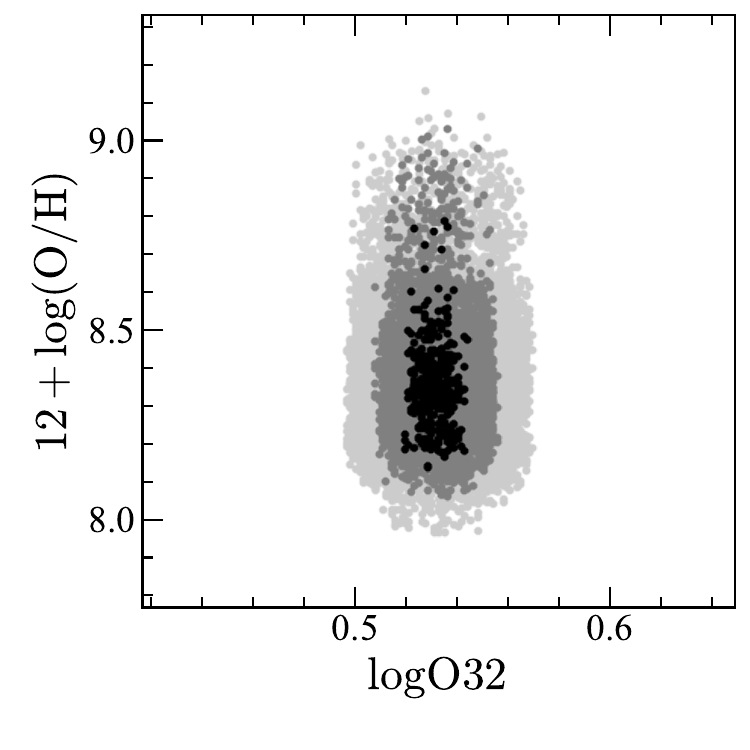}
\end{center}
\end{minipage}
\vspace{-0.2cm}
\end{minipage}
\begin{minipage}{0.85\hsize}
\centering
\begin{minipage}{0.65\hsize}
\begin{center}
\includegraphics[width=0.99\hsize, bb=7 7 571 291,clip]{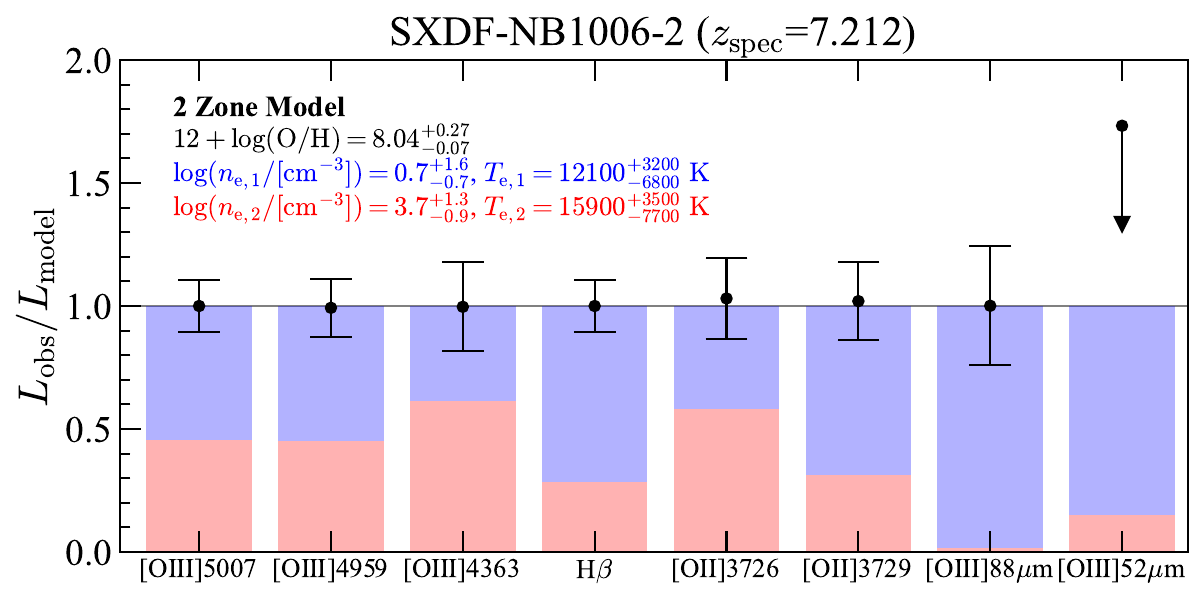}
\end{center}
\end{minipage}
\begin{minipage}{0.3\hsize}
\begin{center}
\includegraphics[width=0.99\hsize, bb=5 24 354 424]{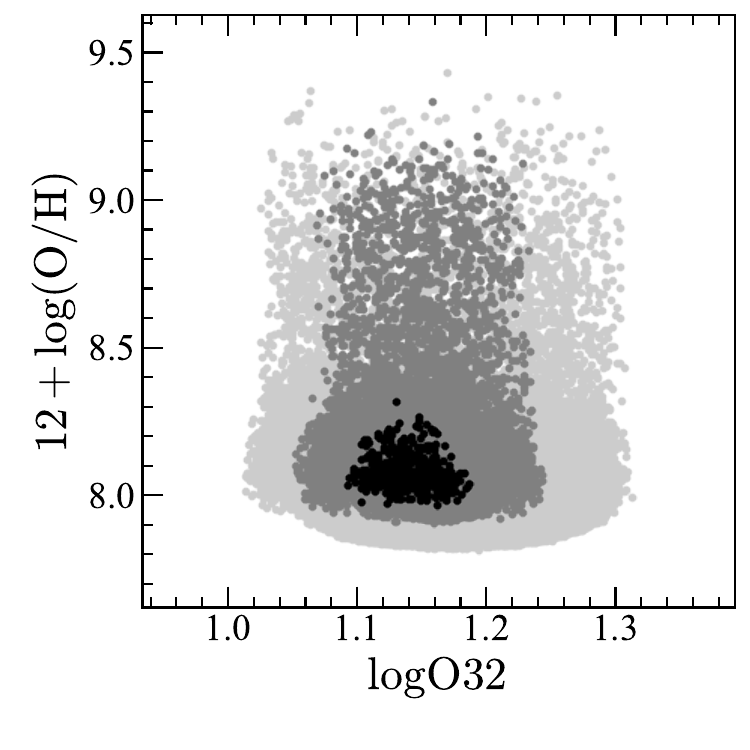}
\end{center}
\end{minipage}
\vspace{-0.2cm}
\end{minipage}
\begin{minipage}{0.85\hsize}
\centering
\begin{minipage}{0.65\hsize}
\begin{center}
\includegraphics[width=0.99\hsize, bb=7 7 571 291,clip]{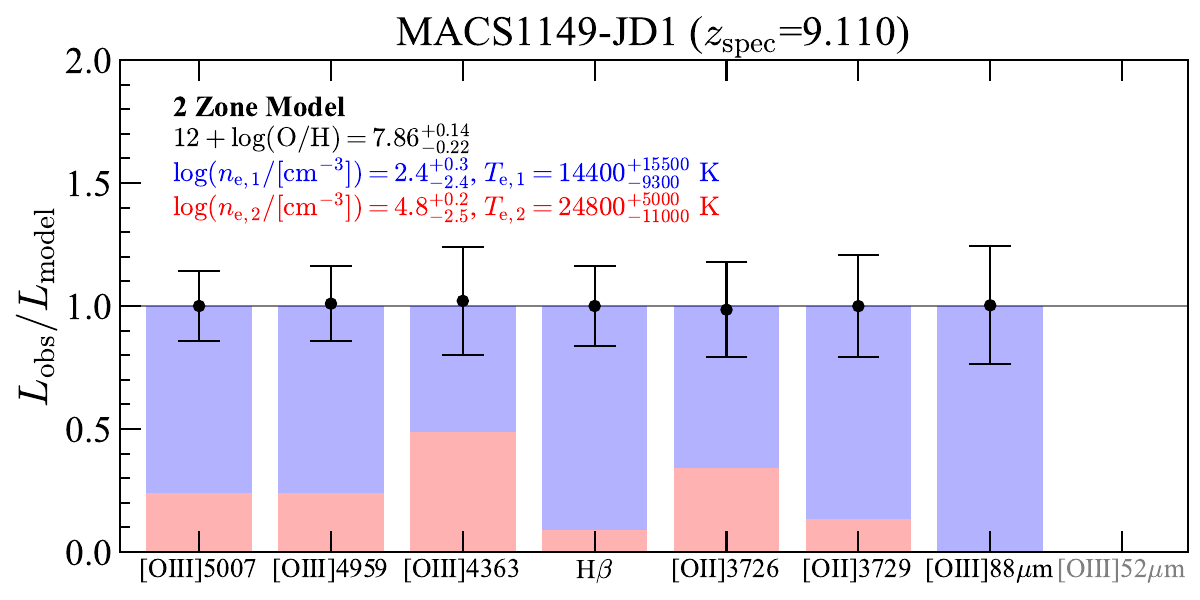}
\end{center}
\end{minipage}
\begin{minipage}{0.3\hsize}
\begin{center}
\includegraphics[width=0.99\hsize, bb=5 24 354 424]{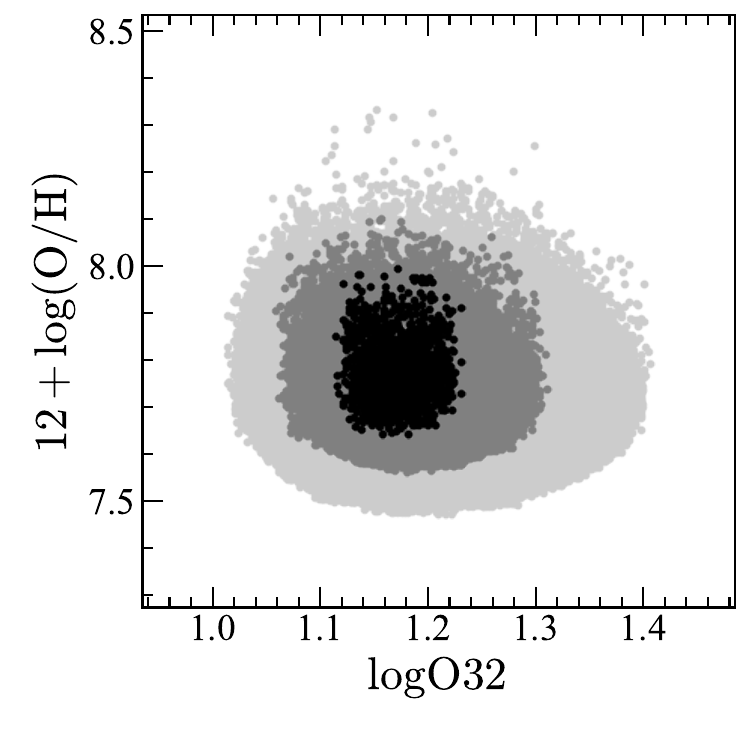}
\end{center}
\end{minipage}
\end{minipage}
\caption{
Results of the 2-zone model fitting with the low-density and high-density gas for optical and FIR emission lines observed with JWST and ALMA.
The left-hand panels show comparisons between the observed and predicted luminosities, with the black circles representing the ratios of them.
With a high temperature in the high-density region and a low temperature in the low-density region, the 2-zone model can successfully reproduce all of the emission lines from optical to FIR.
The blue and red bars indicate the fractions of the luminosities emitted from the low-density and high-density regions, respectively.
The FIR {\sc[Oiii]}88$\mu$m and {\sc[Oiii]}52$\mu$m emission lines primarily arise from the low-density region due to their low critical densities (510 and 3600 cm$^{-3}$, respectively).
The {\sc[Oiii]}$\lambda$4363 luminosity is dominated by emission from the high-density and high-temperature region, as its emissivity strongly depends on the electron temperature.
The right-hand panels show the constraints on the metallicity ($Z$) and the O32 ratio from the 2-zone model.
Data points in the $1\sigma$, $2\sigma$, and $3\sigma$ confidence intervals are indicated with the black, gray, and light-gray dots, respectively.
}
\label{fig_2zone}
\end{figure*}

\begin{deluxetable*}{lccccccccccc}
\tablecaption{ISM Properties Constrained with the 2-Zone Model using the JWST and ALMA Emission Lines}
\label{tab_Z2}
\tablehead{\colhead{Name} & \colhead{$12+\m{log(O/H)}$} & \colhead{$\m{log}n_\m{e,1}$} & \colhead{$\m{log}n_\m{e,2}$} & \colhead{$T_\m{e,1}$} & \colhead{$T_\m{e,2}$} & \colhead{$\m{logO_{32}}$} & \colhead{$\m{log}(L_\m{[OIII]5007,1}/L_\m{[OIII]5007,2})$}  \\
\colhead{} & \colhead{} & \colhead{(cm$^{-3}$)} & \colhead{(cm$^{-3}$)} & \colhead{(K)} & \colhead{(K)} & \colhead{}& \colhead{}
} 
\startdata
J0217-0208 & $9.03^{+0.08}_{-0.10}$  & $2.7^{+0.2}_{-0.3}$  & $3.7^{+1.3}_{-0.5}$  & $6200^{+800}_{-1200}$  & $15800^{+12900}_{-3600}$  & $0.59^{+0.04}_{-0.03}$  & $0.0^{+0.8}_{-0.7}$ \\
COS-3018555981 & $8.45^{+0.34}_{-0.31}$  & $2.4^{+0.4}_{-2.4}$  & $3.9^{+1.1}_{-0.8}$  & $8600^{+2200}_{-3600}$  & $23900^{+3300}_{-8700}$  & $0.53^{+0.01}_{-0.01}$  & $-0.1^{+1.7}_{-0.3}$ \\
SXDF-NB1006-2 & $8.04^{+0.27}_{-0.07}$  & $0.7^{+1.6}_{-0.7}$  & $3.7^{+1.3}_{-0.9}$  & $12100^{+3200}_{-6800}$  & $15900^{+3500}_{-7700}$  & $1.15^{+0.04}_{-0.05}$  & $-0.1^{+1.9}_{-0.5}$ \\
MACS1149-JD1 & $7.86^{+0.14}_{-0.22}$  & $2.4^{+0.3}_{-2.4}$  & $4.8^{+0.2}_{-2.5}$  & $14400^{+15500}_{-9300}$  & $24800^{+5000}_{-11000}$  & $1.16^{+0.07}_{-0.05}$  & $-0.5^{+62.2}_{-5.3}$ \\
\enddata
\tablecomments{Errors are $1\sigma$.
}
\end{deluxetable*}

\subsection{2-Zone Model Fitting}\label{ss_2zone}

Motivated by the failures of the 1-zone modeling in Section \ref{ss_1zone}, we introduce a 2-zone model and relax the single-density and single-temperature assumptions as a first step toward comprehensively understanding the ISM properties of these galaxies \redc{(see also \citealt{2025ApJ...991L..38U} for an independent analysis  on a $z\sim6.8$ galaxy using a similar approach)}.
In our 2-zone model, to keep it as simple as possible, we assume that the two ionized gas regions have different densities and temperatures but share the same metallicity and O32 ratio (i.e., the same ionization parameter in the low-density limit), where $\m{O32}={\textsc{[Oiii]}}\lambda5007/{\textsc{[Oii]}}\lambda\lambda3726,3729$.  
Systematic uncertainties arising from the assumption of the same metallicity and O32 ratio will be discussed in Section \ref{ss_sys}.  
Thus, we have seven free parameters in the 2-zone model: the densities in the two regions ($n_\m{e,1}$ and $n_\m{e,2}$), the temperatures in the two regions ($T_\m{e,1}$ and $T_\m{e,2}$), the metallicity ($\m{12+log(O/H)}$), the O32 ratio (O32), and the ratio of the {\sc[Oiii]}$\lambda$5007 line luminosities between the two regions ($L_\m{[OIII]5007,1}/L_\m{[OIII]5007,2}$), where subscripts 1 and 2 refer to the low- and high-density regions, respectively.  
It is reasonable to mix the {\sc[Oiii]} and {\sc[Oii]} emission in both regions in our 2-zone modeling because the aperture used to obtain our spectra contains many H{\sc ii} regions within a galaxy.  
Moreover, as discussed at the end of Section \ref{ss_1zone}, a model with spatially distinct {\sc[Oiii]} and {\sc[Oii]} regions without mixing cannot reproduce the observed {\sc[Oiii]} lines of J0217-0208.

\begin{figure*}
\centering
\begin{center}
\includegraphics[width=0.7\hsize, bb=7 13 461 393,clip]{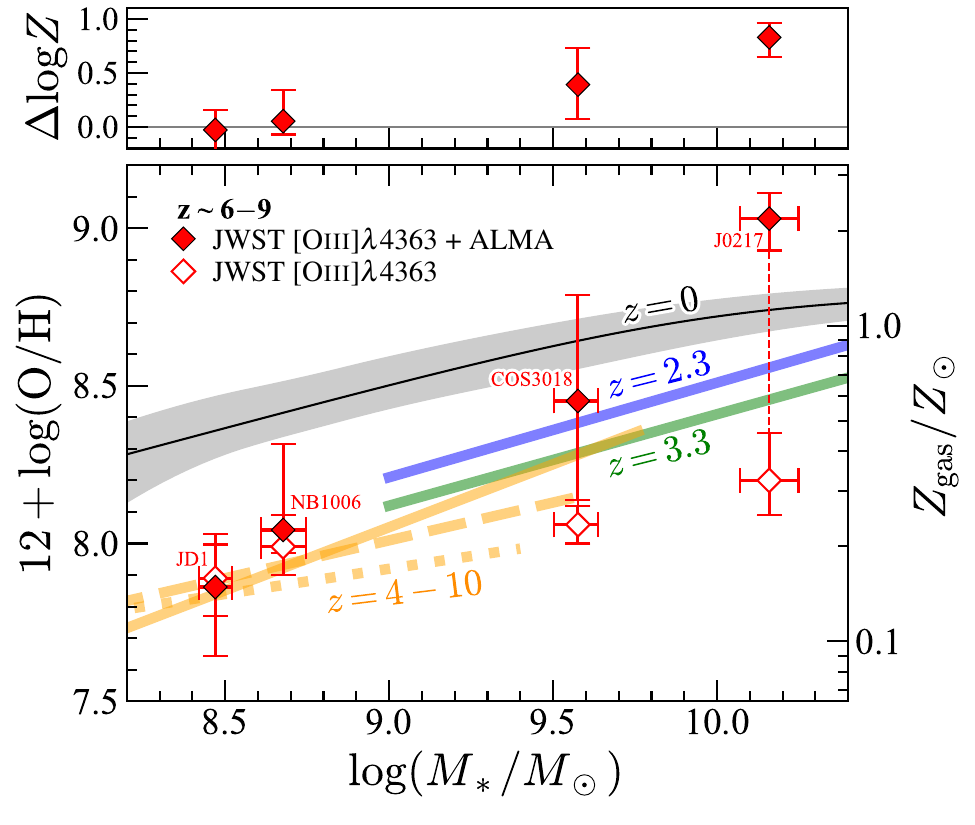}
\end{center}
\vspace{-0.5cm}
\caption{
The bottom panel shows the mass-metallicity relation.
The red-filled diamonds represent our fiducial metallicities measured with the 2-zone model fitting (see Section \ref{ss_2zone}) using the emission lines observed with JWST and ALMA.
The red open diamonds indicate metallicities measured with the direct-$T_\m{e}$ method using {\sc[Oiii]}$\lambda$4363, which fail to reproduce the ALMA FIR {\sc[Oiii]} line luminosities (see Section \ref{ss_1zone}).
As shown in the top panel, metallicities measured with the direct-$T_\m{e}$ method can be significantly underestimated by up to $\sim0.8$ dex, compared to those derived from the JWST and ALMA joint analyses.
The lines in the bottom panel show the mass-metallicity relations at $z\sim0$ in \citet[][the black line]{2013ApJ...765..140A}, at $z\sim2.3$ and 3.3 in \citet[the blue and green lines, respectively]{2021ApJ...914...19S}, and at $z\sim4-10$ in \citet[][the orange dashed line]{2023ApJS..269...33N}, \citet[][the orange dotted line]{2024A&A...684A..75C}, and \citet[][the orange solid line]{2025arXiv250110559R}.
}
\label{fig_Z_Ms}
\end{figure*}

We fit the dust-corrected luminosities of the {\sc [Oii]}$\lambda\lambda$3726,3729, {\sc[Oiii]}$\lambda$4363, H$\beta$, {\sc [Oiii]}$\lambda\lambda$4959,5007, {\sc[Oiii]}52$\mu$m (if available), and {\sc[Oiii]}88$\mu$m lines with the 2-zone model.  
Since the {\sc[Oiii]}$\lambda$5007/{\sc[Oiii]}$\lambda$4959 ratio is set by quantum physics and does not depend on these parameters, effectively we have seven emission lines, which are insufficient to fully determine the seven free parameters.  
As a result, most of the obtained parameters exhibit significant degeneracies, although some parameters, such as metallicity, are well constrained.  
In the 2-zone model, the luminosities of the {\sc[Oiii]} emission lines are calculated with {\sf PyNeb} for each parameter set in each region, and those of the H$\beta$ and {\sc[Oii]}$\lambda\lambda$3726,3729 lines are derived from the metallicity, O32 ratio, and density in each region.  
We use the MCMC method with 200,000 steps to estimate the best-fit parameters.  
We assume $n_\m{e,1}<n_\m{e,2}$ and flat priors for densities (in log space) and temperatures (in linear space) in the ranges of $0<\m{log}(n_\m{e,1}/\m{[cm^{-3}]})<5$, $0<\m{log}(n_\m{e,2}/\m{[cm^{-3}]})<5$, $5000<T_\m{e,1}/\m{[K]}<30000$, and $5000<T_\m{e,2}/\m{[K]}<30000$.

Figure \ref{fig_2zone} compares the observed and predicted luminosities with the 2-zone model, and Table \ref{tab_Z2} summarizes the estimated parameters.  
Although there are large degeneracies among densities, temperatures, and the {\sc[Oiii]}$\lambda$5007 luminosity ratio due to the limited number of available emission lines, the metallicities are relatively well constrained, with uncertainties of $\sim0.1-0.3$ dex.  
Although the lower limit of $n_\m{e,1}$ and the upper limit of $n_\m{e,2}$ are not constrained, we find that the low- and high-density regions have electron densities of $n_\m{e,1}<1000\ \m{cm}^{-3}$ and $n_\m{e,2}>1000\ \m{cm}^{-3}$, respectively.  
Since the density of the high-density region exceeds the critical densities of the FIR {\sc[Oiii]}88$\mu$m and {\sc[Oiii]}52$\mu$m lines, the FIR {\sc[Oiii]} luminosities are dominated by emission from the low-density region.  
Consequently, the density of the low-density gas is comparable to those measured with the {\sc[Oiii]}52$\mu$m/{\sc[Oiii]}88$\mu$m or {\sc[Oiii]}88$\mu$m/{\sc[Oiii]}$\lambda$5007 ratios ($n_\m{e,FIR}\lesssim500$ cm$^{-3}$).  
The luminosities of the {\sc [Oii]}$\lambda\lambda$3726,3729 lines receive contributions from both low- and high-density regions, resulting in a moderately high effective density of $n_\m{e,optical}\sim1000$ cm$^{-3}$.  
We also find that the electron temperature of the low-density gas is lower than that of the high-density gas ($T_\m{e,1}<T_\m{e,2}$) for J0217-0208 and COS-3018555981, which we will further discuss in Section \ref{ss_Z_Z}.  
Since emission from the low-density region dominates the H$\beta$ luminosity, given its weak dependence on temperature and density, the results also suggest that the volume of the low-density region is $>1000$ times larger than that of the high-density region, highlighting the importance of FIR lines in tracing the low-density gas.

\subsection{Comparison of Metallicities}\label{ss_Z_Z}

Since metallicity is one of the most important parameters for describing the ISM properties of galaxies, we test the robustness of direct-$T_\m{e}$ metallicities assuming a 1-zone ISM by comparing them with metallicities derived from 2-zone modeling.  
Figure \ref{fig_Z_Ms} compares the metallicities estimated with the 2-zone model (Section \ref{ss_2zone}) and those derived with the direct-$T_\m{e}$ method (Section \ref{ss_Te}) in the mass-metallicity plane.  
We adopt the metallicities from the 2-zone modeling as the fiducial values because the 2-zone model self-consistently reproduces all the emission line luminosities observed from the optical to FIR, whereas the temperature derived from {\sc[Oiii]}$\lambda$4363 in the direct-$T_\m{e}$ method fails to reproduce the FIR {\sc[Oiii]} luminosities.  
We find that the direct-$T_\m{e}$ metallicities are consistent with the 2-zone metallicities in the two less massive galaxies, MACS1149-JD1 and SXDF-NB1006-2, but are significantly underestimated in the two more massive galaxies, COS-3018555981 and J0217-0208, by 0.4 and 0.8 dex, respectively.  
\redc{Interestingly, we find a possible trend of a larger discrepancy in more metal-rich galaxies, which is consistent with predictions from \citet{2005A&A...434..507S}.}
These differences originate from the different temperatures in the low- and high-density regions.  
For MACS1149-JD1 and SXDF-NB1006-2, the estimated temperatures in the low- and high-density regions are comparable within the errors, allowing the 2-zone model with different densities but the same temperature to reproduce both the optical and FIR emission lines, resulting in a metallicity comparable to that from the direct-$T_\m{e}$ method.  
For J0217-0208 and COS-3018555981, the estimated temperature in the high-density region is higher than that in the low-density region.  
In this configuration with a high-density, high-temperature region and a low-density, low-temperature region, as shown in Figure \ref{fig_2zone}, the {\sc[Oiii]}$\lambda$4363 emission from the high-density, high-temperature region dominates the observed luminosity because its emissivity strongly depends on electron temperature.  
As a result, the temperature estimated from {\sc[Oiii]}$\lambda$4363 is overestimated compared to the effective temperature of the whole system, and the derived metallicity is significantly underestimated.  
These results clearly indicate that a joint analysis using both optical and FIR lines is crucial for accurately measuring the gas-phase metallicity of high-redshift galaxies.

\begin{figure}
\centering
\begin{center}
\includegraphics[width=0.99\hsize, bb=7 8 461 354,clip]{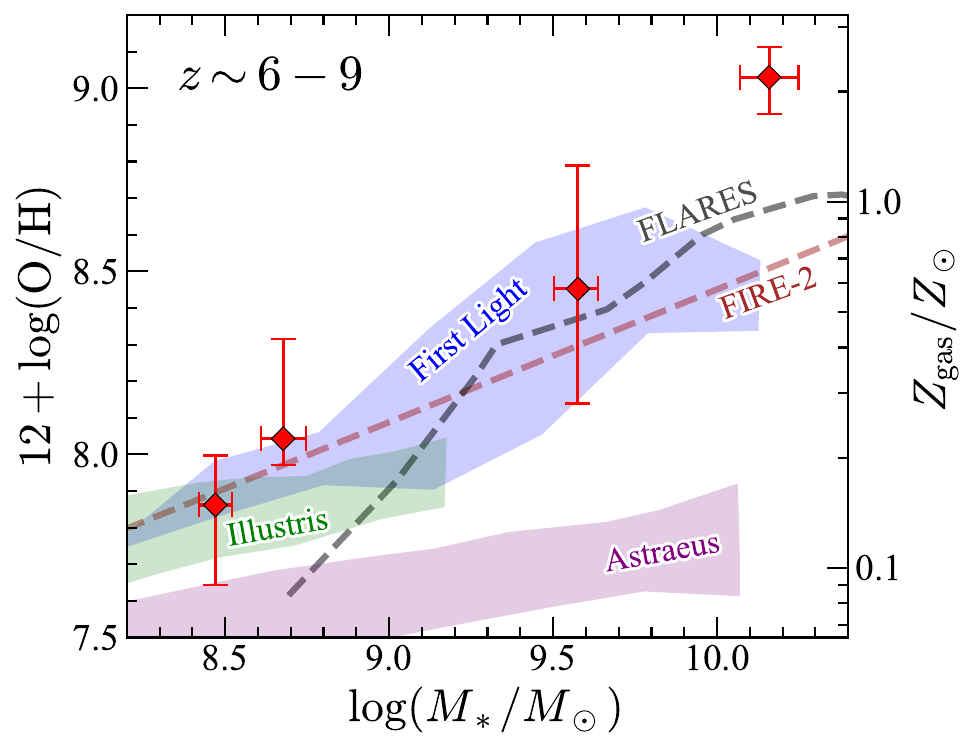}
\end{center}
\vspace{-0.5cm}
\caption{
Comparison of the mass-metallicity relations with simulations.
The red-filled diamonds represent our fiducial metallicities measured with the 2-zone model fitting (same as Figure \ref{fig_Z_Ms}).
We show predictions from cosmological simulations: {First Light} in blue \citep{2023ApJ...953..140N}, {Illustris TNG} in green \citep{2019MNRAS.484.5587T}, Astraeus in purple \citep{2023MNRAS.518.3557U}, FLARES in gray \citep{2023MNRAS.518.3935W}, and FIRE-2 in brown \citep{2024ApJ...967L..41M}.
All predictions are at $z=7$, except for the one at $z=6$ from Illustris TNG.
Our inferred high metallicities in massive galaxies are consistent with the steep slope of the mass-metallicity relations predicted by some simulations.
}
\label{fig_Z_Ms_sim}
\end{figure}

As shown in Figure \ref{fig_Z_Ms_sim}, our inferred metallicities for the two massive galaxies, J0217-0208 and COS-3018555981, are higher than extrapolations of the mass-metallicity relations observationally constrained with low-mass galaxies in \citet{2023ApJS..269...33N} and \citet{2024A&A...684A..75C}, and are more comparable to recent determinations for massive galaxies at $z\sim6-8$ in \citet{2025arXiv250110559R}.  
In Figure \ref{fig_Z_Ms_sim}, we compare our obtained metallicities with predictions from cosmological simulations.  
Although a larger sample is needed for firm conclusions, we find that our high metallicities in these massive galaxies are more aligned with some simulations predicting steep slopes in the relation, such as First Light \citep{2023ApJ...953..140N}, FLARES \citep{2023MNRAS.518.3935W}, and FIRE-2 \citep{2024ApJ...967L..41M}.

\subsection{Systematic Uncertainty of the 2-Zone Model}\label{ss_sys}

In the 2-zone modeling, we assume the same ionization parameters and metallicities in the high- and low-density regions to keep the model as simple as possible.  
However, given their very different densities and temperatures, it is possible that the gas is not well-mixed and that these two regions have different ionization parameters and/or metallicities.  
Here, we quantify the effect of these assumptions on the metallicity estimates, focusing particularly on J0217-0208 and COS-3018555981, which show significantly higher metallicities than those derived with the direct-$T_\m{e}$ method assuming a 1-zone ISM.  
As we discuss below, even when accounting for the systematic uncertainty introduced by these assumptions, the inferred metallicities from the JWST and ALMA joint analysis for J0217-0208 and COS-3018555981 remain higher than those obtained with the direct-$T_\m{e}$ method.

\subsubsection{Effect of Different Ionization Parameters}

First, we consider a situation with different ionization parameters.  
The ionization parameter ($U_\m{ion}$ or $q_\m{ion}$) is expressed as follows:  
\begin{equation}
U_\m{ion}=\frac{q_\m{ion}}{c}=\frac{Q_0}{4\pi R^2_\m{s} n_\m{H}}\propto (Q_0 n_\m{H})^{\frac{1}{3}},
\end{equation}  
where $c$, $Q_0$, and $n_\m{H}$ are the speed of light, the hydrogen-ionizing photon production rate, and the hydrogen density, respectively, and $R_\m{s}$ is the Str{\"o}mgren radius, which is proportional to $Q_0^{\frac{1}{3}} n_\m{H}^{-\frac{2}{3}}$.  
Thus, we expect that the ionization parameter in the high-density region is higher than that in the low-density region, assuming the same hydrogen-ionizing photon production rates in the two regions.

To understand the effect of different ionization parameters on the metallicity estimates in the 2-zone modeling, we conduct the MCMC fitting again, allowing different O32 ratios in the high- and low-density regions with $\m{O32}_\m{1}<\m{O32}_\m{2}$, corresponding to different ionization parameters with $U_\m{ion,1}<U_\m{ion,2}$, where subscripts 1 and 2 refer to the low- and high-density regions, respectively, as in Section \ref{ss_2zone}.  
We find that the obtained metallicities for J0217-0208 and COS-3018555981 are always higher than the best estimates in Section \ref{ss_2zone}, which assume the same ionization parameters in the two regions.

We can understand this result as follows.  
If we assume a higher ionization parameter in the high-density region, the fraction of the {\sc[Oiii]}$\lambda$5007 luminosity from the high-density region among the total luminosity becomes larger.  
In contrast, the {\sc[Oiii]}$\lambda$4363 luminosity from the high-density region does not significantly change, as its total luminosity is already dominated by the emission from the high-density region.  
As a result, the {\sc[Oiii]}$\lambda$4363/{\sc[Oiii]}$\lambda$5007 ratio becomes lower, and the inferred temperature in the high-density region becomes lower.  
Similarly, the {\sc[Oiii]}$\lambda$5007 luminosity from the low-density region becomes lower if we assume a lower ionization parameter in the low-density region.  
The density in the low-density region does not significantly change because the FIR {\sc[Oiii]} luminosities are dominated by emission from the low-density region.  
Consequently, the {\sc[Oiii]}88$\mu$m/{\sc[Oiii]}$\lambda$5007 ratio becomes higher, resulting in a lower temperature in the low-density region.  
Because the temperatures in both regions become lower, the obtained metallicity is higher than the estimate in Section \ref{ss_2zone}.

\subsubsection{Effect of Different Metallicities}

Next, we consider the effect of different metallicities in the low-density and high-density regions.  
In the 2-zone model, the total metallicity (O/H) can be expressed as follows,  
\begin{equation}
\m{\frac{O}{H}=\frac{O(1)+O(2)}{H(1)+H(2)}}, \label{eq_2zone}
\end{equation}  
where $\m{O(1)}$ and $\m{H(1)}$ ($\m{O(2)}$ and $\m{H(2)}$) are the numbers of oxygen and hydrogen atoms, respectively, in the low-density (high-density) region.  
By ignoring higher ionized ions such as $\m{O^{3+}}$, the numerator, $\m{O(1)+O(2)}$, is expressed as  
\begin{align}\label{eq_2zone_O}
\m{O(1)}&+\m{O(2)}=\m{O^{+}(1)+O^{2+}(1)+O^{+}(2)+O^{2+}(2)}\notag\\
&=\frac{1}{n_\m{e,1}}\frac{L_\m{[OII],1}}{\epsilon_\m{[OII]}(T_\m{e,1},n_\m{e,1})}+\frac{1}{n_\m{e,1}}\frac{L_\m{[OIII]5007,1}}{\epsilon_\m{[OIII]5007}(T_\m{e,1},n_\m{e,1})}\notag\\
&+\frac{1}{n_\m{e,2}}\frac{L_\m{[OII],2}}{\epsilon_\m{[OII]}(T_\m{e,2},n_\m{e,2})}+\frac{1}{n_\m{e,2}}\frac{L_\m{[OIII]5007,2}}{\epsilon_\m{[OIII]5007}(T_\m{e,2},n_\m{e,2})},
\end{align}  
where $L_\m{[OII],1}$ and $L_\m{[OII],2}$ ($L_\m{[OIII]5007,1}$ and $L_\m{[OIII]5007,2}$) are the predicted {\sc[Oii]}$\lambda\lambda$3726,3729 ({\sc[Oiii]}$\lambda$5007) luminosities in the low- and high-density regions, and $\epsilon_\m{[OII]}(T_\m{e},n_\m{e})$ ($\epsilon_\m{[OIII]5007}(T_\m{e},n_\m{e})$) is the emissivity of {\sc[Oii]}$\lambda\lambda$3726,3729 ({\sc[Oiii]}$\lambda$5007) in units of $\m{erg\ s^{-1}\ cm^{3}}$.  
As shown in Equation (\ref{eq_2zone_O}), the oxygen abundances in each region ($\m{O^{+}(1)}$, $\m{O^{2+}(1)}$, $\m{O^{+}(2)}$, $\m{O^{2+}(2)}$) depend on parameters other than metallicity (i.e., $n_\m{e,1}$, $n_\m{e,2}$, $T_\m{e,1}$, $T_\m{e,2}$, O32, and $L_\m{[OIII]5007,1}/L_\m{[OIII]5007,2}$).  
In the 2-zone model fitting, these parameters are effectively determined based on the observed luminosities of the {\sc[Oiii]} and {\sc[Oii]} lines, and the metallicity is scaled to match the observed H$\beta$ luminosity.  
Thus, even if we change the metallicities in the low- and high-density regions from the current assumption of having the same metallicity, the oxygen abundances do not change.

The denominator, $\m{H(1)+H(2)}$, is inversely scaled with metallicity and is expressed as  
\begin{equation}\label{eq_2zone_H}
\m{H(1)+H(2)}= \frac{1}{n_\m{e,1}}\frac{L_\m{H\beta,1}}{\epsilon_\m{H\beta}(T_\m{e,1})}+\frac{1}{n_\m{e,2}}\frac{L_\m{H\beta,2}}{\epsilon_\m{H\beta}(T_\m{e,2})},
\end{equation}  
where $L_\m{H\beta,1}$ and $L_\m{H\beta,2}$ are the predicted H$\beta$ luminosities in the low- and high-density regions, and $\epsilon_\m{H\beta}(T)$ is the emissivity of H$\beta$.  
In the 2-zone model assuming the same metallicities in the low- and high-density regions, $L_\m{H\beta,1}$ and $L_\m{H\beta,2}$ are determined so that the metallicities in the two zones are equal ($\m{O(1)/H(1)}=\m{O(2)/H(2)}$).  
To consider the effect of different metallicities, we discuss how the total hydrogen abundance changes if we change $L_\m{H\beta,1}$ and $L_\m{H\beta,2}$ while conserving the total observed H$\beta$ luminosity ($L_\m{H\beta,1}+L_\m{H\beta,2}$).

\begin{figure*}
\centering
\begin{center}
\includegraphics[width=0.99\hsize, bb=50 221 1308 524,clip]{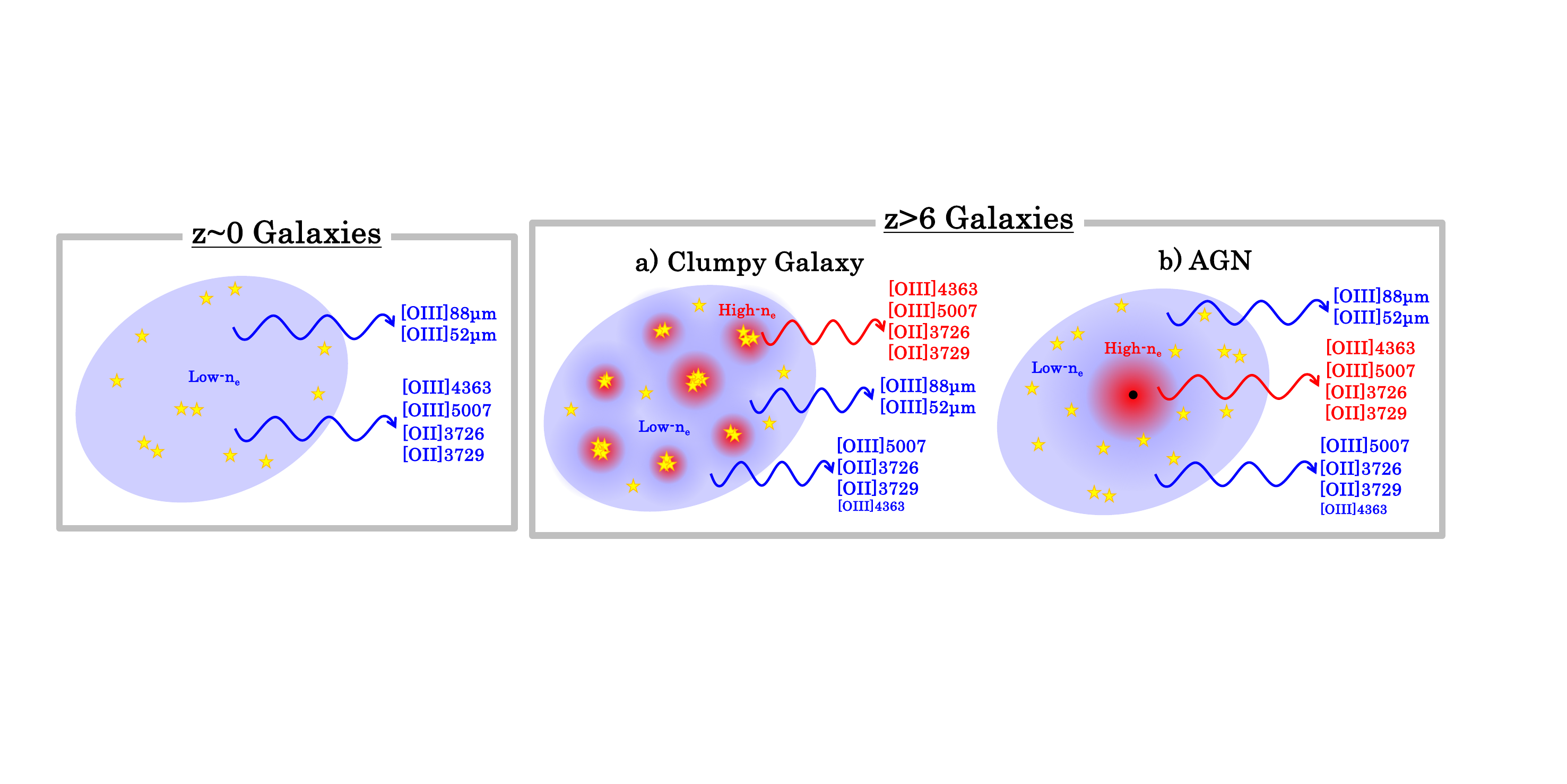}
\end{center}
\vspace{-0.5cm}
\caption{
Schematic illustrations of galaxies at $z\sim0$ (left) and $z>6$ (right).  
At $z\sim0$, galaxies are low-density on average, and the {\sc[Oiii]} and {\sc[Oii]} luminosities are dominated by emission from low-density gas (blue; $n_\mathrm{e}\sim10-100\ \mathrm{cm^{-3}}$).  
At $z>6$, we propose two scenarios: (a) clumpy galaxy and (b) AGN.  
In the clumpy galaxy scenario (a), the optical {\sc[Oiii]} and {\sc[Oii]} emission lines ({\sc[Oiii]}$\lambda\lambda$4959,5007, {\sc[Oii]}$\lambda\lambda$3726,3729, and {\sc[Oiii]}$\lambda$4363) are emitted from both the low-density gas (blue) and the high-density gas in clumps (red), and sometimes the {\sc[Oiii]}$\lambda$4363 luminosity is dominated by emission from the high-density clumps due to its strong sensitivity to the electron temperature.  
The FIR {\sc[Oiii]} emission ({\sc[Oiii]}88$\mu$m and {\sc[Oiii]}52$\mu$m) arises from the low-density gas due to their low critical densities.  
In the AGN scenario (b), the optical {\sc[Oiii]} and {\sc[Oii]} emission originates both from the high-density gas in the AGN narrow-line region and/or outflowing gas around the AGN, and from the low-density gas in the host galaxy.  
In both scenarios, since the {\sc[Oii]}$\lambda\lambda3726,3729$ emission is contributed by both low-density ($n_\mathrm{e}\sim100\ \mathrm{cm^{-3}}$) and high-density ($n_\mathrm{e}\sim10000\ \mathrm{cm^{-3}}$) gas, the electron density measured with the {\sc[Oii]} line ratio at $z>6$ ($n_\mathrm{e}\sim1000\ \mathrm{cm^{-3}}$) represents an intermediate value between those of the low- and high-density components.
}
\label{fig_galaxy}
\end{figure*}

Observations of local galaxies show a clear anti-correlation between electron temperature and metallicity \citep[e.g.,][]{2014ApJ...790...75N}.  
Since the temperature in the low-density region is lower, it is reasonable to assume that the metallicity in the low-density region is higher than that in the high-density region.  
In this situation, we decrease $L_\m{H\beta,1}$ and increase $L_\m{H\beta,2}$ from the values assumed under the condition of equal metallicities.  
For J0217-0208 and COS-3018555981, the densities in the two regions differ by a factor of $>10$.  
Since the H$\beta$ emissivity decreases by only 0.6 dex from $T_\m{e}=5000$ K to $25000$ K, the product of density and emissivity (the denominator in Equation (\ref{eq_2zone_H})) in the low-density region is always lower than that in the high-density region ($n_\m{e,1}\epsilon_\m{H\beta}(T_\m{e,1})<n_\m{e,2}\epsilon_\m{H\beta}(T_\m{e,2})$).  
As a result, if we decrease $L_\m{H\beta,1}$ and increase $L_\m{H\beta,2}$ while conserving $L_\m{H\beta,1}+L_\m{H\beta,2}$, the total hydrogen abundance becomes lower, corresponding to a higher total metallicity than the estimate in Section \ref{ss_2zone}.

\section{Discussion}\label{ss_dis}

\subsection{Evolution of the ISM Structure}\label{ss_dis_z06}

In Sections \ref{ss_ne_ne} and \ref{ss_ne_z}, we find that the densities derived from the optical {\sc[Oii]} lines ($n_\m{e,optical}$) and the FIR {\sc[Oiii]} lines ($n_\m{e,FIR}$) are significantly different.  
Furthermore, in Sections \ref{ss_1zone} and \ref{ss_2zone}, we show that the 2-zone model with low-density and high-density regions can self-consistently reproduce both the optical and FIR emission line luminosities, whereas the 1-zone model cannot.  
Since such discrepancies between $n_\m{e,optical}$ and $n_\m{e,FIR}$ are not observed in local galaxies, where the densities are low in both $n_\m{e,optical}$ and $n_\m{e,FIR}$, these results suggest an evolution in the ISM structure from low redshift to high redshift.  
Here, we discuss two scenarios that can explain this evolution: (a) clumpy galaxies and (b) AGN activity.  
Figure \ref{fig_galaxy} shows schematic illustrations of these two scenarios.

\begin{itemize}

\item[(A)] {\it Clumpy Galaxy.}
Many studies report that high-redshift galaxies are clumpy, with the fraction of clumpy galaxies increasing from $z\sim0$ to higher redshifts.  
For example, \citet{2025ApJ...980..138H} discuss that $\gtrsim60\%$ of UV-bright galaxies at $z\sim7$ show clumpy structures extending up to $\sim5$ kpc.  
Faint galaxies at $z\sim5-7$ also show a high fraction of clumpy morphologies \citep{2024MNRAS.52711372A}.  
\citet{2024arXiv240218543F} report a lensed galaxy at $z=6$ that contains $\geq15$ small ($\sim10-50$ pc) star-forming clumps, which cannot be resolved in NIRCam images without lensing magnification.  
Among the galaxies at $z>6$ discussed in this study, J1211-0118 (Figure \ref{fig_snap_hscjwst}), COS-3018555981 \citep{2024arXiv241107695S}, and MACS1149-JD1 \citep{2024MNRAS.533.2488M,2024A&A...686A..85A} show multiple sub-components.  
Although J0217-0208 and SXDF-NB1006-2 do not display clumpy morphologies in our NIRCam images, it is possible that they host star-forming clumps much smaller than the NIRCam resolution, similar to the lensed galaxy reported by \citet{2024arXiv240218543F}.  
The gas structure of these clumpy galaxies can be well described with the 2-zone model, where the gas in the star-forming clumps is dense, and the diffuse gas surrounding the clumps is less dense.  
Thus, in this clumpy galaxy scenario, the FIR {\sc[Oiii]} emission ({\sc[Oiii]}88$\mu$m and {\sc[Oiii]}52$\mu$m) arises from the low-density gas around the clumps, due to their low critical densities.  
In contrast, the optical {\sc[Oiii]} and {\sc[Oii]} emission lines ({\sc[Oiii]}$\lambda\lambda$4959,5007, {\sc[Oii]}$\lambda\lambda$3726,3729, and {\sc[Oiii]}$\lambda$4363) are emitted from both the low-density gas and the high-density gas within the clumps, resulting in moderately high electron densities inferred from the {\sc[Oii]}$\lambda$3729/{\sc[Oii]}$\lambda$3726 ratio ($n_\m{e,optical}\sim1000$ cm$^{-3}$).  
Since we do not detect the low-density component suggested by $n_\m{e,FIR}$ in the density maps based on the {\sc[Oii]}$\lambda$3729/{\sc[Oii]}$\lambda$3726 ratio for the four galaxies analyzed here (Figure \ref{fig_map_ne}; \citealt{2024MNRAS.533.2488M,2024arXiv241107695S}), the physical scales of the clumps may be much smaller than the spatial resolution of JWST.

\item[(B)] {\it AGN.}
Recent JWST studies have found many high-redshift AGNs at $z>4$, with abundances higher than expected before the launch of JWST \citep[e.g.,][]{2023ApJ...954L...4K,2023ApJ...959...39H,2023ApJ...957L...7K,2024ApJ...963..129M,2024A&A...691A.145M,2024arXiv240815615M}.  
Although we do not observe any obvious AGN signatures, such as broad hydrogen Balmer lines, in the galaxies studied here, some of them (e.g., J0217-0208) may host type-2 AGNs given their luminous nature and strong emission lines \citep[e.g.,][]{2017ApJ...851...40L,2025ApJ...982...27Z}.  
The electron densities of the narrow-line regions of AGNs and the outflowing gas driven by AGN activity are typically high, $\sim10^3-10^6$ cm$^{-3}$ \citep[e.g.,][]{1990agn..conf.....B,2006agna.book.....O,2013peag.book.....N,2018A&A...618A...6K}, which is comparable to the estimated density of the high-density region in the 2-zone model discussed in Section \ref{ss_2zone}.  
Indeed, the Seyfert galaxy NGC1068 exhibits a discrepancy between $n_\m{e,optical}$ and $n_\m{e,FIR}$ similar to that seen in galaxies at $z>6$ (Figure \ref{fig_ne_ne}), and spatially resolved spectroscopic observations targeting NGC1068 reveal that its density reaches $\gtrsim1000$ cm$^{-3}$ around the nucleus \citep[e.g.,][]{2009PASJ...61..259O,2018A&A...618A...6K}.  
Thus, it is possible that some of the galaxies studied here host AGNs, where the dense gas around the AGN and the gas in the host galaxy correspond to the high-density and low-density regions in the 2-zone model, respectively.  
In this AGN scenario, because the densities of the AGN narrow-line region and the outflowing gas are significantly higher than the critical densities of the FIR {\sc[Oiii]} lines, most of the FIR {\sc[Oiii]} emission originates from the host galaxy, which has a lower density ($\lesssim1000$ cm$^{-3}$).  
The optical {\sc[Oiii]} and {\sc[Oii]} emission lines are contributed by both the dense gas around the AGN and the diffuse gas in the host galaxy, resulting in an electron density of $n_\m{e,optical}\sim1000$ cm$^{-3}$ inferred from the {\sc[Oii]}$\lambda$3729/{\sc[Oii]}$\lambda$3726 ratio.  
Even if the AGN narrow-line region has a very high density, the inferred density would still be around $\sim1000-10000$ cm$^{-3}$ due to the critical densities of $3300$ and $14000$ cm$^{-3}$ for {\sc[Oii]}$\lambda$3726 and {\sc[Oii]}$\lambda$3729, respectively.

\end{itemize}

These two scenarios can explain the evolution of electron densities discussed in Section \ref{ss_ne_z}, considering the redshift evolution of the clumpy galaxy fraction \citep[e.g.,][]{2024MNRAS.52711372A,2025ApJ...980..138H} and AGN abundance \citep[e.g.,][]{2023ApJ...959...39H,2024A&A...691A.145M}.  
At $z\sim0$, \redc{although there are some clumpy dwarf galaxies \citep[e.g.,][]{2019ApJ...886...74M,2023A&A...670A..92M,2024A&A...685A..46R}}, most galaxies are low-density, not clumpy, and AGNs are rare.
\redc{Indeed, the two local galaxies studied \citet{2023NatAs...7..771C,2024arXiv240518476C} show compact morphologies.}
At higher redshifts, the increasing fractions of clumpy galaxies and AGNs cause the emission from high-density regions (star-forming clumps in clumpy galaxies or dense gas around AGNs) to become more dominant, resulting in higher densities probed by optical {\sc[Oii]}$\lambda\lambda$3726,3729 or {\sc[Sii]}$\lambda\lambda$6716,6731 emission lines.  
In contrast, the FIR {\sc[Oiii]} lines  are always tracing the low-density regions of galaxies due to their low critical densities, resulting in low densities of $n_\m{e,FIR}\lesssim1000$ cm$^{-3}$ at all redshifts.

\redc{In addition to these two possibilities, other physical mechanisms could also contribute.
For example, porous, density-bounded {\sc Hii} regions with high Lyman continuum escape have been proposed to explain both optical and FIR line properties \citep[e.g.,][]{2022A&A...667A..35R}.  
Shocks, which can strongly affect the density structure of the ISM, may also play a role \citep[e.g.,][]{1976ApJ...209..395D,1977ApJS...33..437D,1977ApJ...214..179D,1983A&A...127...15C,1985A&A...149..109A}.  
Moreover, diffuse ionized gas could influence both the observed line ratios and the inferred electron densities.  
It could be possible that a combination of these processes shapes the observed discrepancies.}

\subsection{Selection Bias}

The galaxies studied here are limited to those with ALMA {\sc[Oiii]}88$\mu$m detections.  
In this section, we discuss the effect of this selection on the results we have obtained.  
Among 21 spectroscopically confirmed galaxies at $z>6$ observed with ALMA reported so far \citep{2016Sci...352.1559I,2017ApJ...837L..21L,2017A&A...605A..42C,2018Natur.557..392H,2019PASJ...71...71H,2019ApJ...874...27T,2020ApJ...896...93H,2020IAUS..341..309S,2021A&A...646A..26B,2022MNRAS.515.1751W,2024MNRAS.527.6867A,2022ApJ...934...64A,2022ApJ...929..161W,2024ApJ...977L...9Z,2024arXiv240920533C,2024arXiv240920549S}\footnote{Although \citet{2023ApJ...950...61Y} reported no clear detection of {\sc[Oiii]}88$\mu$m in a galaxy at $z\sim10$, GHZ1, the frequency range of the ALMA observations in \citet{2023ApJ...950...61Y}, corresponding to $10.10<z<11.14$, does not cover the spectroscopic redshift of GHZ1 ($z=9.875$) recently determined with JWST in \citet{2024arXiv241010967N}.}, the {\sc[Oiii]}88$\mu$m line is detected in 20 galaxies.  
The remaining galaxy is z7\_GSD\_3811 at $z=7.7$, which was spectroscopically confirmed with the Ly$\alpha$ emission line in \citet{2016ApJ...826..113S}.  
Although \citet{2021A&A...646A..26B} reported non-detection of {\sc[Oiii]}88$\mu$m in this galaxy, we find a $\sim3\sigma$ signal at the expected {\sc[Oiii]}88$\mu$m frequency with the recently determined systemic redshift in \citet{2025MNRAS.536.2355J} using JWST.  
These results indicate that detections of the {\sc[Oiii]}88$\mu$m line are common in high-redshift galaxies, and that our galaxy selection based on {\sc[Oiii]}88$\mu$m detections does not bias the results obtained compared to the entire galaxy sample at $z>6$.

\begin{figure*}
\centering
\begin{minipage}{0.4\hsize}
\includegraphics[height=0.99\hsize, bb=8 8 350 355,clip]{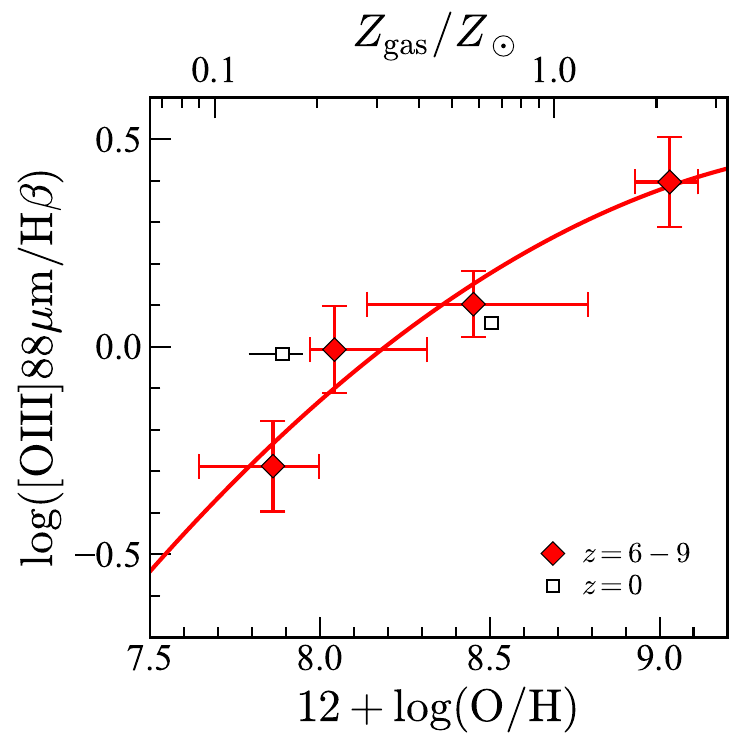}
\end{minipage}
\begin{minipage}{0.4\hsize}
\includegraphics[height=0.99\hsize, bb=8 8 350 355,clip]{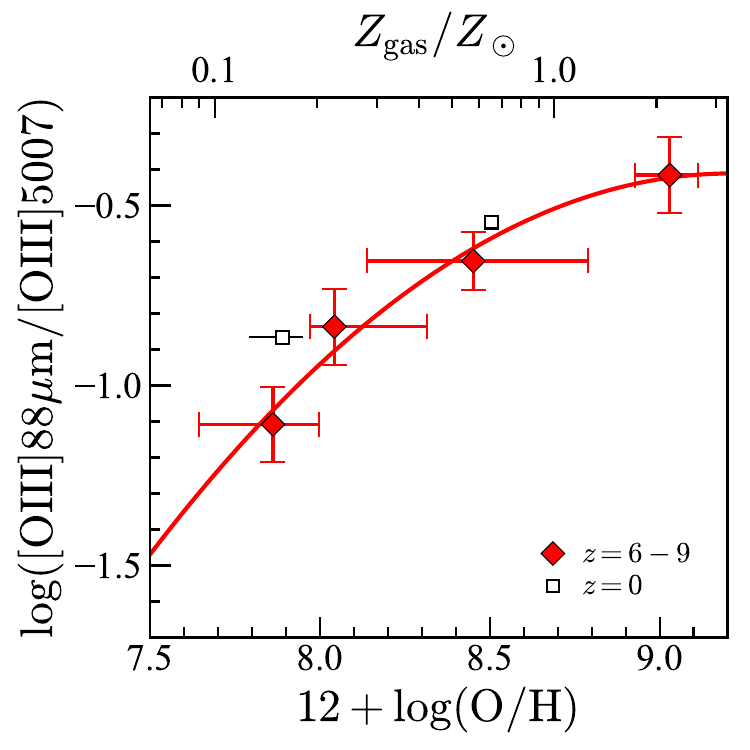}
\end{minipage}
\caption{
Relation between the metallicity and the {\sc[Oiii]}88$\mu$m/H$\beta$ (left) and {\sc[Oiii]}88$\mu$m/{\sc[Oiii]}$\lambda$5007 (right) line ratios.  
The red diamonds show the observed line ratios and metallicities constrained with the 2-zone model at $z\sim6-9$ in this study, and the black open squares are results at $z\sim0$ \citep{2023NatAs...7..771C,2024arXiv240518476C}.  
The red curves represent the best-fitting relations presented in Equations (\ref{eq_Z_1}) and (\ref{eq_Z_2}).
}
\label{fig_Z_OIIIr}
\end{figure*}

\subsection{JWST+ALMA Metallicity Estimator}

In Section \ref{ss_Z_Z}, we find that the commonly used direct-$T_\m{e}$ method sometimes significantly underestimates the metallicity compared to the fiducial value obtained with 2-zone modeling using the optical and FIR lines detected with JWST and ALMA.  
Thus, if a galaxy is observed with JWST and ALMA, the most reliable way to estimate the metallicity is to conduct the 2-zone modeling described in Section \ref{ss_2zone}.  
Here, we discuss an alternative method to estimate the metallicity using the JWST and ALMA lines.

In Figure \ref{fig_Z_OIIIr}, we plot the emission line ratios of {\sc[Oiii]}88$\mu$m/H$\beta$ and {\sc[Oiii]}88$\mu$m/{\sc[Oiii]}$\lambda$5007 as a function of the metallicity obtained via the 2-zone modeling for the four galaxies analyzed here.  
We find that these line ratios correlate well with the metallicity, and obtain the best-fitting relations as follows:
\begin{align}
\m{log\left(\frac{\textsc{[Oiii]}88\mu m}{H\beta}\right)}&=0.265+0.428x-0.209x^2 \label{eq_Z_1}, \\
\m{log\left(\frac{\textsc{[Oiii]}88\mu m}{\textsc{[Oiii]}\lambda5007}\right)}&=-0.509+0.376x-0.362x^2, \label{eq_Z_2}
\end{align}
where $x=12+\m{log(O/H)}-8.69$.
Note that these relations cannot be well reproduced with a 1-zone photoionization model such as with {\sf Cloudy} \citep{2017RMxAA..53..385F}, because the emission lines do not originate from a homogeneous 1-zone gas.  
Although the parameters in these relations are not tightly constrained due to the limited number of galaxies with 2-zone metallicity estimates, these relations can serve as a guideline to determine whether the direct-$T_\m{e}$ method is underestimating the metallicity or not.  
If a galaxy is observed with JWST and ALMA, one can compare the metallicity inferred from these relations with the metallicity obtained via the direct-$T_\m{e}$ method.  
If the direct-$T_\m{e}$ metallicity is significantly lower than that inferred from these relations, it is necessary to conduct 2-zone modeling to determine the true metallicity.  
Future JWST and ALMA observations targeting a statistical sample of high-redshift galaxies will enable us to better constrain these relations.

\section{Summary}\label{ss_summary}
In this paper, we investigate the ISM properties of high-redshift galaxies in detail.  
We use our new JWST and ALMA observations, as well as data from the literature, and conduct a joint JWST and ALMA analysis to self-consistently reproduce all of the observed emission lines.  
Our major findings are summarized below:

\begin{enumerate}

\item
Our JWST/NIRSpec IFU spectroscopy targeting three galaxies at $z=6-7$ successfully detects rest-frame optical emission lines such as {\sc[Oii]}$\lambda\lambda$3726,3729, H$\gamma$, H$\beta$, and {\sc[Oiii]}$\lambda\lambda$4959,5007 (Figure \ref{fig_spec_jwst}).  
We also identify the {\sc[Oiii]}$\lambda$4363 line in two galaxies (Figure \ref{fig_spec_jwst_OIII4363}), and derive their metallicities to be $12+\m{log(O/H)}=8.0-8.2$ using the direct-$T_\m{e}$ method.  
The moderately high spectral resolutions of G235H and G395H allow us to measure the line ratios of the {\sc[Oii]}$\lambda\lambda$3726,3729 lines, resulting in an average electron density of $n_\m{e,optical}\sim1000$ cm$^{-3}$.  
Dust \redc{attenuation} measured from the Balmer decrement is low, with $E(B-V)\simeq0.0-0.1$.

\item
Our new ALMA Band 9 and 10 observations detect the {\sc[Oiii]}52$\mu$m line in one galaxy at $z=6.2$, but not in the other two galaxies (Figure \ref{fig_spec_ALMA}).
Using the density-sensitive {\sc[Oiii]}52$\mu$m/{\sc[Oiii]}88$\mu$m line ratios, we obtain the constraints of $n_\m{e,FIR}\lesssim500$ cm$^{-3}$, systematically lower than those measured with the optical {\sc[Oii]} lines (Figure \ref{fig_ne_ne}).  
Two additional galaxies at $z=7-9$ also show a similar trend, where the densities measured with the FIR {\sc[Oiii]} line are systematically lower than those from the optical {\sc[Oii]} lines.

\item 
We find that the electron densities of our galaxies measured with the optical {\sc[Oii]} lines are consistent with the redshift evolution trend of increasing density toward higher redshift reported in previous studies (Figure \ref{fig_ne_z}).  
In contrast, the electron densities of galaxies at $z\sim6-14$ derived from the FIR {\sc[Oiii]} lines are low and comparable to those at $z\sim0$, showing no redshift evolution.  
These contrasting evolution trends can be interpreted by scenarios involving an increasing fraction of clumpy galaxies or AGNs toward higher redshifts (Figure \ref{fig_galaxy}).

\item 
We conduct a joint JWST and ALMA analysis for four galaxies (two from our sample and two from the literature) well observed with both telescopes to constrain their ISM properties.  
The 1-zone model, which assumes a homogeneous temperature and density derived from the JWST optical lines, fails to reproduce the ALMA-observed FIR {\sc[Oiii]} line luminosities (Figure \ref{fig_1zone}), even when hypothetical heavily dust-obscured gas is considered.  
Instead, we introduce a 2-zone model with high-density ($n_\m{e}>1000$ cm$^{-3}$) and low-density ($n_\m{e}<1000$ cm$^{-3}$) regions, which successfully reproduces all of the emission line luminosities observed with JWST and ALMA (Figure \ref{fig_2zone}).

\item
The best-fit 2-zone models for the four galaxies suggest that the FIR {\sc[Oiii]}88$\mu$m and {\sc[Oiii]}52$\mu$m luminosities are dominated by emission from the low-density region due to their low critical densities, resulting in the observed low densities ($n_\m{e,FIR}\lesssim500$ cm$^{-3}$) from the FIR {\sc[Oiii]} lines.  
In contrast, the optical {\sc[Oiii]} and {\sc[Oii]} emissions arise from both the low-density and high-density regions, leading to moderately high densities ($n_\m{e,optical}\sim1000$ cm$^{-3}$) derived from the optical {\sc[Oii]} lines.

\item
We adopt the metallicities estimated with the 2-zone modeling as the fiducial values because the 2-zone model self-consistently reproduces both the optical and FIR lines, while the 1-zone model based on the {\sc[Oiii]}$\lambda$4363-derived temperature does not.  
Compared to the fiducial metallicities, the direct-$T_\m{e}$ metallicities are significantly underestimated up to 0.8 dex in two of the four galaxies analyzed in this study (Figure \ref{fig_Z_Ms}), implying a steeper slope in the mass-metallicity relation.  
In these two galaxies, the {\sc[Oiii]}$\lambda$4363 emission from the high-density and high-temperature regions dominates the observed luminosity due to its strong temperature dependence, resulting in an overestimation of the electron temperature based on {\sc[Oiii]}$\lambda$4363.

\end{enumerate}

While based on a small sample, our results highlight the importance of combining optical and FIR emission lines for accurately determining the gas-phase metallicity of high-redshift galaxies, including those with little dust.
We provide strong-line metallicity calibrations using optical and FIR lines (Figure \ref{fig_Z_OIIIr}), although they are currently constrained by only four galaxies.
Future JWST and ALMA observations targeting a larger sample of high-redshift galaxies, along with spatially resolved data, will enable more precise calibrations and a deeper understanding of the ISM properties in early galaxies.

\begin{acknowledgments}
\redc{We thank the anonymous referee for careful reading and valuable comments that improved the clarity of the paper.}
We are grateful to Rodrigo Herrera-Camus, Daichi Kashino, Ken Tadaki, and Tommaso Treu for useful comments and discussions.
This paper makes use of the following ALMA data: ADS/JAO.ALMA\#2022.1.00012.S, ADS/JAO.ALMA\#2023.1.00022.S.
ALMA is a partnership of ESO (representing its member states), NSF (USA) and NINS (Japan), together with NRC (Canada), NSTC and ASIAA (Taiwan), and KASI (Republic of Korea), in cooperation with the Republic of Chile. The Joint ALMA Observatory is operated by ESO, AUI/NRAO and NAOJ.
This work is based on observations made with the NASA/ESA/CSA James Webb Space Telescope. The data were obtained from the Mikulski Archive for Space Telescopes at the Space Telescope Science Institute, which is operated by the Association of Universities for Research in Astronomy, Inc., under NASA contract NAS 5-03127 for JWST.
These observations are associated with programs GTO-, GO-1657 and GO-1840. 
The authors acknowledge the GTO- and GO-1840 teams led by Nora L\"utzgendorf, Javier Alvarez-Marquez, and Takuya Hashimoto, respectively, for developing their observing programs.
The JWST data presented in this article were obtained from the Mikulski Archive for Space Telescopes (MAST) at the Space Telescope Science Institute. The specific observations analyzed can be accessed via \dataset[10.17909/x1r7-h433]{https://doi.org/10.17909/x1r7-h433}.
This publication is based upon work supported by the World Premier International Research Center Initiative (WPI Initiative), MEXT, Japan, the Japan Society for the Promotion of Science (JSPS) Grant-in-Aid for Scientific Research (21K13953,  24H00245), the JSPS Core-to-Core Program (JPJSCCA20210003), and the JSPS International Leading Research (22K21349).
This work was supported by the joint research program of the Institute for Cosmic Ray Research (ICRR), University of Tokyo.
YH acknowledges support from the Sumitomo Foundation, the Ito Science Promotion Society, and the Yamaguchi Scholarship Foundation.
RLS and MG acknowledge support from NASA grant JWST-GO-01657.011-A.
RSE acknowledges generous financial support from the Peter and Patricia Gruber Foundation
TJ acknowledges support from the NASA under grant 80NSSC23K1132, and from a UC Davis Chancellor’s Fellowship.

\software{Prospector \citep{2021ApJS..254...22J}, PyNeb \citep{2015A&A...573A..42L}}

\end{acknowledgments}

\clearpage
\bibliography{apj-jour,reference}
\bibliographystyle{apj}


\end{document}